\documentclass{nature_max}

\usepackage{graphicx}
\usepackage{graphicx}
\usepackage{amsmath}
\usepackage{bm}
\usepackage{color}
\usepackage{mathbbol}
\usepackage{rotating}
\usepackage{caption}

\title{Carbon nanotubes as excitonic insulators}

\author{Daniele Varsano$^1$, Sandro Sorella$^2$,
        Davide Sangalli$^3$, Matteo Barborini$^{1,4}$, 
        Stefano Corni$^{1,5}$, Elisa Molinari$^{1,6}$
        \& Massimo Rontani$^1$}

\begin{document}

\maketitle

\begin{affiliations}
 \item CNR-NANO, Via Campi 213a, 41125 Modena, Italy.
 \item SISSA \& CNR-IOM Democritos, Via Bonomea 265, 34136 Trieste, Italy.
 \item CNR-ISM, Division of Ultrafast Processes in Materials (FLASHit),
       Area della Ricerca di Roma 1, 00016 Monterotondo Scalo, Italy.
 \item Present address: Physics \& Materials Science Research Unit,
       University of Luxembourg, 162a avenue de la Fa\"{\i}encerie,
       1511 Luxembourg, Luxembourg.
 \item Present address: Dipartimento di Scienze Chimiche, Universit\`a
       degli Studi di Padova, Via Marzolo 1, 35131 Padova, Italy.
 \item Dipartimento di Scienze Fisiche, Informatiche e Matematiche (FIM), 
       Universit{\`a} degli Studi di Modena e Reggio Emilia, 41125 Modena, 
       Italy.
\end{affiliations}

\begin{abstract}
  Fifty years ago Walter Kohn speculated that a zero-gap
  semiconductor might be unstable
  against the spontaneous generation of excitons---electron-hole
  pairs bound together by Coulomb attraction.
  The reconstructed ground state would then open a gap
  breaking the symmetry of the underlying
  lattice, a genuine consequence of electronic correlations.
  Here we show that this excitonic insulator is realized in
  zero-gap carbon nanotubes by performing first-principles
  calculations through many-body perturbation theory as well as
  quantum Monte Carlo.
  The excitonic order modulates the charge between
  the two carbon sublattices opening an
  experimentally observable gap, which scales as the inverse
  of the tube radius and weakly depends on the axial
  magnetic field.
  Our findings call into question the Luttinger liquid paradigm for
  nanotubes and provide tests to experimentally
  discriminate between excitonic and Mott insulator.
\end{abstract}

\newpage

Long ago Walter Kohn speculated that grey tin---a zero-gap 
semiconductor---could be unstable against the tendency of
mutually attracting electrons and holes to form bound pairs, 
the excitons\cite{Sherrington1968}.
Being neutral bosoniclike particles, the excitons
would spontaneously occupy the same macroscopic wave function,
resulting in a reconstructed insulating ground state with a
broken symmetry inherited from the exciton character\cite{Keldysh1964,Cloizeaux1965,Kohn1967,Halperin1968}.
This excitonic insulator (EI) would share intriguing 
similarities with the Bardeen-Cooper-Schrieffer (BCS) superconductor ground 
state\cite{Kohn1967,Lozovik1976,Portengen1996b,Balatsky2004,Rontani2005a,Su2008,Littlewood2008},
the excitons---akin to Cooper pairs---forming
only below a critical temperature and collectively enforcing
a quasiparticle gap. 
The EI was intensively sought after 
in systems as diverse as mixed-valence 
semiconductors and semimetals\cite{Bucher1991,Rontani2013}, 
transition metal chalcogenides\cite{DiSalvo1976,Rossnagel2011}, 
photoexcited semiconductors at quasi equilibrium\cite{Rice1977,Keldysh1995},
unconventional ferroelectrics\cite{Ikeda2005},
and, noticeably, semiconductor bilayers in the   
presence of a strong magnetic field that quenches the kinetic 
energy of electrons\cite{Spielman2000,Nandi2012}.
Other candidates include electron-hole bilayers\cite{DePalo2002,Kunes2015},
graphene\cite{Khveshchenko2001,Vafek2008,Drut2009,Gamayun2009}
and related two dimensional structures\cite{Lozovik2008,Dillenschneider2008,Min2008,Zhang2008,Rodin2013,Fogler2014,Stroucken2015}, 
where the underscreened Coulomb interactions might reach the 
critical coupling strength stabilizing the EI.  
Overall, the observation of the EI remains elusive.

Carbon nanotubes, which are rolled
cylinders of graphene whose low-energy electrons are
massless particles\cite{Dresselhaus1998,McEuen2010}, 
exhibit strong excitonic effects, due to 
ineffective dielectric screening and enhanced interactions 
resulting from one
dimensionality\cite{Ando1997,Maultzsch2005,Wang2005,Wang2007}.
As single tubes can be
suspended to suppress the effects of 
disorder and screening by the nearby substrate or 
gates\cite{Waissman2013,Laird2015,Aspitarte2017},
the field lines of Coulomb attraction between electron and hole
mainly lie unscreened in the vacuum (Fig.~1a).
Consequently, the
interaction is truly long-ranged and in principle---even for zero gap---able 
of binding
electron-hole pairs
close to the Dirac point in momentum space (Fig.~1b).
If the binding energy is finite,
then the ground state is unstable against the
spontaneous generation of excitons having negative excitation energy, 
$\varepsilon_{\text{u}}<0$.
This is the analogue of the Cooper instability that
heralds the transition
to the superconducting state---the
excitons replacing the Cooper pairs.

\begin{figure}[htbp]
\setlength{\unitlength}{1 cm}
\begin{center}
\includegraphics[trim=2cm 7.5cm 9cm 7.0cm,clip=true,width=16.4cm]{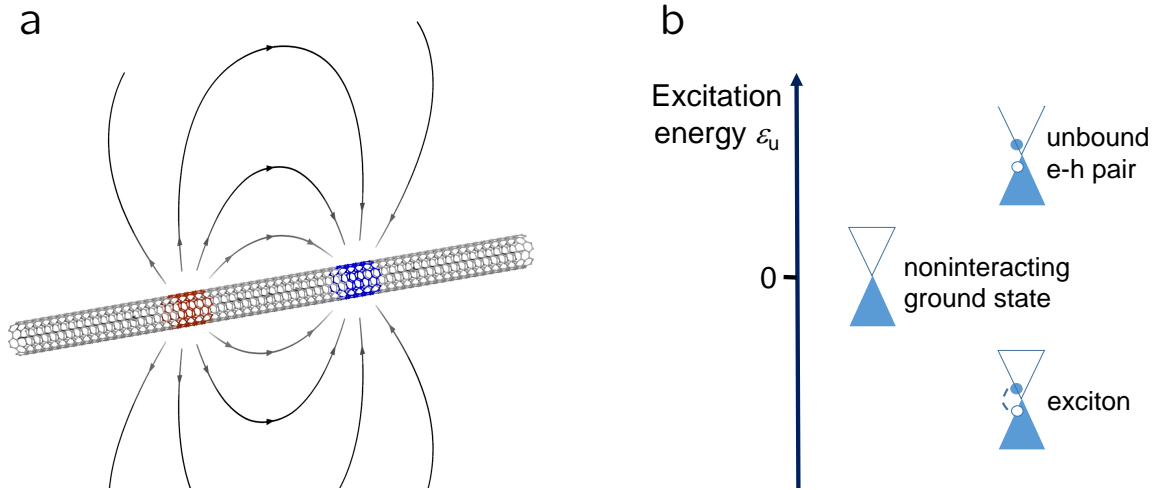}
\end{center}
\caption{{\bf Excitonic instability in carbon nanotubes.}
{\bf a,} Sketch of a suspended armchair carbon nanotube. 
The field lines of the Coulomb force between electron and hole 
lie mainly in the vacuum, hence screening is heavily suppressed.
{\bf b,} Excitonic instability in the armchair carbon nanotube. 
The scheme represents the excitation energy $\varepsilon_{\text{u}}$
of an electron-hole (e-h) pair relative to                
the noninteracting ground state, 
a zero-gap semiconductor.
In the absence of interaction, the 
excitation energy $\varepsilon_{\text{u}}$ 
of an e-h pair is positive.
The long-range interaction may bind
e-h pairs close to the Dirac point in momentum space.
If an exciton forms, then its excitation energy $\varepsilon_{\text{u}}$ is
negative. This instability leads to the reconstruction
of the ground state into an excitonic insulator.
}
\end{figure}

Here we focus on the armchair family of zero-gap carbon nanotubes,
because symmetry prevents their gap
from opening as an effect of curvature or bending\cite{Charlier2007}.
In this paper we show that armchair tubes are predicted to
be EIs by first-principles calculations.
The problem is challenging, because the key quantities 
controlling this phenomenon---energy band differences and 
exciton binding energies---involve many-body corrections beyond 
density functional theory that are of the order of a few meV, 
which is close to the limits of currently available methods. 
In turn, such weak exciton binding reflects in the extreme
spatial extension of the exciton wave function, hence its 
localization in reciprocal space requires very high sampling accuracy. 
To address these problems, we perform state-of-the-art many-body 
perturbation theory calculations within the $GW$ and Bethe-Salpeter 
schemes\cite{Onida2002}. We find that bound excitons exist in 
the (3,3) tube with finite negative excitation energies. 
We then perform unbiased quantum Monte Carlo 
simulations\cite{Foulkes2001} to prove that the reconstructed
ground state is the EI, its signature being the broken symmetry
between inequivalent carbon sublattices---reminescent of the exciton
polarization.
Finally, to investigate the trend with the size of the system,
which is not yet in reach of first-principles calculations, 
we introduce an effective-mass model, which shows   
that both EI gap and critical temperature fall in the meV range
and scale with the inverse of the tube radius. 
Our findings are in contrast with the widespread belief
that electrons in undoped armchair tubes form a Mott insulator---a 
strongly correlated Luttinger liquid\cite{Balents1997,Kane1997b,Egger1997,Krotov1997,Yoshioka1999,Nersesyan2003,Chen2008}.
We discuss the physical origin of this conclusion and
propose independent experimental tests to discriminate
between excitonic and Mott insulator.

\begin{figure}[htbp]
\setlength{\unitlength}{1 cm}
\begin{center}
\includegraphics[trim=2cm 8cm 9cm 8cm,clip=true,width=16.4cm]{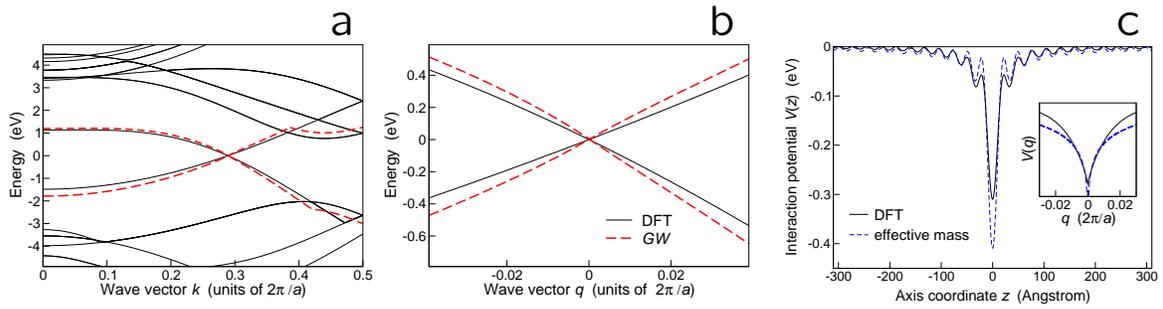}
\end{center}
\caption{{\bf Electronic properties 
from many-body perturbation theory.} 
{\bf a,} $GW$ (dashed lines) and
DFT (solid lines) band structure of the armchair carbon nanotube (3,3).
{\bf b,} Zoom close to the Dirac point K.
The momentum $q$ is referenced from K.
{\bf c,} Long-range part of electron-hole 
interaction $V(z)$ along the tube axis according
to: DFT (solid line), effective-mass model (dashed line).
Inset: interaction $V(q)$ in momentum space.
$V$ is integrated over the mesh of the
$q$ grid and projected onto the conduction and 
valence bands shown in panel b, 
with $\left|q\right|< 0.09 (2\pi)/a$.
The graphene lattice constant is $a=2.46$ \AA.
}
\end{figure}

\section*{Results}

\section*{Exciton binding and instability}

For the sake of computational convenience
we focus on the smallest (3,3) armchair tube, 
which was investigated several times from first principles\cite{Liu2002,Machon2002,Cabria2003,Spataru2004,Bohnen2004,Connetable2005,Dumont2010,Lu2013}. 
We first check whether the structural optimization of the tube  
might lead to deviations from the ideal cylindrical shape,
affecting the electronic states. 
Full geometry relaxation  
(Methods) yields an equilibrium structure with 
negligible corrugation. Thus, contrary to a previous   
claim\cite{Lu2013}, corrugation cannot be responsible of gap opening. 
We find that the average length of C-C bonds along the tube axis,
1.431 \AA, is shorter than around the circumference, 1.438 \AA,
in perfect agreement with the literature\cite{Liu2002}.

We use density functional theory (DFT) to compute the band structure
(solid lines in Fig.~2a), which provides the expected\cite{Charlier2007}
zero gap at the Dirac point K. 
In addition, we adopt the $G0W0$ approximation 
for the self-energy operator\cite{Onida2002} to evaluate 
many-body corrections to Kohn-Sham eigenvalues.
The highest valence and lowest conduction bands are shown as dashed lines. 
The zoom near K (Fig.~2b) shows that electrons remain massless,
with their bands stretched by $\sim$ 28\% with respect to DFT 
(farther from K the stretching is $\sim$ 13\%, 
as found previously\cite{Spataru2004}).
Since electrons and holes in these bands have linear dispersion, they cannot
form a conventional Wannier exciton, whose binding energy is proportional 
to the effective mass. However, the screened
e-h Coulomb interaction $V(z)$ along the tube axis, projected onto the 
same bands, has long range (Fig.~2c)---a remarkable effect 
of the topology of the tube holding even for vanishing gap.
Consequently, $V(q)$ exhibits a singularity in reciprocal space
at $q=0$ (smoothed by numerical discretization in the inset of Fig.~2c), 
which eventually binds the exciton.

\begin{figure}[htbp]
\setlength{\unitlength}{1 cm}
\begin{center}
\includegraphics[trim=2cm 7.0cm 9cm 5.5cm,clip=true,width=16.4cm]{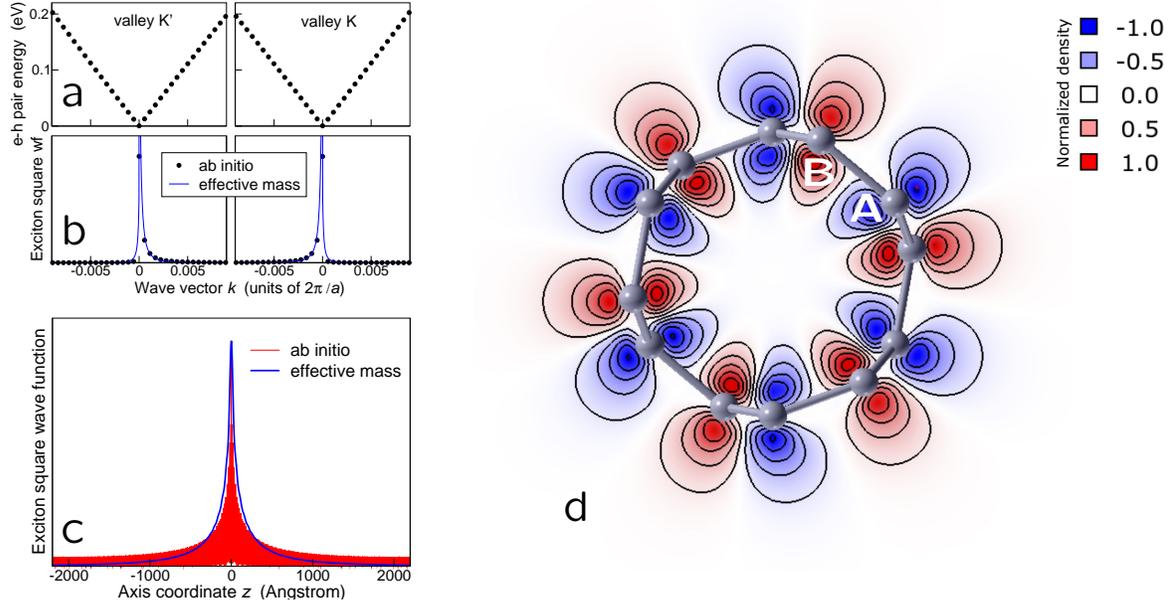}
\end{center}
\caption{{\bf Wave function of the lowest-energy 
exciton of the (3,3) tube.} 
{\bf a,} $GW$ excitation spectrum of 
free e-h pairs with zero center-of-mass momentum 
in the two Dirac valleys.
{\bf b,} Square modulus of the triplet 
exciton wave function vs momentum $k$. 
Both first-principles (dots) and effective-mass (solid lines)
probability weights 
accumulate asymmetrically close to Dirac points.
The effective-mass model includes the dressed long-range interaction,
the short-range intervalley exchange,
and the small asymmetry of 
Dirac cones (cf.~Supplementary Notes 1-3; a previous phenomenological
theory\cite{Rontani2014} by one of the authors,  
which ignored the key-role of long-range interaction,
is ruled out by the present work).  
{\bf c,} Square modulus of the triplet exciton wave function 
vs e-h distance along the axis, $z$,
according to first-principles (red curve) and effective-mass 
(blue curve) calculations. 
The Bohr diameter is larger than 2 $\mu$m.
{\bf d,} Cross-sectional contour map of the 
transition density of the singlet exciton,
$\varrho_{\text{tr}}(\mathbf{r})$, obtained from first principles.
The blue / red colour points to the deficit / surplus of charge,
the isolines are equally spaced, the
normalization of $\varrho_{\text{tr}}(\mathbf{r})$
is such that its maximum value is one, and
letters label sublattices.
}
\end{figure}

We solve the Bethe-Salpeter equation over an ultradense grid of  
1800 $k$-points, which is computationally very demanding but essential 
for convergence.
We find several excitons with negative excitation energies 
$\varepsilon_{\text{u}}$, 
in the range of 1--10 meV (Table 1).
\begin{table}[htbp]
\begin{center}
\begin{tabular}{ c c c }
\hline
    & Triplet & Singlet \\
\hline
Lowest      &  -7.91  &  -6.10  \\
1st excited &  -6.40     &  -5.10  \\
2st excited &   6.65    &  8.82  \\
\hline
\end{tabular}
\end{center}
\caption{
{\bf Excitation energies $\varepsilon_{\text{u}}$ of low-lying excitons 
of the (3,3) tube obtained from first-principles many-body 
perturbation theory in units of meV.}}
\end{table}
The exciton spectral weight is concentrated 
in a tiny neighbourhood of K and K$'$ points in reciprocal 
space (Fig.~3b), hence the excitons are extremely shallow, spread over
microns along the axis (Fig.~3c). 
Only e-h pairs with negative $k$
in valley K and 
positive $k$ in valley K$'$ 
contribute to the exciton wave function,
which is overall symmetric under time reversal but not under 
axis reflection within one valley,
$k\rightarrow -k$,
as shown in Fig.~3b (the axis origin is at Dirac point). 
On the contrary, the wave 
functions of excitons reported so far in
nanotubes\cite{Ando1997,Spataru2004,Maultzsch2005}
are symmetric in $k$-space.
The reason of this unusual behavior originates from 
the vanishing energy gap, since then e-h pairs cannot be backscattered
by Coulomb interaction due to the orthogonality of initial 
and final states\cite{Ando1998b}. 
In addition, pair energies are not degenerate for $k\rightarrow -k$,
as Dirac cones are slightly asymmetric
(Supplementary Discussion and Supplementary Fig.~10).

The exciton with the lowest negative $\varepsilon_{\text{u}}$ 
makes the system unstable against the EI. 
The transition density, $\varrho_{\text{tr}}(\mathbf{r})
=\left<\text{u}\right|\hat{\varrho}
(\mathbf{r})\left|0\right>$,
hints at the broken symmetry 
of the reconstructed ground state, as
it connects the noninteracting ground state, $\left|0\right>$, 
to the exciton state, $\left|\text{u}\right>$, 
through the charge fluctuation
operator $\hat{\varrho}$ (Fig.~3d).
Here we focus on the simpler charge order (spin singlet excitons)
and neglect magnetic phenomena (spin triplet), as the only 
relevant effect of spin-orbit coupling in real 
tubes\cite{Kuemmeth2008,Steele2013}
is to effectively mix both symmetries.
Figure 3d may be regarded as a snapshot 
of the polarization charge oscillation 
induced by the exciton, breaking the inversion symmetry between  
carbon sublattices A and B. 
Note that this
originates from the opposite symmetries of
$\left|0\right>$ and $\left|\text{u}\right>$ under A $\leftrightarrow$ B 
inversion and not from the
vanishing gap. This charge displacement between
sublattices is the generic signature of the EI, as 
its ground state may be regarded as a BCS-like condensate of 
excitons $\left|\text{u}\right>$ 
(see the formal demonstration in Supplementary Note 5).

\section*{Broken symmetry of the excitonic insulator}

We use quantum Monte Carlo to verify the excitonic nature
of the many-body ground state, by
defining an order parameter characteristic of the EI, $ \varrho_{\text{AB}}$.
In addition, we  
introduce an alternative order parameter, 
$\varrho_{\text{Transl}}$, peculiar to a dimerized charge density wave 
(CDW) similar to the Peierls CDW
predicted by some authors\cite{Bohnen2004,Connetable2005,Dumont2010}
for the smallest armchair tubes. 
The EI order parameter measures the uniform charge displacement between
A and B sublattices,
$ \varrho_{\text{AB}}=  ( \sum_{i\in \text{A}}  
n_i -\sum_{i\in \text{B}} n_i ) / N_{\text{atom}} $,
whereas $\varrho_{\text{Transl}}$ detects any deviation
from the periodicity of the undistorted structure by evaluating the
charge displacement between adjacent cells,
$\varrho_{\text{Transl}}=  
\sum_i  n_i (-1)^{i_z} / N_{\text{atom}} $ (Figs.~4b-e).
Here the undistorted structure is made of a unit cell of
twelve C atoms repeated along the $z$ direction with a period of
2.445 {\AA} and labeled by the integer $i_z$,
$n_i$ is the operator counting the 
electrons within a sphere of radius 
1.3 a.u.~around the $i$th atom, and $N_{\text{atom}}$ is the total number
of atoms in the cluster. 
Both order parameters
$ \varrho_{\text{AB}}$ and $\varrho_{\text{Transl}}$ vanish
in the symmetric ground state of the undistorted structure,
which is invariant under sublattice-swapping inversion 
and translation symmetries. 

\begin{figure}[htbp]
\setlength{\unitlength}{1 cm}
\begin{center}
\includegraphics[trim=2cm 7.5cm 9cm 7.5cm,clip=true,width=16.4cm]{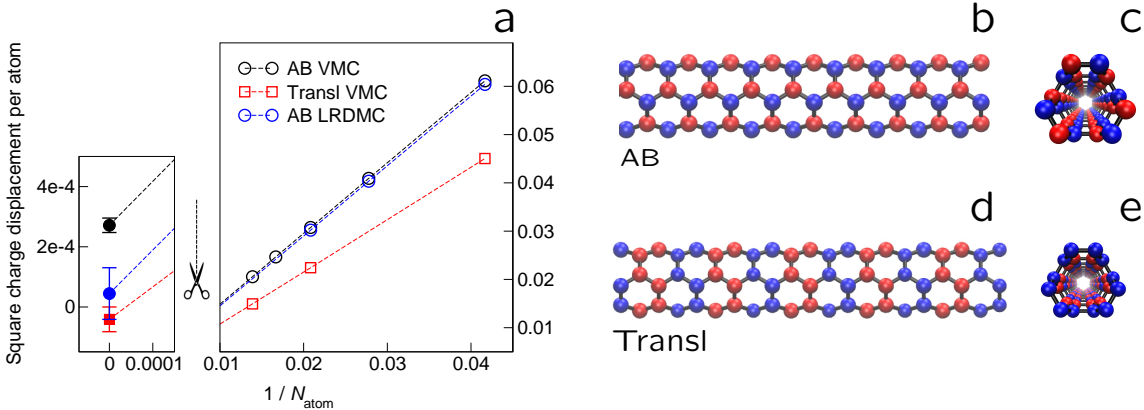}
\end{center}
\caption{{\bf Broken symmetry of the ground state 
from quantum Monte Carlo.}
{\bf a,} The square of the charge
displacement per atom (empty circles and squares for `AB' and
`Transl' order parameters, respectively)
is plotted vs the inverse of the number of atoms, $N_{\text{atom}}$, 
as obtained by variational (VMC)
and lattice-regularized diffusion (LRDMC) quantum Monte Carlo.
The filled symbols are linear extrapolations to the $N_{\text{atom}}=
\infty$ limit. The error bars 
are estimated by means of the  
jackknife method using more than 30 independent samples for each 
independent twist (Methods).
The error bars of empty symbols are
not visible on the scale of the plot.
{\bf b-e,} The sketches of the tube  
illustrate the two possible broken symmetries, with the blue / red colour
pointing to the deficit / surplus of charge.
The AB order parameter, peculiar to the EI, is a 
uniform charge displacement between the two carbon sublattices (panels b
and c show respectively the lateral and cross-sectional views of the tube).
The Transl parameter is a charge displacement between two
adjacent unit cells, signaling a charge density wave order breaking the 
translational symmetry (panels d and e).
}
\end{figure}

We then perform variational Monte Carlo (VMC), 
using a correlated Jastrow-Slater ansatz that has
proved\cite{Capello2005} 
to work well in 1D correlated systems (Methods),
as well as it is able to recover the 
excitonic correlations present in the mean-field EI wave 
function\cite{Keldysh1964,Cloizeaux1965,Kohn1967,Halperin1968}
(Supplementary Discussion).
We plot VMC order parameters in Fig.~4a.
Spontaneously broken symmetry occurs in the thermodynamic limit 
if the square order parameter, 
either $\varrho_{\text{AB}}^2$ or 
 $\varrho_{\text{Transl}}^2$,  
scales as $1/N_{\text{atom}}$
and has a non vanishing limit value for
$N_{\text{atom}}\rightarrow \infty$. 
This occurs for $\varrho_{\text{AB}}^2$ (black circles in Fig.~4a),  
confirming the prediction of the EI,
whereas $\varrho_{\text{Transl}}^2$ vanishes (red squares),
ruling out the CDW instability
(see Supplementary Discussion as well as the
theoretical literature\cite{Bohnen2004,Connetable2005,Chen2008,Dumont2010}
for the Peierls CDW case).
We attribute the simultaneous
breaking of sublattice symmetry and protection
of pristine translation symmetry 
to the effect of long-range interaction. 

The vanishing of $\varrho_{\text{Transl}}$ validates the ability
of our finite-size scaling analysis to discriminate
between kinds of order in the bulk.
Though the value of $\varrho_{\text{AB}}$
after extrapolation is small,
$\varrho_{\text{AB}}=0.0165 \pm 0.0007$, it is non zero within
more than twenty standard deviations.
Besides, the quality of the fit of Fig.~4a appears good, because
the data for the five largest clusters are compatible with the
linear extrapolations of both $\varrho_{\text{AB}}^2$ and
$\varrho_{\text{Transl}}^2$ within
an acceptable statistical error.
The more accurate diffusion Monte Carlo (LRDMC) values 
(obtained with the lattice regularization), 
shown in Fig.~4a as blue circles,
confirm the accuracy of the variational calculation. 
However, as their cost is on the verge 
of present supercomputing capabilities,
we were unable to treat clusters larger that $N_{\text{atom}}=48$,
hence the statistical errors are too large to support 
a meaningful non zero value in the 
thermodynamic limit. Nevertheless,  
we obtain a non zero LRDMC value smaller   
than the one estimated by VMC but compatible with it
within a few standard deviations. 

\section*{Trends}

As the extension of our 
analysis to systems larger than the (3,3) tube is beyond reach,  
we design an effective-mass theory 
to draw conclusions about trends in the armchair tube family,
in agreement with 
first-principles findings.
We solve the minimal Bethe-Salpeter equation for the massless 
energy bands $\varepsilon(k)=\pm \gamma \left|k\right|$ 
(Fig.~2b and Supplementary Note 1) and the 
long-range Coulomb interaction $V(q)$,
the latter diverging logarithmically in one dimension
for small momentum transfer $q$, 
$V(q)=(2e^2/A\kappa_{\text{r}})\ln(\left|q\right|R)$ 
(inset of Fig.~2c and Supplementary Note 2).
Here $\gamma$ is graphene tight-binding parameter including 
$GW$ self-energy corrections, 
$k$ is the wave vector along the axis, $A$ is the tube length, 
$R$ is the radius, and $\kappa_{\text{r}}$ 
accounts for screening beyond the effective-mass approximation.
By fitting the parameters $\gamma=0.5449$ eV$\cdot$nm and 
$\kappa_{\text{r}}=10$ to
our first-principles data, we obtain a 
numerical solution of Bethe-Salpeter equation recovering 
approximately 60\% of the lowest exciton energy $\varepsilon_{\text{u}}$
reported in Table 1 (Supplementary Note 3). 
Moreover, the wave function agrees with the 
one obtained from first principles (Fig.~3b, c). Importantly,
$\varepsilon_{\text{u}}$ smoothly converges in an energy
range that---for screened interaction---is significantly smaller 
than the extension of the Dirac cone,
with no need of ultraviolet cutoff (Supplementary Fig.~9).
Therefore, 
the exciton has an intrinsic length (binding energy), 
which scales like $R$ ($1/R$).

\begin{figure}[htbp]
\setlength{\unitlength}{1 cm}
\begin{center}
\includegraphics[trim=2cm 8.5cm 9cm 7.5cm,clip=true,width=16.4cm]{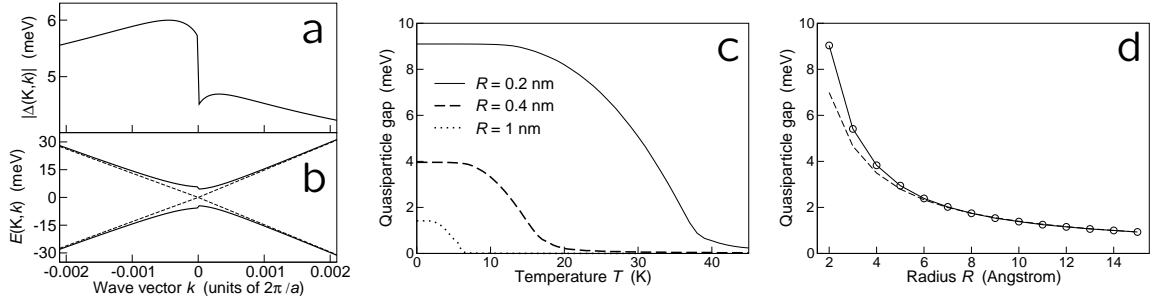}
\end{center}
\caption{{\bf Excitonic insulator behaviour from mean-field theory.}
{\bf a,} Excitonic order parameter, $\left|\Delta(\tau=\text{K},k)
\right|$, vs momentum $k$ within K valley and
({\bf b}) corresponding quasiparticle dispersion, $E(\text{K},k)$,
for the (3,3) armchair carbon nanotube. The data are derived 
by solving self-consistently the gap equation. 
For comparison, the noninteracting bands are indicated
(dashed lines).
The band in the K$'$ valley is obtained by time reversal, as
$\left|\Delta(\text{K}',k)\right|= \left|\Delta(\text{K},-k)\right|$.
{\bf c,} Quasiparticle gap vs temperature $T$ for different radii
[for the (3,3) tube $R=2$ {\AA}].
{\bf d,} Quasiparticle gap vs $R$. 
The dashed curve is a fit proportional to $1/R$ pointing to
the scaling behaviour at large $R$.
}
\end{figure}

We adopt a mean-field theory of the EI as we expect 
the long-range character of excitonic correlations 
to mitigate the effects of quantum fluctuations.
The EI wave function can be described as
\begin{equation}
\left|\Psi_{\text{EI}}\right>=\prod_{\sigma\sigma'\tau k}\left[u_{\tau k}
+ \chi_{\sigma\sigma'} v_{\tau k} e^{i \eta} \,
\hat{c}^{\tau +}_{k,\sigma}
\hat{v}^{\tau }_{k,\sigma'} \right] \left| 0 \right>.
\label{eq:EI}
\end{equation}
Here $\left| 0 \right>$ is the zero-gap ground state
with all valence states filled and conduction states empty,
the operator $\hat{c}^{\tau +}_{k,\sigma}$ ($\hat{v}^{\tau +}_{k,\sigma}$)
creates an electron in the conduction (valence) band 
with wave vector $k$, spin
$\sigma$, valley $\tau=$ K or K$'$,
$\eta$ is an arbitrary phase,
and the $2\times 2$ matrix $\chi_{\sigma\sigma'}$
discriminates between
singlet and triplet spin symmetries of the e-h pair
$\hat{c}^{\tau +}_{k,\sigma}
\hat{v}^{\tau }_{k,\sigma'}\left| 0 \right>$ (Fig.~1b).
The positive variational quantities $u_{\tau k}$
and $v_{\tau k}$ are the population amplitudes
of valence and conduction levels, respectively,
with $u_{\tau k}^2
+ v_{\tau k}^2=1$.
Whereas in the zero-gap state $u_{\tau k}=1$ and $v_{\tau k}=0$,
in the EI state both $u_{\tau k}$ and $v_{\tau k}$ are finite 
and ruled by the EI order parameter $\Delta(\tau k)$,
according to 
$u_{\tau k}\,v_{\tau k}=\left|\Delta(\tau k)\right|/2E(\tau k)$,
with $E(\tau k)=[\varepsilon(\tau k)^2 + \left|\Delta(\tau k)\right|^2]^{1/2}$.
The parameter $\Delta(\tau k)$ obeys the self-consistent equation
\begin{equation}
\left|\Delta(\tau k)\right|=\sum_{\tau' q} 
V^{\tau\tau'}\!(k,k+q)
\,u_{\tau' k+q}\,v_{\tau' k+q},
\end{equation}
which is solved numerically by recursive iteration
(here $V$ includes both long- and short-range interactions 
as well as form factors, see Supplementary Note 4).
As shown in Fig.~5a, in each valley $\left|\Delta(\tau k)\right|$ 
is asymmetric around the Dirac point, 
a consequence of the peculiar character
of the exciton wave function of Fig.~3b.
The electrons or holes added to the neutral ground state 
are gapped quasiparticle excitations of the EI,
whose energy bands $\pm E(\tau k)$ are shown in Fig.~5b.
The order parameter at the Dirac point, 
$\left|\Delta (\tau, k=0)\right|$, is half the many-body gap.
This gap is reminescent of the exciton binding energy,
since in the ground state all electrons and holes
are bound, so one needs to ionize an exciton-like collective state 
to create a free electron-hole pair.   
The gap strongly depends on temperature, 
with a low-temperature plateau, a steep 
descent approaching the critical temperature, and a milder tail
(Fig.~5c).
The gap approximately scales as $1/R$ for different tubes (circles
in Fig.~5d): 
whereas at large $R$ such scaling is exact
(cf.~dashed curve), 
at small $R$ the gap is enhanced by
short-range intervalley interaction (the decay of $\Delta$ 
will be mitigated if $\kappa_{\text{r}}$ is sensitive to $R$).

In experiments, many-body gaps are observed in
undoped, ultraclean suspended tubes\cite{Deshpande2009}, whereas
Luttinger liquid signatures emerge in doped tubes\cite{Charlier2007,McEuen2010}. 
Though it is difficult to compare with the measured 
many-body gaps\cite{Deshpande2009}, 
as the chiralities of the tubes are unknown
and the radii estimated indirectly, 
the measured range of 10--100 meV
is at least one order of magnitude larger than our predictions.
By doping the tube, we expect that the enhanced screening  
suppresses the EI order,
quickly turning the system into a Luttinger liquid. We are confident that 
advances in electron spectroscopies will allow
to test our theory. 

The broken symmetry associated with the EI ground
state depends on the exciton spin\cite{Halperin1968}. 
For spin singlet 
($\chi_{\sigma\sigma'}=\delta_{\sigma\sigma'}$)
and order parameter real  
($\eta=0,\pi$), $\left|\Psi_{\text{EI}} \right>$ breaks
the charge symmetry between A and B carbon sublattices.
The charge displacement per electron, $\Delta e/e$, 
at each sublattice site  is 
\begin{equation}
\frac{\Delta e}{e} = \pm \cos{\eta}\frac{a}{A}
\sum_{\tau k}\frac{\left|\Delta(\tau k)\right|}{2E(\tau k)},
\label{eq:chargeunb}
\end{equation}
where the positive (negative) sign refers to the A (B) sublattice
(Supplementary Note 6).
For the (3,3) tube
this amounts to $\varrho_{\text{AB}} = 0.0068$,
which compares well with Monte Carlo estimates of 0.0067 and 0.0165
from LRDMC and VQMC, respectively. Note that
assessing the energy difference between EI and zero-gap
ground states is beyond the current capability
of quantum Monte Carlo: the mean-field estimate of the
difference is below 10$^{-6}$
Hartree per atom, which is less than the noise threshold of the 
method (10$^{-5}$ Hartree per atom).

\begin{figure}[htbp]
\setlength{\unitlength}{1 cm}
\begin{center}
\includegraphics[trim=2cm 7.5cm 9cm 7.5cm,clip=true,width=16.4cm]{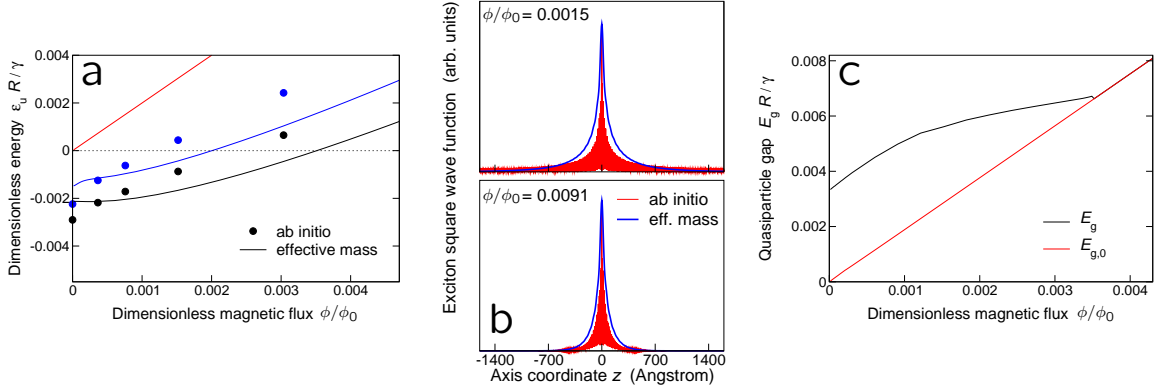}
\end{center}
\caption{{\bf Effect of an axial magnetic field.}
{\bf a,} Excitation energies, $\varepsilon_{\text{u}} R/\gamma$,
of low-lying excitons vs magnetic flux, $\phi / \phi_0$.
Both first-principles (dots) and effective-mass 
(solid lines) data are reported. 
The black (blue) colour labels the 
triplet (singlet) spin symmetry. The red 
line is the noninteracting gap 
and the dashed line is the instability threshold. 
{\bf b,} Square modulus of the wave function of the lowest exciton 
vs e-h distance along the axis, $z$, for increasing values of
magnetic flux. Both ab initio (red lines)
and effective-mass (blue lines) data are reported.
{\bf c,}
Total quasiparticle gap $E_{\text{g}} R/\gamma$ vs $\phi / \phi_0$. 
This observable may be accessed through Coulomb blockade spectroscopy.
The red line is the noninteracting gap, $E_{\text{g},0}$.
}
\end{figure}

\section*{Effect of magnetic field}

The EI is sensitive to the opening of a noninteracting gap,
$E_{\text{g},0}$, tuned by the magnetic field parallel to the tube axis, $B$.
The ratio of the flux piercing the cross section,
$\phi=\pi R^2B$, to the flux 
quantum, $\phi_0=ch/e$, amounts 
to an Aharonov-Bohm phase displacing the position of
the Dirac point along the transverse direction\cite{Ajiki1993},
$k_{\perp}=(\phi/\phi_0)R^{-1}$. 
Consequently, 
$E_{\text{g},0}=2\gamma\left|k_{\perp}\right|$
is linear with $\phi/\phi_0$ (red line in Fig.~6a, c). 
Figure 6a shows the evolution of 
low-lying singlet (blue lines) and triplet (black lines) 
excitons of the (3,3) tube.
In addition, we have implemented  
a full first-principles description of $B$ building on 
a previous method\cite{Sangalli2011}. 
First-principles (circles)
and model (solid lines) calculations show a fair agreement, which
validates the effective-mass theory 
since all free parameters
have been fixed at zero field.
Here we rescale energies by $R/\gamma$ since we expect the plot 
to be universal, except for small corrections due to
short-range interactions.
Excitation energies obtained within the effective-mass model crossover
from a low-field region, where $\varepsilon_{\text{u}}$ is almost constant, 
to a high-field region, where $\varepsilon_{\text{u}}$ increases linearly
with $\phi/\phi_0$.
Exciton wave functions are effectively squeezed by the field in
real space (Fig.~6b), whereas in reciprocal space they loose
their asymmetric character: the amplitudes become evenly distributed 
around the Dirac points (Supplementary Discussion and Fig.~11) 
and similar to those reported in 
literature\cite{Ando1997,Spataru2004,Maultzsch2005}. 
At a critical flux $\phi_{\text{c}}/\phi_0\approx 0.035$           
the excitation energy $\varepsilon_{\text{u}}$
becomes positive, hence the tube exits the EI phase and $\Delta$ vanishes
in a BCS-like fashion. 
We point out that the critical field intensity,
$B_{\text{c}} \approx 460$ T $\cdot (R$ [{\AA}]$)^{-2}$, is out of reach
for the (3,3) tube but feasible for larger tubes.
The total transport gap, $E_{\text{g}} = (E_{\text{g},0}^2 +
4\left|\Delta\right|^2)^{1/2}$, first scales with $\phi/\phi_0$
as $E_{\text{g},0}$, then its slope decreases
up to the critical threshold $\phi_{\text{c}}/\phi_0$, 
where the linear dependence
on $\phi/\phi_0$ is restored (Fig.~6c). 
This behaviour is qualitatively
similar to that observed by Coulomb blockade spectroscopy
in narrow-gap tubes close to the 
`Dirac' value of $B$, which counteracts the effect of
$E_{\text{g},0}$ on the transport gap, fully suppressing the noninteracting 
contribution\cite{Deshpande2009}.

\section*{Discussion}

The observed\cite{Deshpande2009} many-body gap of armchair tubes 
was attributed to the Mott insulating state. The  
system was modeled as a strongly interacting
Luttinger liquid with a gap enforced by 
short-range interactions\cite{Balents1997,Krotov1997}, 
whereas the long tail of the interaction was cut off 
at an extrinsic, setup-dependent 
length\cite{Kane1997b,Egger1997,Yoshioka1999,Nersesyan2003,Chen2008}.
This model thus neglects the crucial effect of long-range 
interaction, which was highlighted in Fig.~1:
Were any cutoff length smaller than the intrinsic exciton length,  
which is micrometric and scales with $R$, excitons could not bind. 

Whereas armchair carbon nanotubes are regarded as quintessential realizations
of the Luttinger liquid, since their low-energy properties
are mapped into those of two-leg ladders\cite{Balents1997}, 
we emphasize that
this mapping is exact for short-range interactions only. 
Among e-h pair collective modes with total 
momentum $q = 0$, Luttinger liquid theory 
routinely describes plasmons\cite{Schulz1993} but not excitons.
Contrary to conventional wisdom, armchair tubes are EIs.

The excitonic and Mott insulators are qualitatively different. 
The EI exhibits long-range charge order, which does not affect
the translational symmetry of the zero-gap tube.
In the Mott insulator, charge and spin correlations 
may or may not decay, but always
add a $2\pi/(2k_{\text{F}})$ [or $2\pi/(4k_{\text{F}})$] periodicity to the
pristine system, $k_{\text{F}}$ being the Fermi 
wave vector\cite{Yoshioka1999,Nersesyan2003}.
The EI gap scales like $1/R$ (Fig.~5d), 
the Mott gap like $1/R^{1/(1-g)}$, with 
predicted\cite{Kane1997b,Yoshioka1999,Nersesyan2003,Chen2008}
values of $g$ pointing to a faster decay, $g<1$.
The EI order parameter is suppressed at high temperature 
(Fig.~5c) and strong magnetic field (Fig.~6c); 
the Mott gap is likely independent of both fields  
(the Aharonov-Bohm phase does not affect Hubbard-like Coulomb integrals).
Importantly, the EI gap is very sensitive to the dielectric
environment\cite{Zittartz1968a}, whereas the Mott gap is not.
This could explain the dramatic variation of narrow transport gaps 
of suspended tubes submerged in different
liquid dielectrics\cite{Aspitarte2017}. 

We anticipate that armchair tubes exhibit 
an optical absorption spectrum in the THz range
dominated by excitons, which provides an independent test of the EI phase.
Furthermore, we predict they
behave as `chiral electronic ferroelectrics', displaying a
permanent electric polarization $\mathbf{P}$ of
purely electronic origin\cite{Portengen1996b}, whereas
conventional ferroelectricity originates from ionic
displacements.
In fact, the volume average of $\mathbf{P}$ is zero but its circulation
along the tube circumference is finite. Therefore,
a suitable time-dependent field 
excites the ferroelectric resonance\cite{Portengen1996b}
associated with the oscillation of $\mathbf{P}$.
The special symmetry of armchair tubes\cite{Ando1998b} is expected
to protect this collective (Goldstone) mode of
oscillating electric dipoles from phase-locking mechanisms.
The resulting soft mode---a displacement current along the tube
circumference---is a manifestation of 
the long-debated\cite{Kohn1970,Guseinov1973,Lozovik1976,Portengen1996b,Rontani2005a,Balatsky2004,Su2008,Littlewood2008} exciton superfluidity.

In conclusion, our calculations demonstrated that
an isolated armchair carbon nanotube at charge
neutrality is an excitonic insulator, owing to the strong e-h
binding in quasi-1D, and the almost unscreened long-range 
interactions. The emergence
of this exotic state of matter, 
predicted fifty years ago,
does not fit the common picture of
carbon nanotubes as Luttinger liquids.
Our first-principles calculations
provide tests to discriminate between the excitonic insulator
and the Luttinger liquid
at strong coupling, the Mott insulator state.
We expect a wide family of narrow-gap carbon nanotubes
to be excitonic insulators.
Carbon nanotubes are thus invaluable systems
for the experimental investigation of this phase of matter.

\begin{methods}

\subsection{Many-body
perturbation theory from first principles.}

The ground-state calculations for the
(3,3) carbon nanotube were performed by using a DFT approach,
as implemented in the Quantum ESPRESSO package\cite{Giannozzi2009}.
The generalized gradient approximation (GGA) PW91
parametrization\cite{Perdew1992}
was adopted together with plane wave basis set and norm-conserving
pseudopotentials to model the electron-ion interaction. The kinetic
energy cutoff for the wave functions was set to 70 Ry.
The Brillouin zone was sampled by using a 200 $\times$ 1 $\times$
1 $k$-point grid.
The supercell side perpendicular to the tube was set to 38 Bohr
and checked to be large enough to avoid spurious interactions
with its replica.

Many-body perturbation theory\cite{Onida2002} calculations
were performed using the Yambo code\cite{Marini2009}.
Many-body corrections to the Kohn-Sham eigenvalues were
calculated within the $G0W0$ approximation to the self-energy operator,
where the dynamic dielectric function was obtained within the
plasmon-pole approximation.
The spectrum of excited states was then computed by solving the
Bethe-Salpeter equation (BSE). The static screening in the direct
term was calculated within the random-phase approximation with
inclusion of local field effects; the Tamm-Dancoff approximation
for the BSE Hamiltonian was employed after having verified that
the correction introduced by coupling the resonant and antiresonant
part was negligible.
Converged excitation energies, $\varepsilon_{\text{u}}$, were obtained
considering respectively 3 valence and 4 conduction bands in the
BSE matrix.
For the calculations of the $GW$ band structure and the
Bethe-Salpeter matrix the Brillouin zone was sampled with a
1793 $\times$ 1 $\times$ 1 $k$-point grid.
A kinetic energy cutoff of 55 Ry was used for the evaluation of
the exchange part of the self energy and 4 Ry for the screening
matrix size. Eighty unoccupied bands were used in the integration
of the self-energy.

The effect of the magnetic field parallel to the axis
on the electronic structure of the nanotube ground state
(eigenvalues and eigenfunctions) was investigated following the
method by Sangalli \& Marini\cite{Sangalli2011}. For each value of the field,
the eigenvalues and eigenfunctions were considered to build
the screening matrix and the corresponding excitonic Hamiltonian.

To obtain the equilibrium structure, we first considered possible
corrugation effects. We computed the total energy for a set of
structures obtained by varying the relative positions of A and B carbon
atoms belonging to different sublattices, so that they were
displaced one from the other
along the radial direction by the corrugation length $\Delta$
and formed two cylinders,
as in Fig.~1(b) of Lu {\it et al.}\cite{Lu2013}.
Then, we fitted the total energy per carbon atom
with an elliptic paraboloid in the two-dimensional parameter space
spanned by $\Delta$ and the carbon bond length. In agreement with
Lu {\it et al.}\cite{Lu2013}, we find a corrugated structure with a bond length
of 1.431 {\AA} and a corrugation parameter $\Delta = 0.018$ {\AA}.
Eventually, starting from this structure, we performed a full geometry
relaxation of the whole system allowing all carbon positions to change
until the forces acting on all atoms became less than 5$\cdot 10^{-3}$
eV$\cdot${\AA}$^{-1}$. 
After relaxation, the final structure presents a negligible
corrugation ($\Delta < 10^{-5}$ {\AA}) and an average length of C-C bonds along the tube axis, 1.431 {\AA}, slightly shorter than the C-C bonds
around the tube circumference, 1.438 {\AA}. The average radius and
translation vector of the tube are respectively 2.101 {\AA} and
2.462 {\AA}, in perfect agreement with the literature\cite{Liu2002}.
The obtained equilibrium coordinates of C atoms in the unitary cell
are shown in Supplementary Table 1.

\subsection{Quantum Monte Carlo method.}

We have applied the quantum Monte Carlo method
to carbon nanotubes by using standard pseudopotentials for the 1$s$ core
electrons of the carbon atom\cite{Burkatzki2007}.
We minimize the total energy expectation value of the
first-principles Hamiltonian, within the Born-Oppheneimer approximation,
by means of a correlated wave function, $J \left|\text{SD}
\right>$. This is made of a Slater
determinant, $\left|\text{SD}
\right>$, defined in a localized GTO VDZ basis\cite{Burkatzki2007} 
($5s5p1d$)
contracted into six hybrid orbitals per carbon atom\cite{Sorella2015}, 
multiplied by a Jastrow term, $J$. The latter, $J=J_1 J_2$,
 is the product of two
factors: a one-electron one term,
$J_1= \prod_{i} \exp{ \! \left [u_{\text{1body}}(\textbf{r}_i) \right] }$,
and a two-electron correlation factor,
$J_2 = \prod_{i<j} 
\exp{\!\left[u(\textbf{r}_i,\textbf{r}_j)\right]}$.
The two-body Jastrow factor $J_2$
depends explicitly on the $N_{\text{e}}$ electronic positions, 
$\{\textbf{r}_i \}$,
and, parametrically, on the $N_{\text{C}}$
carbon positions, $\textbf{R}_I$, $I=1,\cdots N_{\text{C}}$.
The pseudopotential functions, $u$ and $u_{\text{1body}}$, are written as:
\begin{equation} 
u(\textbf{ r},\textbf{ r}^\prime) = 
  u_{\text{ee}} ( | \textbf{ r} - \textbf{ r}^\prime| ) 
  + \sum_{\mu>0,\nu>0} u_{\mu\nu}\, \chi_\mu(\textbf{r}) 
\chi_\nu(\textbf{r}^\prime)  ,
\end{equation}
\begin{equation}
 u_{\text{1body}}(\textbf{r})=\sum_{\mu>0} u_{\mu0}\, \chi_\mu(\textbf{r}),
\end{equation}
where $u_{\text{ee}}= 2^{-1}r /( 1 + b_{\text{ee}} r)$ is a simple function,
depending on the single variational parameter $b_{\text{ee}}$, which allows
to satisfy the electron-electron cusp condition,
and $u_{\mu \nu}$ is a symmetric matrix of finite dimension.
For non-null indices, $\mu,\nu>0$,
the matrix $\bm{u}$ describes the variational freedom of $J_2$
in a certain finite atomic basis, $\chi_\mu(\textbf{r})$,
which is localized around the atomic centers
$\textbf{R}_{I(\mu)}$ and is made of $3s2p$ GTO orbitals per atom.
Note that the one-body Jastrow term $J_1$ is expanded over the same 
atomic basis
and its variational freedom is determined by the first column of the matrix, 
$u_{\mu0}$.

We use an orthorombic unit cell $L_x \times L_y \times L_z$
containing twelve atoms
with $L_x=L_y= 36$ {\AA} and
$L_z=2.445$ {\AA}.
This cell is repeated along the $z$ direction for
$n=1,2,3,4,5,6$ times,
up to $72$ carbon atoms in the supercell.
Periodic images in the $x$  and $y$ directions are far enough that
their mutual interaction can be safely neglected.
Conversely, in the $z$ direction we apply twisted periodic boundary
conditions and we integrate over that with a number $n_\theta$
of twists, $n_\theta=80,40,30,20,20,20$ for $n=1,2,3,4,5,6$,
respectively, large enough to have converged results for each supercell.

The initial Slater determinant was taken by performing a standard
LDA calculation. The molecular orbitals, 
namely their expansion coefficients in the GTO localized basis set, as well as
the matrix $\bm{u}$ determining the Jastrow factor,  
were simultaneously optimized with well
established methods developed in 
recent years\cite{Dagrada2016,Umrigar2007}, 
which allows us to consider up to $3000$ independent 
variational parameters in a very stable and efficient way.
Note that the two-body Jastrow term $J_2$ can be chosen to 
explicitly recover
the EI mean-field wave function (\ref{eq:EI}), as shown in
Supplementary Discussion. 
After the stochastic optimization the correlation functions
/ order parameters can be computed in a
simple way within variational Monte Carlo (VMC).

We also employ lattice regularized diffusion Monte Carlo (LRDMC) within the
fixed-node approximation, using a lattice mesh of
$a_{\text{mesh}}=0.2$ and $a_{\text{mesh}}=0.4$ a.u., respectively, in order
to check the convergence for $a_{\text{mesh}}\to 0$.
The fixed-node approximation is necessary for fermions for
obtaining statistically meaningful ground-state properties.
In this case the correlation functions / order parameters,
depending only on local (i.e., diagonal in the basis) operators,
such as the ones presented in this work,
are computed with the forward walking technique\cite{Calandra1998}, which
allows the computation of pure expectation values on the fixed-node
ground state.

\end{methods}

\section*{Code availability}

Many-body perturbation
theory calculations were performed by means of the codes Yambo 
(http://www.yambo-code.org/) and Quantum ESPRESSO 
(http://www.quantum-espresso.org), which are both 
open source software. Quantum Monte Carlo calculations
were based on TurboRVB code (http://trac.sissa.it/svn/TurboRVB),
which is available from S.S.~upon 
reasonable request (email: sorella@sissa.it).

\section*{Data availability}

The data that support the findings of this study are available from the
corresponding author upon reasonable request.

\section*{Supplementary Note 1}

\section*{Effective-mass theory of armchair 
carbon nanotubes} 

In this Note we recall the effective-mass
theory of electronic $\pi$-states in single-wall carbon nanotubes,
focusing on the lowest conduction and highest valence 
band of undoped armchair tubes\cite{Ajiki1993,Ando2005,Rontani2014}.
Carbon nanotubes may be thought of as wrapped sheets of graphene,
hence nanotube electronic states are built from those of graphene
through a folding procedure, after quantizing the transverse wave vector.  
Low-energy graphene states belong to one of the two Dirac cones,
whose apexes intersect the 
degenerate K and K$'$ points, respectively, at the corners of
graphene first Brillouin zone. At these two points the energy gap is zero.

Close to Brillouin zone corners $\tau=\text{K},\text{K}'$, a nanotube 
state $\psi(\mathbf{r})$ is the superposition of slowly-varying
envelope functions $F^{\tau\eta}\!(\mathbf{r})$ multiplied by the Bloch states
$\psi_{\tau\eta}(\mathbf{r})$, the latter having two 
separate components localized on sublattices
$\eta=A$ and $\eta=B$,
respectively (cyan and red dots in Supplementary Fig.~1):
\begin{equation}
\psi(\mathbf{r})=\sum_{\tau=\text{K},\text{K}'}
\sum_{\eta=A,B}
F^{\tau\eta}\!(\mathbf{r})\,\psi_{\tau\eta}(\mathbf{r}).
\label{eq:envelope}
\end{equation}
The effective-mass approximation of Supplementary 
Eq.~\eqref{eq:envelope} goes 
beyond the usual one-valley
treatment, as below we explicitly consider intervalley coupling due to 
Coulomb interaction. The relative phases of different Bloch state components 
$\psi_{\tau\eta}$ are fixed by symmetry considerations, as detailed 
in Supplementary Note 7. 
The envelope $F^{\tau\eta}$ is 
a pseudospinor with respect to valley and
sublattice indices, $\bm{F}\equiv(F^{\text{K}A},
F^{\text{K}B},F^{\text{K}'A},F^{\text{K}'B})^T$. 
In the valley-sublattice product space, $\bm{F}$
obeys the Dirac equation of graphene:
\begin{equation}
\gamma\left[\bm{\sigma}_x\otimes 
\bm{1}_{\tau} \hat{k}_x 
+ \bm{\sigma}_y\otimes \bm{\tau}_z \hat{k}_y
\right]\bm{F}(\mathbf{r})
 = \varepsilon\, \bm{F}(\mathbf{r}).
\label{eq:graphene}
\end{equation}
Here $\bm{\sigma}_x$ and $\bm{\sigma}_y$ are 2 $\times$ 2 Pauli matrices
acting on the sublattice pseudospin, $\bm{\tau}_z$ and the 2 $\times$ 2
identity matrix $\bm{1}_{\tau}$ act on the valley pseudospin,
$\hat{k}_x = -i\partial /\partial x$ is 
is the wave vector operator along the circumference direction $x$ and
$\hat{k}_y = -i\partial /\partial y$ 
acts on the tube axis coordinate $y$,
$\gamma$ is graphene's band parameter,
and $\varepsilon$ is the single-particle energy.
Furthermore, $\bm{F}$ obeys 
the boundary condition along the tube circumference:
\begin{equation}
\bm{F}(\mathbf{r+L}) = \bm{F}(\mathbf{r})
\exp \left( 2\pi i \, \varphi \right),
\end{equation}
where $\mathbf{L}$ is the chiral vector in the circumference direction of the
tube and $\left|\mathbf{L}\right|=L=2\pi R$ is the circumference.
A magnetic field may or may not be applied along the tube axis, 
with $\varphi = \phi / \phi_0$ being the ratio of the magnetic
flux $\phi$ piercing the tube cross section
to the magnetic flux quantum $\phi_0 = ch/e$.
Supplementary Eq.~\eqref{eq:graphene} depends on the 
reference frame. Note that in our effective-mass treatment 
the $x$ 
and $y$ directions are parallel to the
circumference and axis of the tube, respectively,
as shown in Supplementary Fig.~1a, whereas in the main text as well
as in the first-principles treatment the $z$ axis is parallel to the tube.  

\begin{figure}
\setlength{\unitlength}{1 cm}
\begin{picture}(8.5,5.8)
\put(1.5,0.6){\includegraphics[trim=0cm 0 0cm 0cm,clip=true,width=1.8in]{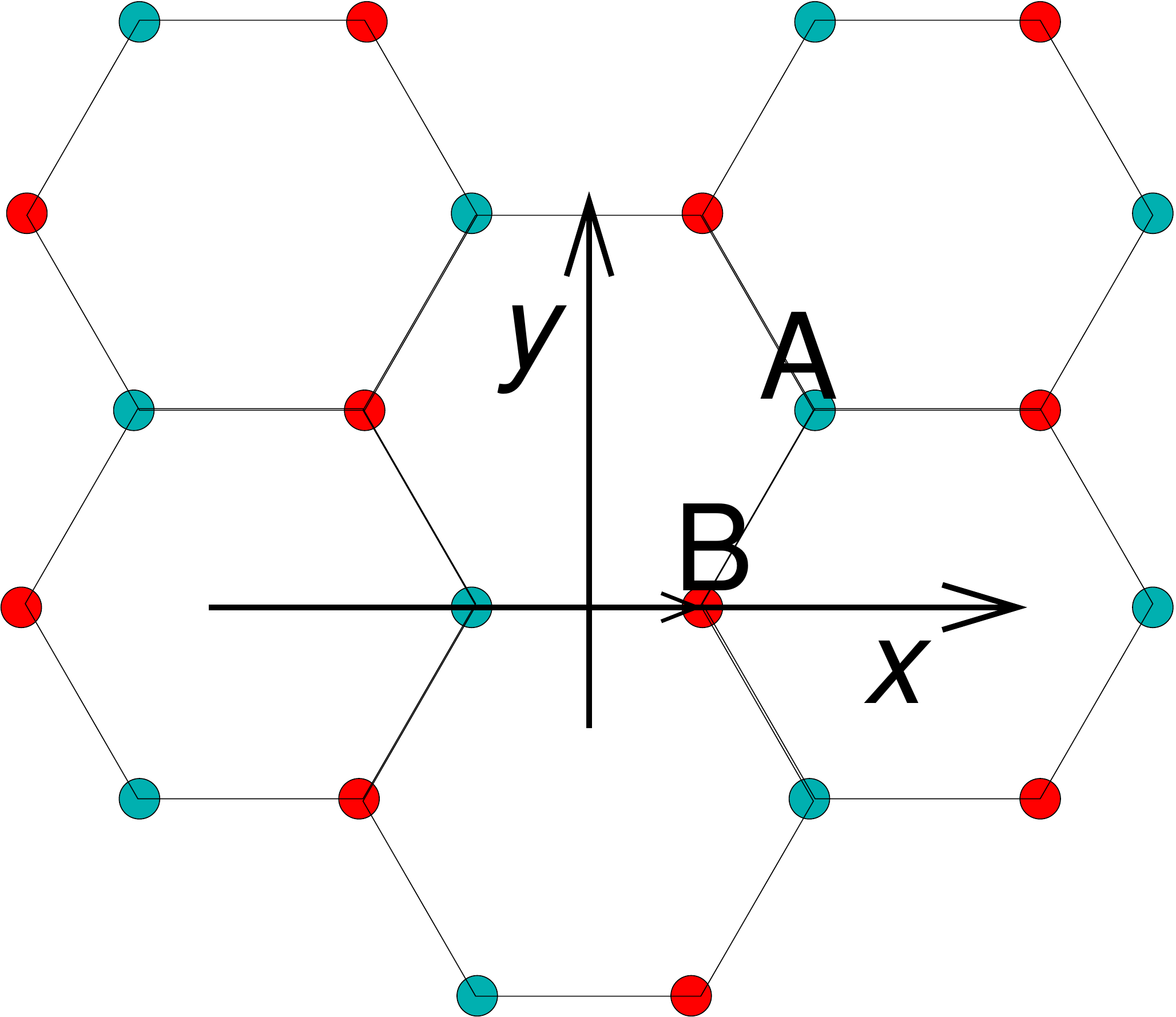}}
\put(9.2,0.5){\includegraphics[trim=0.05cm 0.1cm 0.1cm 0.1cm,clip=true,width=1.8in]{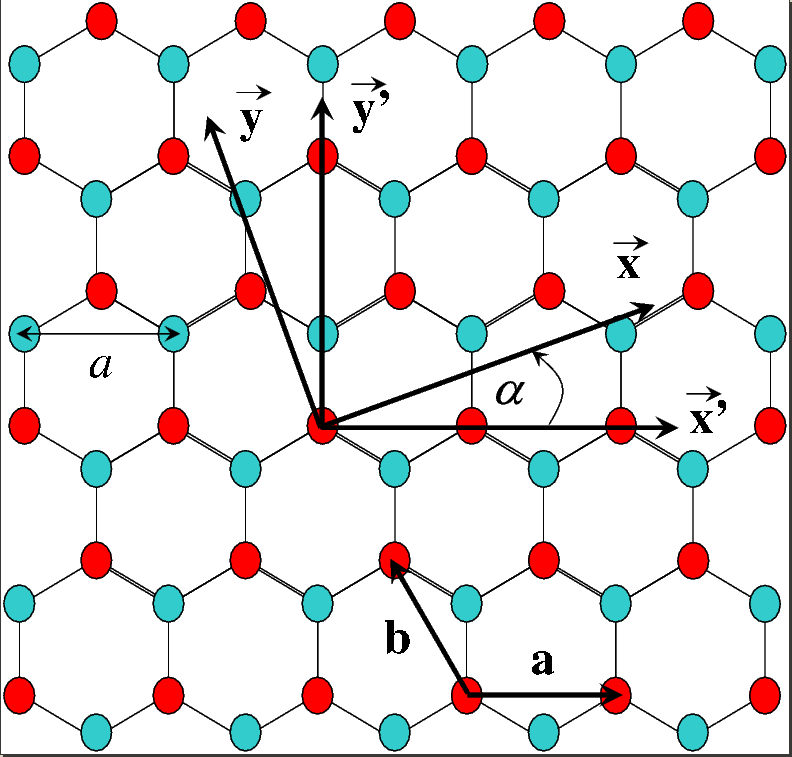}}
\put(10.5,4.29){\includegraphics[trim=0cm 0 0cm 0cm,clip=true,width=0.3cm]{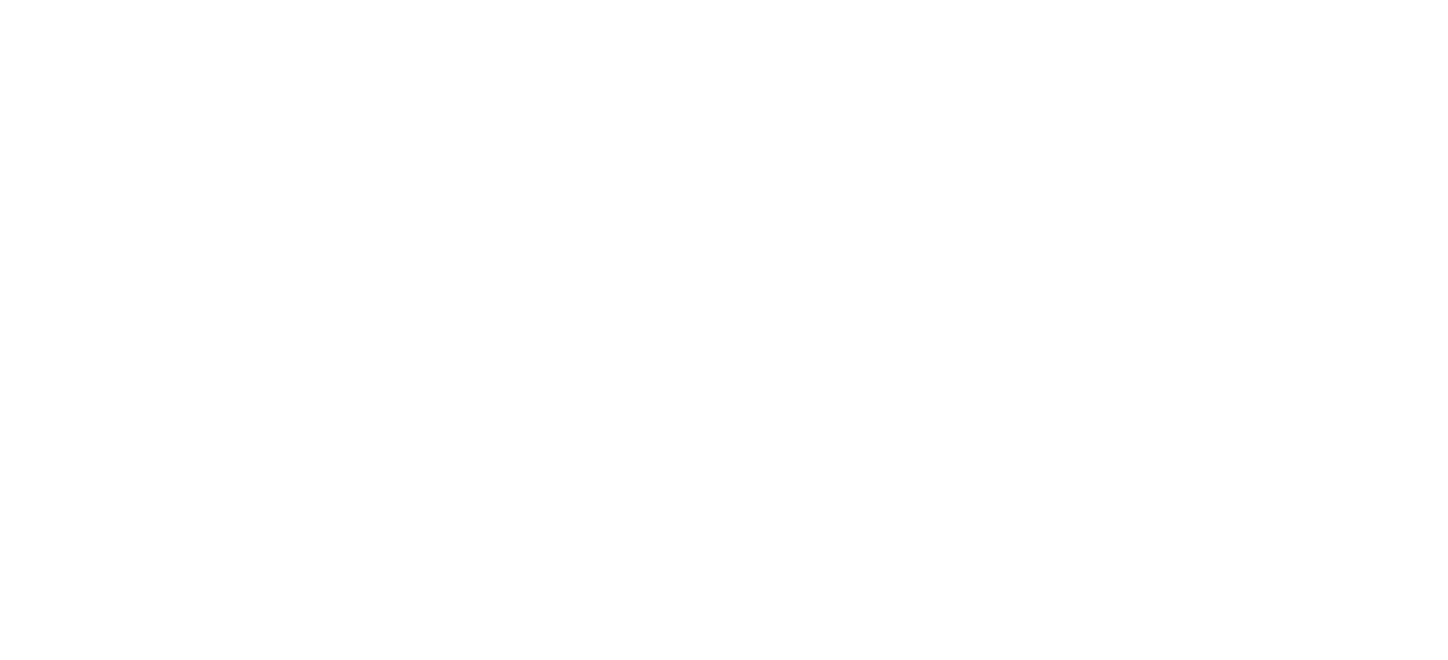}}
\put(11.2,4.35){\includegraphics[trim=0cm 0 0cm 0cm,clip=true,width=0.3cm]{./rectangle.pdf}}
\put(12.7,3.42){\includegraphics[trim=0cm 0 0cm 0cm,clip=true,width=0.3cm]{./rectangle.pdf}}
\put(13.15,2.55){\includegraphics[trim=0cm 0 0cm 0cm,clip=true,width=0.3cm]{./rectangle.pdf}}
\put(1.2,0.7){\bf a}
\put(8.4,0.7){\bf b}
\end{picture}
\caption*{{\bf Supplementary Fig.~1} Carbon nanotube reference frames for the
effective-mass model.
{\bf a} 
Reference frame for the armchair tube used in this work.
The $x$ and $y$ directions are parallel to the 
circumference 
and axis of the tube, respectively.
The small vector is $\mathbf{R}_0^B$, i.e., the basis vector
locating the origin of the B sublattice.
Cyan and red dots point to A and B sublattices, respectively.
{\bf b} 
Ando's reference frame for a generic tube.
The frame origin is located on an atom of the B sublattice.
The tube frame is obtained by rotating
the $x^{\prime}y^{\prime}$
graphene reference frame
by the chiral angle $\alpha$.
The chiral vector $\mathbf{L}$ identifying the tube circumference is
$\mathbf{L}=-m\mathbf{a} -(n+m)\mathbf{b}$ in terms of the conventional 
chiral indices $(n,m)$, 
where $\mathbf{a}$ and $\mathbf{b}$ are the primitive
translation vectors of graphene shown in
the picture. For an equivalent choice of $\mathbf{L}$
one has $\alpha = \pi/6$ for $(n,n)$ armchair tubes and $\alpha = 0$
and for $(n,0)$ zigzag tubes.
$a$ is the lattice constant of graphene.
}
\end{figure}

The energy bands are specified by the valley index $\tau$, 
the valence index
$\alpha = c$, $v$ denoting either the conduction ($\alpha = c$) 
or the valence band ($\alpha = v$), and  
the wave vector $k$ in the axis direction. 
The wave functions in K and K$'$ valleys are
respectively
$\bm{F}\equiv(\bm{F}^{\text{K}}_{\alpha k}  (\mathbf{r})   ,0)^T$
and $(0,\bm{F}^{\text{K}'}_{\alpha k}(\mathbf{r}))^T$, 
with $\bm{F}^{\tau}_{\alpha k}(\mathbf{r})   
\equiv(F^{\tau A}_{\alpha k},
F^{\tau B}_{\alpha k})^T$ being a plane-wave
pseudospinor in the sublattice space,
\begin{equation}
\bm{F}^{\tau}_{\alpha k}  (\mathbf{r}) 
=
\bm{\xi}^{\tau}_{\alpha k}(x)  
\frac{1}{\sqrt{A}} \exp{(iky)} ,
\end{equation}
where $A$ is the tube length and the wave function 
$\bm{\xi}^{\tau}_{\alpha k}(x)$ for the motion along
the circumference direction is  
\begin{equation}
\bm{\xi}^{\tau}_{\alpha k}  (x) =
\frac{1}{\sqrt{L}} \exp{(ik_{\perp} x)} 
\bm{F}_{\tau \alpha k}.  
\label{eq:spinor} 
\end{equation}
The constant pseudospinor $\bm{F}_{\tau \alpha k}$ is  
a unit vector with a $k$-dependent
phase between the two sublattice components,
\begin{equation}
\bm{F}_{\text{K} \alpha k}  =
\frac{1}{\sqrt{2}} 
{  b(k) \choose  s_{\alpha}  } ,\qquad 
\bm{F}_{\text{K}' \alpha k}  =
\frac{1}{\sqrt{2}} 
{  b^*\!(k) \choose  s_{\alpha}  },
\label{eq:F} 
\end{equation}
where
\begin{equation}
b(k) =
\frac{ k_{\perp} -ik  }{\sqrt{ k^2_{\perp} + k^2     }} ,
\label{eq:b} 
\end{equation}
and
$s_{\alpha}=\pm 1$ for conduction and valence bands, respectively.
In Supplementary Eqs.~\eqref{eq:spinor} and \eqref{eq:b} 
the transverse wave vector $k_{\perp}$
is proportional to the magnetic flux $\varphi$,
\begin{equation}
k_{\perp} =
\frac{\varphi}{R}.
\label{eq:k_x} 
\end{equation}
In each valley, the energy is
\begin{equation}
\varepsilon_{\alpha}(k) = s_{\alpha}  \gamma
\sqrt{ k^2_{\perp} + k^2     } ,
\label{eq:energy} 
\end{equation}
where the origin of the $k$ axis is located at the Dirac point K (K$'$).

\begin{figure}
\setlength{\unitlength}{1 cm}
\begin{picture}(8.5,8.0)
\put(4.0,0.0){\includegraphics[trim=0cm 0 0cm 0cm,clip=true,width=2.8in]{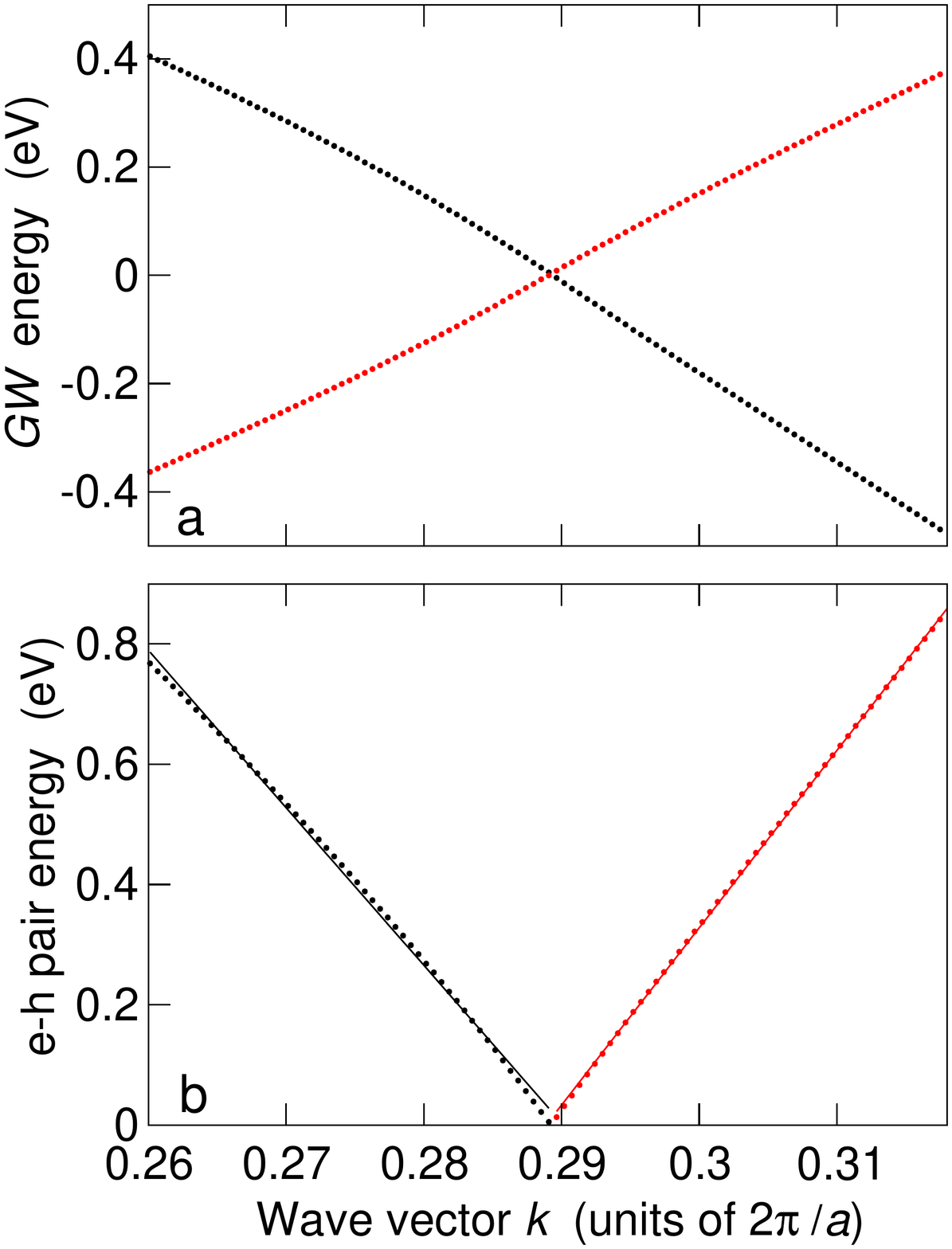}}
\end{picture}
\caption*{{\bf Supplementary Fig.~2} 
$GW$ band structure of the (3,3) tube.
{\bf a} $GW$ band structure vs wave vector $k$ close to the Dirac point K.
Red [grey] and black dots point to chirality indices ${\cal{C}}=+1$ and -1,
respectively. {\bf b} Electron-hole pair excitation energy vs $k$.
The lines are linear fits to the data.} 
\end{figure}

Figure 2a of main text shows the first-principles band structure of the (3,3) 
tube in a range of a few eV around the Dirac point, 
with $k$ scanning half Brillouin zone, between the origin 
($k=0$, $\Gamma$ point)
and $k=\pi/a$ ($a=2.46$ {\AA}
is graphene lattice constant). The negative $k$ axis, containing
the K$'$ point, is obtained by specular reflection. 
The DFT / $GW$ location of the Dirac point is  
K = 0.289 $(2\pi/a)$, 
whereas the effective-mass estimate is K = $1/3$ $(2\pi/a)$
(the discrepancy between DFT and
tight-binding predictions
is well documented in the
literature\cite{Bohnen2004,Connetable2005}).
As seen in Supplementary Fig.~2a,
the $GW$ bands are approximately linear  
in an energy range of at least $\pm$ 0.4 eV around the Dirac point,
which validates the effective-mass model at low energy.

Note that, 
in the absence of the magnetic field,
electron states have a well defined 
chirality\cite{Ando1998a,Ando1998b,McEuen1999}, which is one of 
the two projections, $\cal{C}$, of the sublattice pseudospin onto the 
momentum direction, expressed as the eigenvalues ${\cal{C}}=\pm 1$ 
of the operator 
$\bm{\sigma}_y \otimes \bm{\tau}_z$. The chirality index is highlighted
by red (${\cal{C}}=+1$) and black (${\cal{C}}=-1$) colour in Supplementary 
Fig.~2a.

\section*{Supplementary Note 2}

\section*{Electron-electron interaction: 
Effective-mass vs first-principles description}

Within the effective-mass framework,
the Coulomb interaction $v$ between two electrons on the
carbon nanotube cylindrical surface
located at $\mathbf{r}\equiv(x,y)$ and $\mathbf{r}'\equiv(x',y')$, 
respectively, is\cite{Ando2005}
\begin{equation}
v(\mathbf{r},\mathbf{r}')= \sum_q e^{iq(y-y')}\frac{2e^2}{\kappa_{\text{r}} A}
K_0\!\left(2R\left|q\sin{\!\left(\frac{x-x'}{2R}\right)} \right|\right),
\end{equation}
where $\kappa_{\text{r}}$ is a static dielectric constant that takes 
into account polarization effects due  
to the electrons not included in the 
effective-mass description
plus the contribution of the dielectric background.
The interaction matrix element between 
single-particle states is\cite{Ando2006,Secchi2010}
\begin{eqnarray}
&&V_{(\tau,\alpha, k + q),(\tau',\beta',k');(\tau',\alpha',k'+q)
(\tau,\beta,k)}
\nonumber\\
&&\quad=\quad \int\!\! d\mathbf{r}\!\int \!\! d\mathbf{r}' \;
[\bm{F}^{\tau}_{\alpha k + q}  (\mathbf{r})]^{\dagger}\!\cdot\!
\bm{F}^{\tau}_{\beta k }(\mathbf{r})\,\, v(\mathbf{r},\mathbf{r}')\;
[\bm{F}^{\tau'}_{\beta' k'}  (\mathbf{r}')]^{\dagger}\!\cdot\!
\bm{F}^{\tau'}_{\alpha' k'+q }(\mathbf{r}') \nonumber\\
&&\quad\quad =\quad\frac{1}{A}\,
\bm{F}_{\tau \alpha k + q}^{\dagger}\!\cdot\!
\bm{F}_{\tau\beta k }\,\,\, \bm{F}_{\tau'\beta' k'}^{\dagger}\!\cdot\!
\bm{F}_{\tau'\alpha' k'+q}\,\,v(q),
\label{eq:V}
\end{eqnarray}
where the one-dimensional effective interaction 
resolved in momentum space, 
\begin{equation}
v(q) = \frac{2e^2}{\kappa_{\text{r}}}\,I_0\!\left(R\left|q\right|\right)\,
K_0\!\left(R\left|q\right|\right),
\label{eq:v_q}
\end{equation}
is modulated by a form factor given by overlap terms
between sublattice pseudospinors 
[$I_0(z)$ and $K_0(z)$ are the 
modified Bessel functions of the first and second kind, 
respectively\cite{Abramowitz1972}].
The effect of screening due to the polarization of those 
electrons that are treated within
the effective-mass approximation
is considered by replacing $v(q)$ with 
\begin{equation}
w(q) = \frac{v(q)}{\varepsilon(q)}
\label{eq:vw}
\end{equation}
in the matrix element \eqref{eq:V},
where $\varepsilon(q)$ is the static dielectric function
(to discriminate between screened and unscreened matrix elements 
we use respectively `w' and `v' letters throughout the
Supplementary Information).
It may be shown that dynamical polarization 
effects are negligible 
in the relevant range of small frequencies, 
which is comparable to exciton binding energies.

Note that terms, similar to Supplementary Eq.~\eqref{eq:V}, 
that scatter electrons from one valley to the other are absent
in the effective mass approximation.
These small intervalley terms, as well as the interband exchange terms,
which are both induced by the residual, short-range part of
Coulomb interaction, are discussed in Supplementary Note 3.

{\bf Effect of chiral symmetry.} The chiralities of electron states, 
which is illustrated in Supplementary Fig.~3a 
(solid and dashed lines label ${\cal{C}}=+1$
and ${\cal{C}}=-1$, respectively), signficantly affects Coulomb
interaction matrix elements. This occurs through the form factors of the type
$\bm{F}^{\dagger}\!\cdot\!\bm{F}$ appearing in 
Supplementary Eq.~(\ref{eq:V}), which 
are overlap terms between sublattice pseudospinors.
\begin{figure}
\setlength{\unitlength}{1 cm}
\begin{picture}(14.5,5.9)
\put(0.4,0.3){\bf a}
\put(7.0,0.3){\bf b}
\put(4.3,3.4){$\varepsilon$}
\put(5.2,2.2){$k$}
\put(3.8,1.0){valley K}
\put(1.2,1.0){valley K$'$}
\put(1.0,1.2){\includegraphics[trim=0cm 0 0cm 0cm,clip=true,width=1.9in]{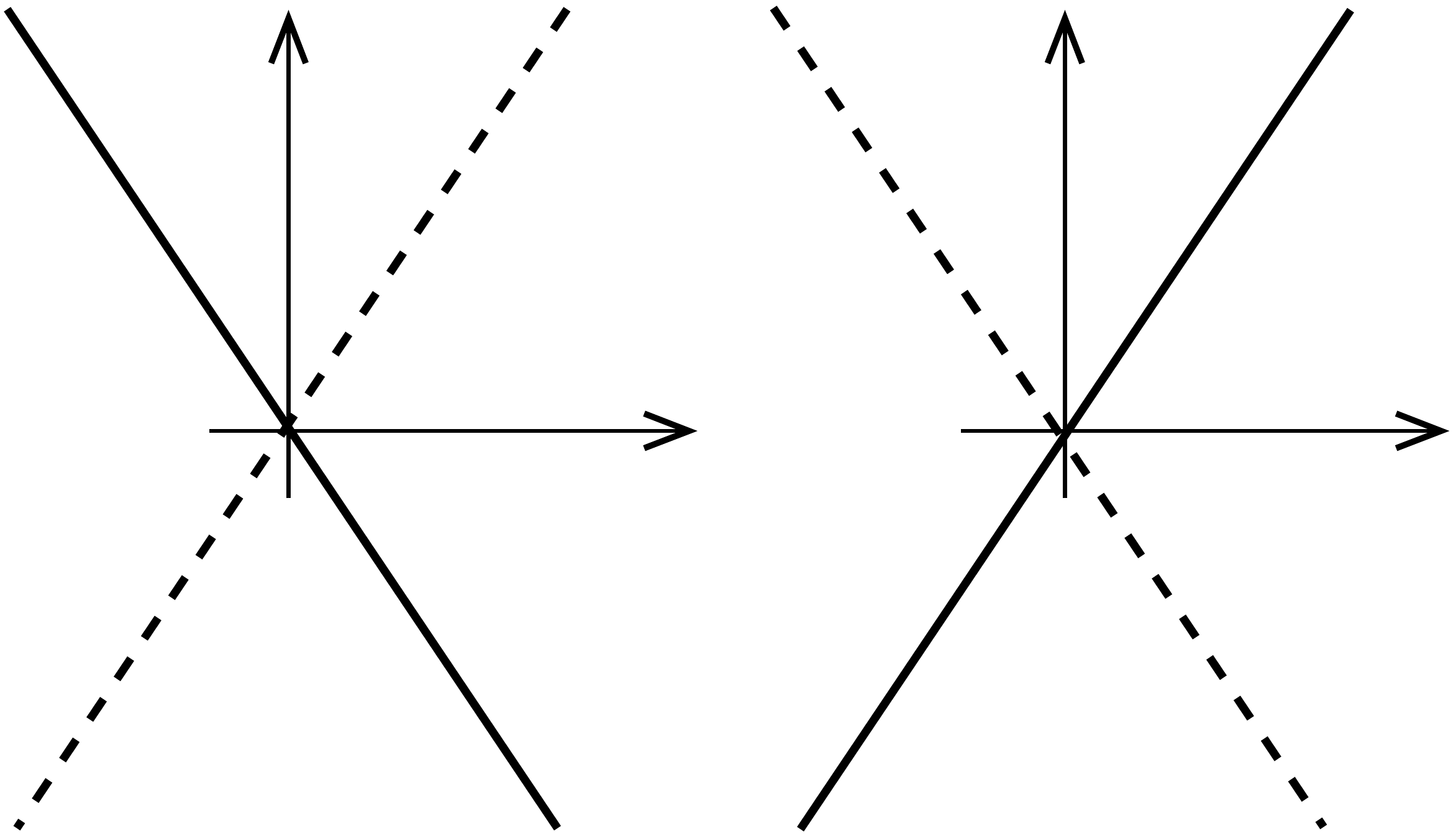}}
\put(7.7,2.7){\includegraphics[trim=0cm 0 0cm 0cm,clip=true,width=1.3in]{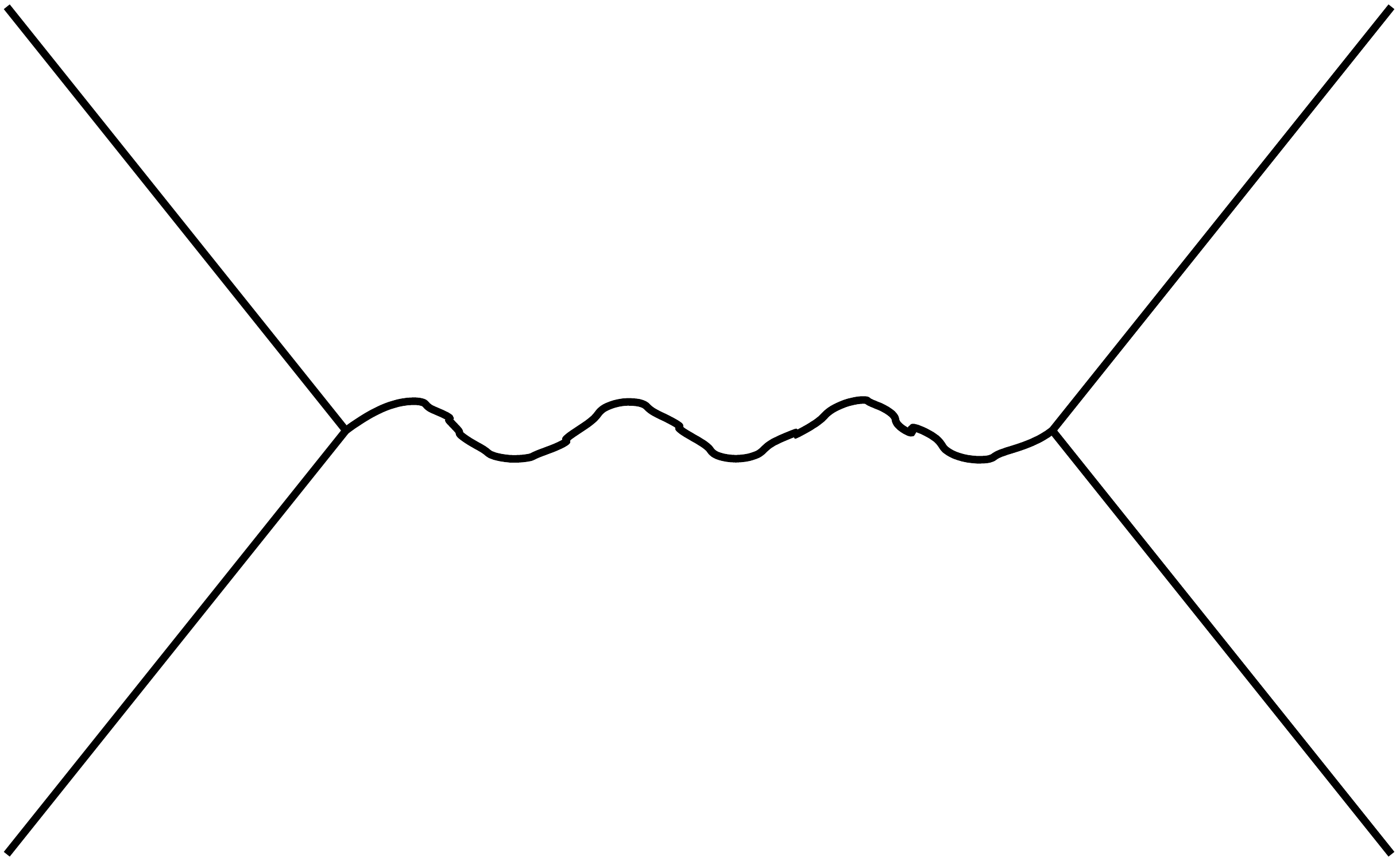}}
\put(12.0,2.7){\includegraphics[trim=0cm 0 0cm 0cm,clip=true,width=1.3in]{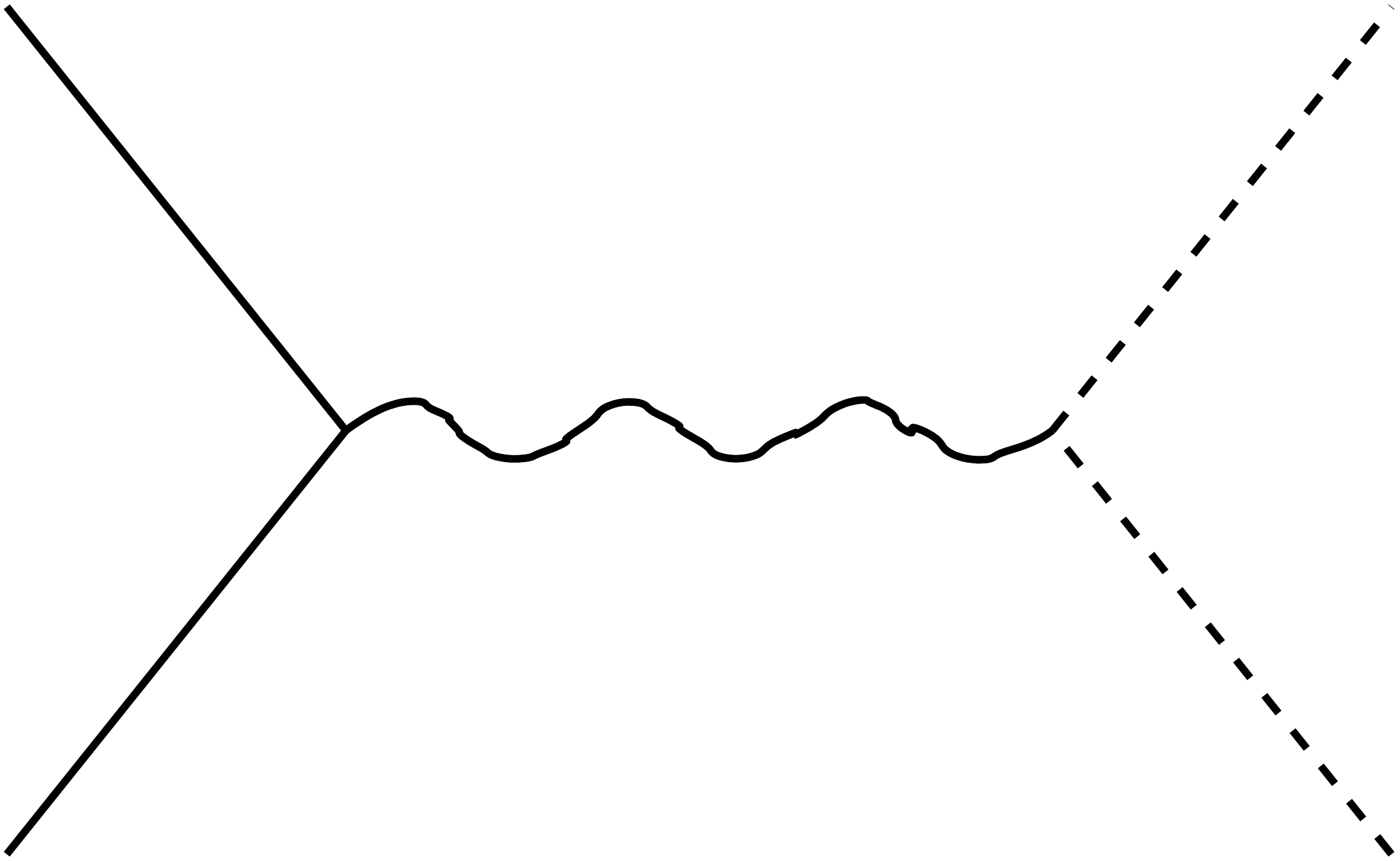}}
\put(7.7,0.0){\includegraphics[trim=0cm 0 0cm 0cm,clip=true,width=1.3in]{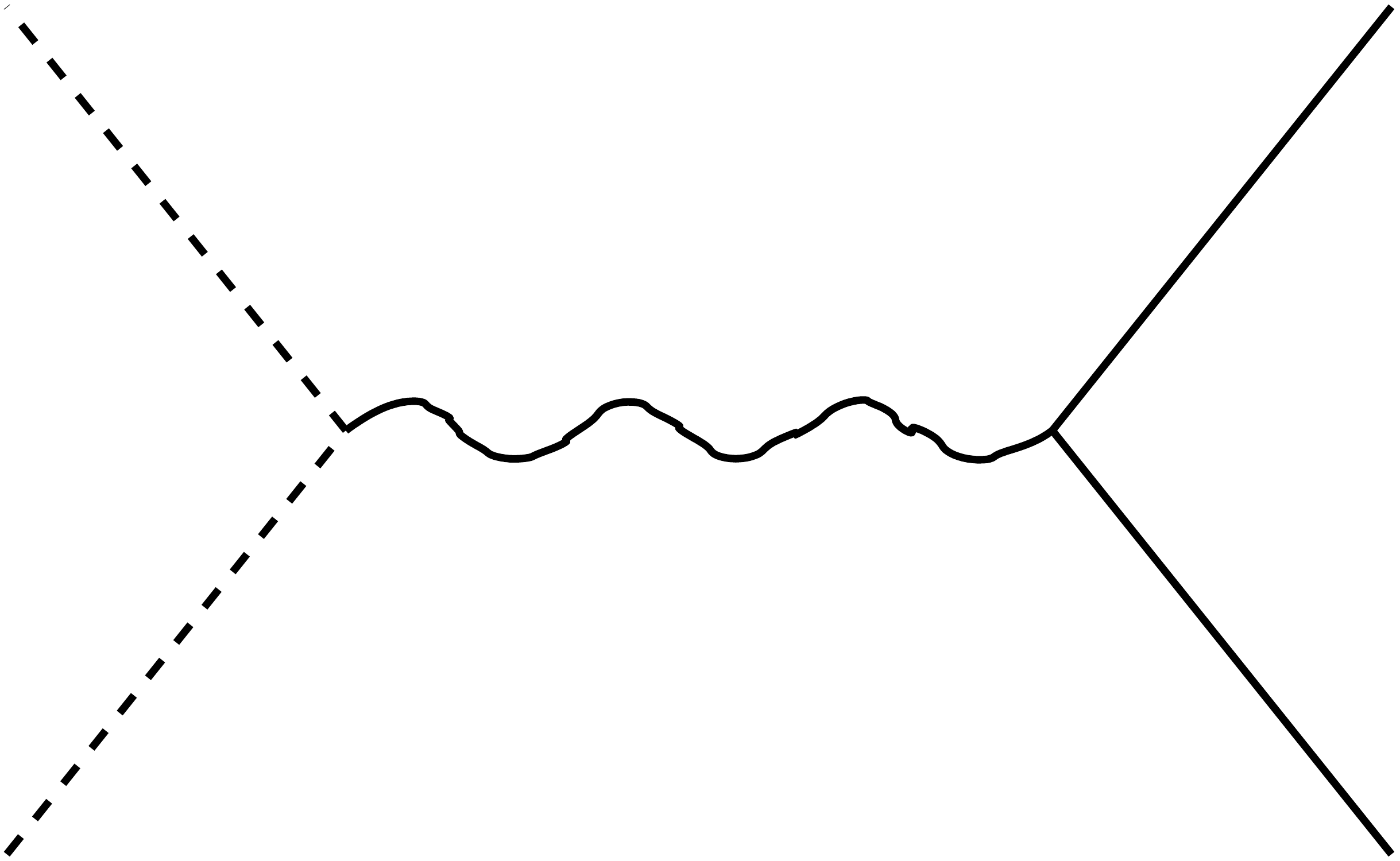}}
\put(12.0,0.0){\includegraphics[trim=0cm 0 0cm 0cm,clip=true,width=1.3in]{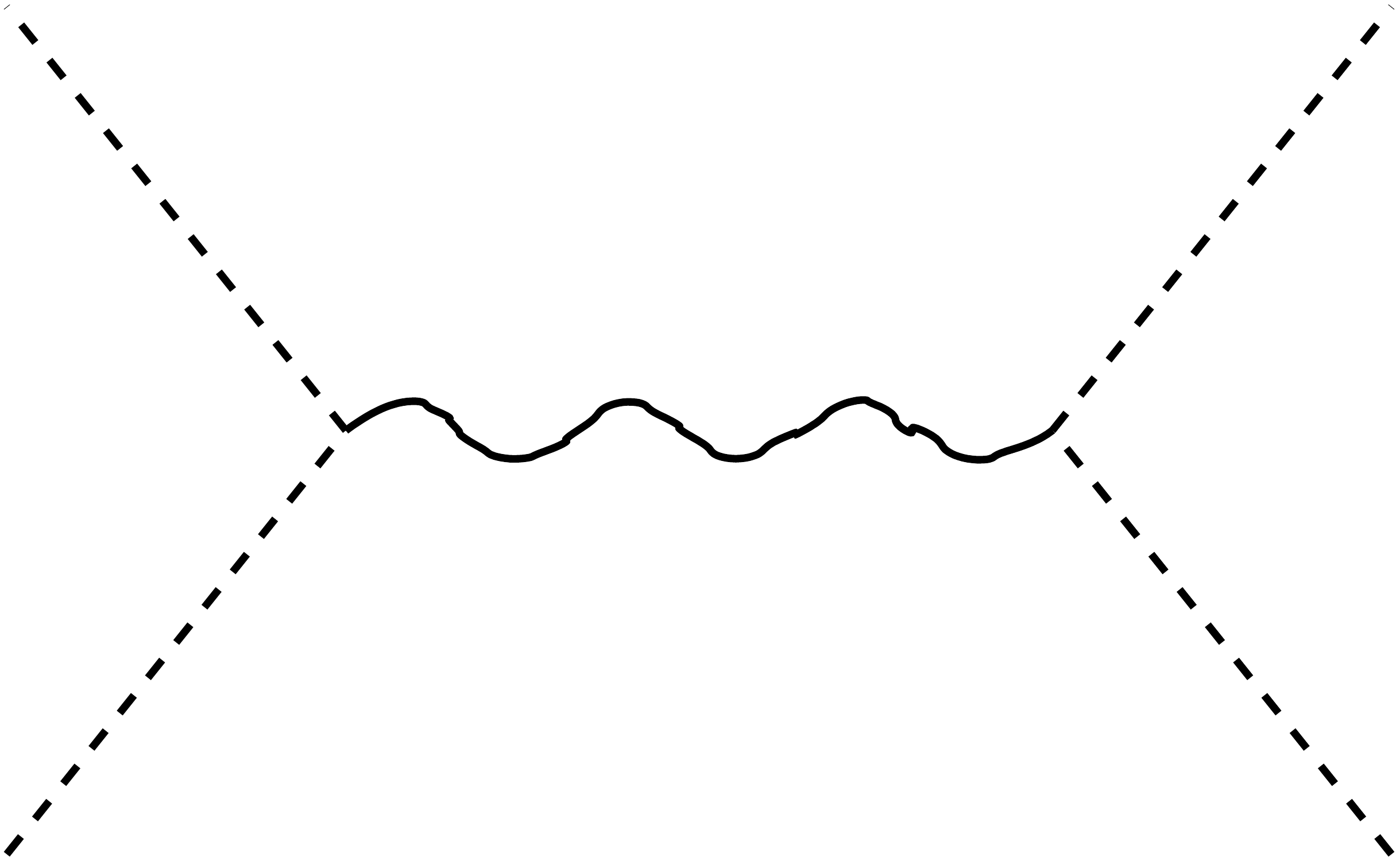}}
\put(8.1,2.9){$\tau, \beta, k$}
\put(9.1,2.35){$\tau', \alpha', k'+q$}
\put(10.1,4.9){$\tau', \beta', k'$}
\put(7.6,4.9){$\tau, \alpha, k+q$}
\put(9.0,3.9){$v(q)$}
\end{picture}
\caption*{{\bf Supplementary Fig.~3} 
Effect of chirality on Coulomb interaction matrix elements.
{\bf a} Energy bands and chiralities of electron states in armchair 
carbon nanotubes
in the absence of the magnetic field. 
Solid and dashed lines highlight chirality ${\cal{C}}=\pm 1$, respectively.
{\bf b} Allowed scattering processes induced 
by long-range Coulomb interaction.
The indices $\tau=$ K, K$'$ and $\alpha=c,v$ label valleys and bands,
respectively.
The chirality is conserved at each vertex of diagrams.
}
\end{figure}
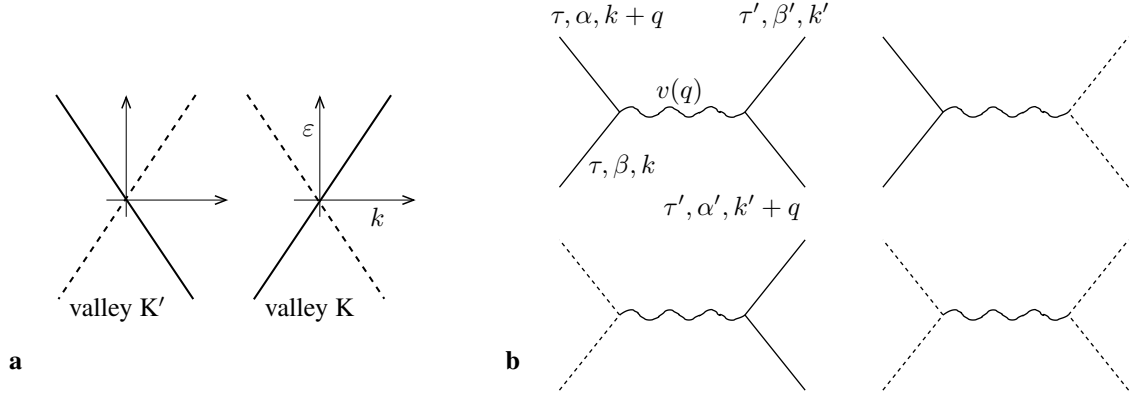
As apparent from their analytical structure,
\begin{equation}
\bm{F}_{\tau \alpha k + q}^{\dagger}\cdot
\bm{F}_{\tau\beta k } = \frac{1}{2}\left[\,\text{sign}(k)\,\text{sign}(k+q)
+ s_{\beta}s_{\alpha}\right],
\label{eq:formfactor}
\end{equation}
the chiral symmetry of the states is conserved at each vertex
of Coulomb scattering diagrams (see Supplementary Fig.~3b), 
hence initial and final states
scattered within the same band must have the same momentum direction. 
This significantly affects the Bethe-Salpeter equation for excitons, 
as we show below. 
We are especially interested in the dominant long-range Coulomb matrix 
element\cite{Halperin1968}
that binds electrons and holes:  
\begin{equation}
V_{(\tau,c, k + q),(\tau,v,k);(\tau,v,k+q)(\tau,c,k)}
\equiv \frac{\tilde{V}(k+q,k)}{A}.
\end{equation}
This term scatters electron-hole pairs from the initial pair state
$(c,k)(v,k)$ to the final state $(c,k+q)(v,k+q)$ within the same valley
$\tau$. Throughout this Supplementary Information we use the
tilde symbol for quantities whose dimension is an energy multiplied 
by a length, like $V=\tilde{V}/A$.

In the first instance we neglect screening, since
for low momentum transfer, $q\rightarrow 0$,
polarization is suppressed hence
$\varepsilon(q)\rightarrow 1$.
In this limit Coulomb interaction diverges logarithmically,
\begin{equation}
v(q) \rightarrow -\frac{2e^2}{\kappa_{\text{r}}}
\ln{\!\left(R\left|q\right|\right)},
\end{equation}
but this is harmless to the Bethe-Salpeter equation,
since $v(q)$ occurs only in the kernel of the scattering term,
hence it is integrated over $q$ for macroscopic lengths $A$,
\begin{equation}
-\frac{1}{A}\sum_q \tilde{V}(k+q,k)\ldots \rightarrow
-\frac{1}{2\pi}\int\! dq\, \tilde{V}(k+q,k)\ldots,
\end{equation}
which removes the divergence. Note that, throughout this Supplementary
Information and opposite to the 
convention of Fig.~2c of main text, we take $V$ as a positive quantity.
In detail, 
we discretize the momentum
space axis, $k\rightarrow k_i$, where $k_i=i2\pi/(Na)$, 
$i=-N/2 + 1,\ldots,0,1, \ldots,N/2$, 
$N=A/a$ is the number of unitary cells,  
and $\Delta k = 2\pi/(Na)$ is the mesh used in the calculation.
Hence, the regularized matrix element, integrated over the mesh, is
\begin{equation}
V(k_j+q_i,k_j)  =  \frac{1}{2\pi}\int_{q_i - \Delta k}^{q_i}
\!\!\!\!dq\,\tilde{V}(k_j+q,k_j).
\label{eq:discrete}
\end{equation}

\begin{figure}
\setlength{\unitlength}{1 cm}
\begin{picture}(14.5,10.2)
\put(1.70,4.02){\includegraphics[trim=0cm 0 0cm 0cm,clip=true,width=5.77in]{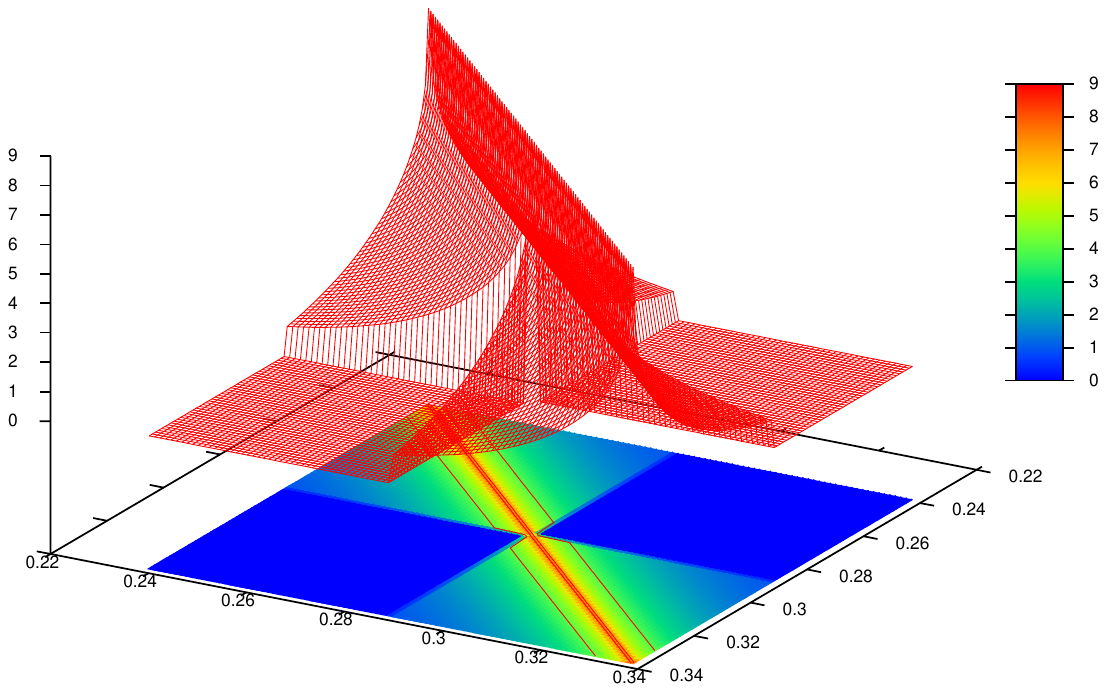}}
\put(1.70,-1.57){\includegraphics[trim=0cm 0 0cm 0cm,clip=true,width=5.77in]{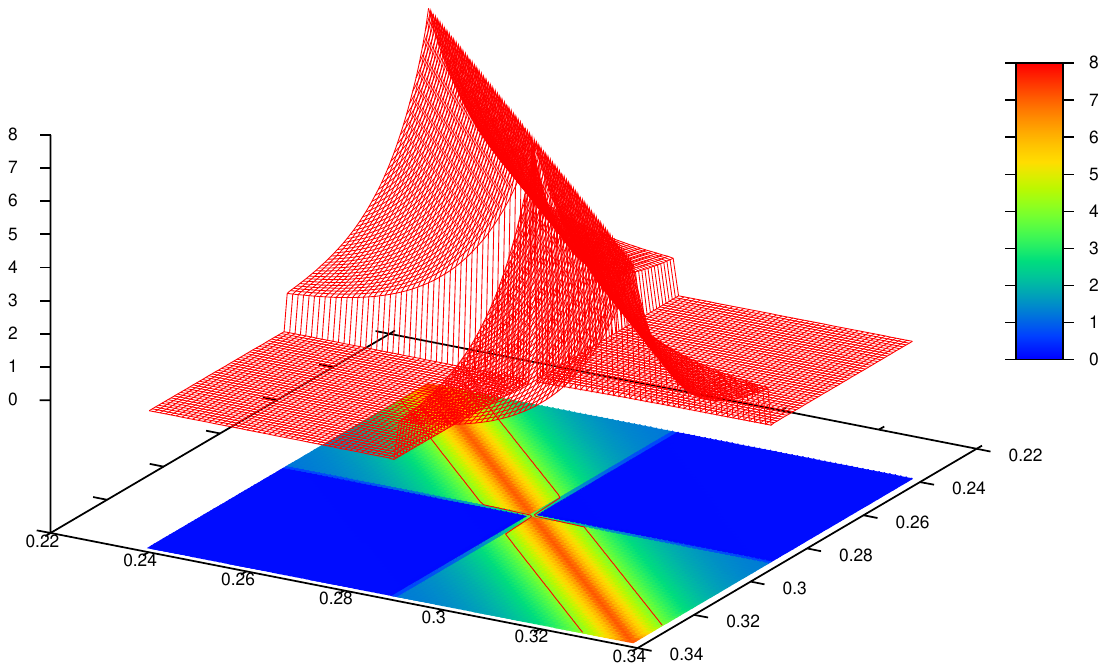}}
\put(10.8,10.2){meV}
\put(9.0,10.0){\bf a}
\put(9.0,4.4){\bf b}
\put(9.5,5.5){\begin{turn}{30}$k$  (units of $2\pi/a$)\end{turn}}
\put(4.5,5.9){\begin{turn}{-10}$k'$  (units of $2\pi/a$)\end{turn}}
\put(3.4,9.8){$V(k,k')$}
\put(9.7,6.1){K}
\put(6.3,5.9){K}
\put(3.4,4.2){$W^{\text{DFT}}(k,k')$}
\put(10.8,4.6){meV}
\put(9.5,-0.1){\begin{turn}{30}$k$  (units of $2\pi/a$)\end{turn}}
\put(4.5,0.3){\begin{turn}{-10}$k'$  (units of $2\pi/a$)\end{turn}}
\put(9.7,0.5){K}
\put(6.3,0.3){K}
\end{picture}
\caption*{{\bf Supplementary Fig.~4} 
Dominant interband Coulomb matrix element 
in the $(k,k')$ space close to the K point.
{\bf a} Effective-mass `bare' matrix element $V(k,k')$, with 
$\kappa_{\text{r}}=10$
and $\varepsilon(k-k')=1$. The isolines of the two-dimensional
contour map point to the heights of 4 and 8 meV, respectively. 
{\bf b} Modulus of DFT screened matrix element $W^{\text{DFT}}(k,k')$
obtained within the random phase approximation for 
the $(3,3)$ armchair tube. Here $N=900$
and K = $0.289(2\pi)/a$.

}
\end{figure}

In Supplementary Fig.~4 we compare $V$
(panel a, $\kappa_{\text{r}}=10$) with the modulus of the screened DFT
matrix element $W^{\text{DFT}}$ 
obtained for the $(3,3)$ tube (panel b).
The two plots are three-dimensional contour maps in a square domain
$(k,k')$ centered around the Dirac point, 
with K = $0.289(2\pi)/a$ and $N=900$.
The two matrix elements agree almost quantitatively, as they both exhibit:
(i) zero or very small values in the second and fourth quadrants,
i.e., $k >$ K and $k'<$ K or $k <$ K
and $k'>$ K  (ii) a logarithmic spike on the domain
diagonal, i.e., $k'\rightarrow k$. 
This behavior has a simple interpretation in terms of exciton scattering,
as an electron-hole pair with zero center-of-mass momentum,
$(c,k)(v,k)$, has a well-defined 
chirality with respect to the noninteracting ground state, i.e., 
$\Delta{\cal{C}} = + 2 = 1 - (-1)$ 
for $k>$ K ($\Delta{\cal{C}} = - 2$ for $k<$ K). 
The chirality of the e-h pair is conserved during Coulomb scattering, i.e.,
as the pair changes its relative momentum from 
$2k = k - (- k)$ to $2k'$.

\begin{figure}
\setlength{\unitlength}{1 cm}
\begin{picture}(14.5,6.0)
\put(0.2,0.0){\includegraphics[trim=0cm 0 0cm 0cm,clip=true,width=3.1in]{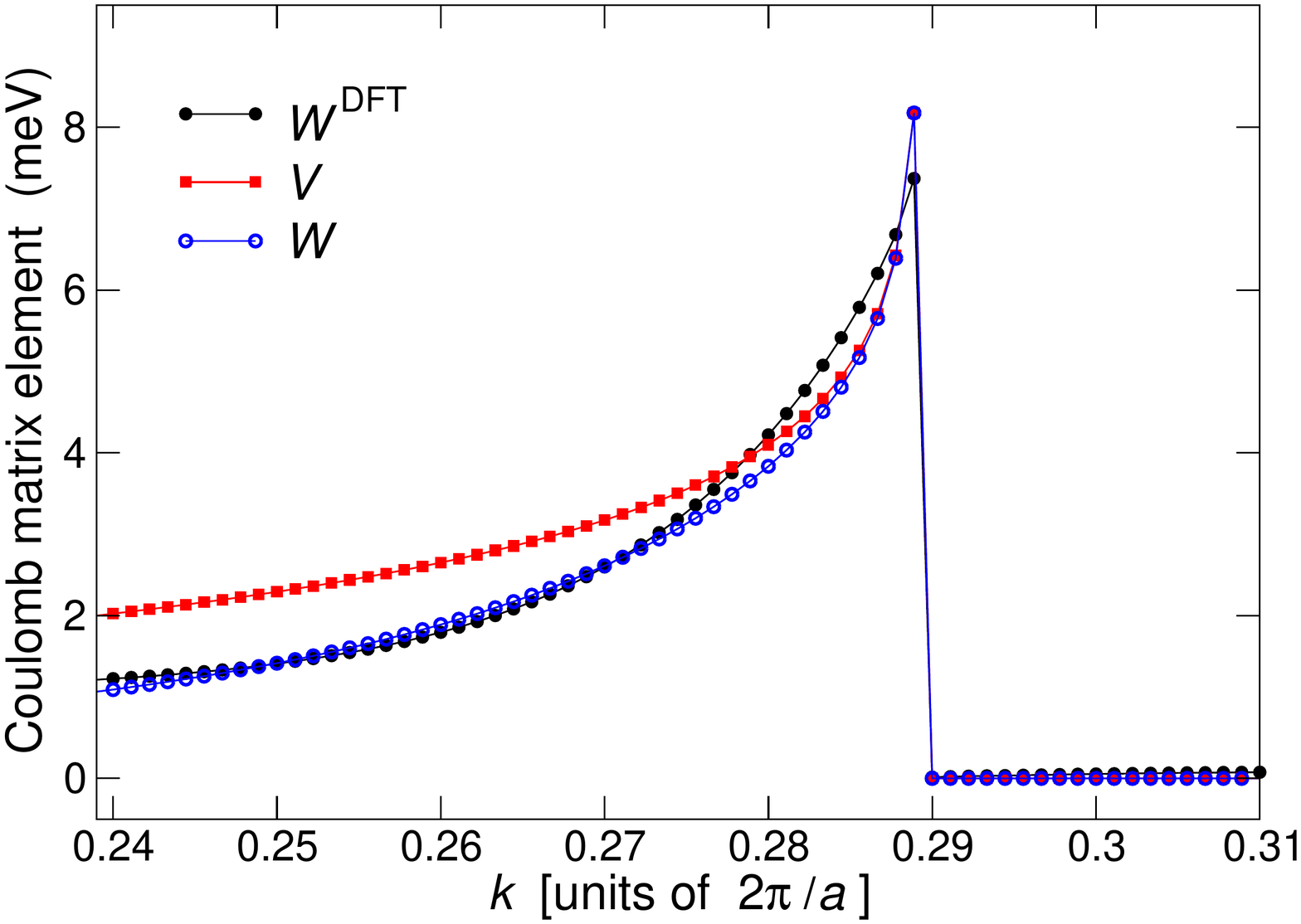}}
\put(8.2,0.0){\includegraphics[trim=0cm 0 0cm 0cm,clip=true,width=3.1in]{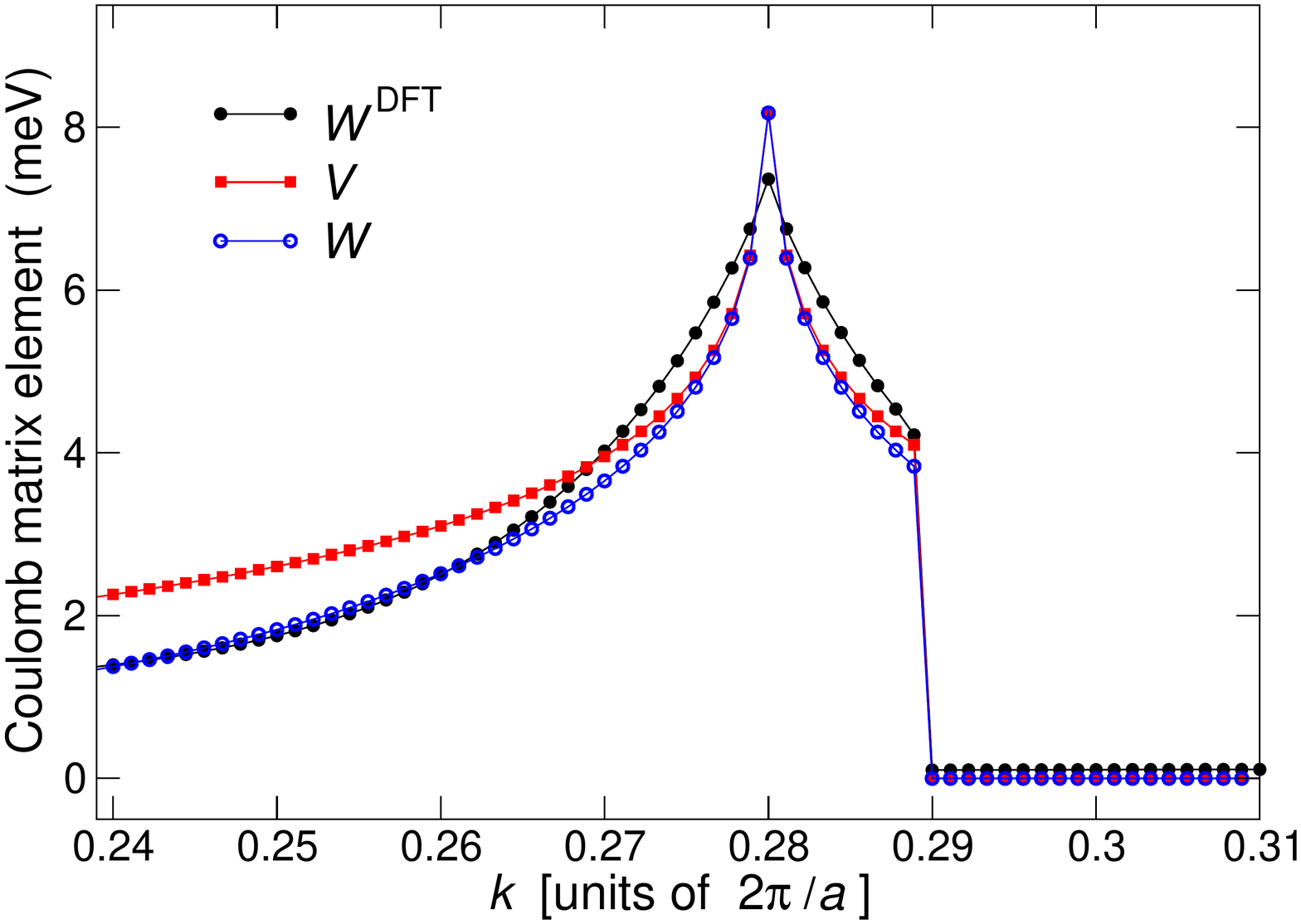}}
\put(6.2,1.3){\bf a}
\put(14.2,1.3){\bf b}
\end{picture}
\caption*{{\bf Supplementary Fig.~5} 
Dominant interband Coulomb matrix elements vs $k$.
Dominant interband Coulomb matrix elements $V(k,k_0)$ 
(squares), $W(k,k_0)$ (empty circles), and
$W^{\text{DFT}}(k,k_0)$ (filled circles) vs $k$, with 
fixed $k_0$. {\bf a}  
$k_0=0.289(2\pi)/a$. {\bf b} $k_0=0.28(2\pi)/a$.
Curves are discontinuous at $\text{K}=0.289(2\pi)/a$,
lines are guides to the eye, $N=900$.
}
\end{figure}

{\bf Effect of electronic polarization.}
In order to appreciate the minor differences 
between $V(k,k')$ and $W^{\text{DFT}}(k,k')$
it is convenient to compare the cuts of the maps of Supplementary 
Fig.~4 along a line $k'=k_0$, 
as shown in Supplementary Fig.~5
for $k_0=0.289 (2\pi)/a$ (panel a) and $0.28 (2\pi)/a$ (panel b), 
respectively. For small momentum transfer, $q=k-k_0\approx 0$,
$V(k,k_0)$ (squares) exhibits a sharper spike than
$W^{\text{DFT}}(k,k_0)$ (filled circles). This is an effect of the 
regularization of the singularitity occurring in the DFT approach,
as in the first-principles calculation the tube is actually
three-dimensional.
As $\left|q\right|$ increases, $V$ is systematically blushifted with respect
to $W^{\text{DFT}}$ since it does not take into account the effect
of the RPA polarization,
$\Pi(q)$, which acquires a finite value. 

Within the effective-mass approximation, $\Pi(q)$ enters the
dressed matrix element $W$ through the dielectric 
function\cite{Ando1997},
\begin{equation}
\varepsilon(q) = 1 + 
\frac{2e^2}{\kappa_{\text{r}}}\,I_0\!\left(R\left|q\right|\right)\,
K_0\!\left(R\left|q\right|\right)\,\Pi(q).
\end{equation}
Here we use the simple ansatz
\begin{equation}
\Pi(q) = A_{\text{ansatz}}(Rq)^2,
\end{equation}
as this choice makes the dressed Coulomb interaction 
scale like the three-dimensional bare Coulomb
potential for large $q$ (i.e., at short distances),
$W \sim 1/q^2$. In Supplementary Fig.~5a, b
the dressed matrix 
element $W$ [empty circles, $A_{\text{ansatz}}=50/(\pi\gamma)$,
$\gamma/a=1.783$ eV]  
quantitatively agrees with its ab initio counterpart, 
$W^{\text{DFT}}$ (filled circles),
in the whole range of $k$ in which 
electrons are massless (cf.~Supplementary Fig.~2).
Note that for $k>\text{K}=0.289(2\pi)/a$
the effective-mass potentials are exactly zero 
whereas $W^{\text{DFT}}$ shows some numerical noise.

\begin{figure}
\setlength{\unitlength}{1 cm}
\begin{picture}(14.5,9.2)
\put(10.8,10.2){meV}
\put(9.0,10.0){\bf a}
\put(9.0,4.4){\bf b}
\put(9.5,5.5){\begin{turn}{30}$k$  (units of $2\pi/a$)\end{turn}}
\put(4.5,5.9){\begin{turn}{-10}$k'$  (units of $2\pi/a$)\end{turn}}
\put(3.4,9.8){$W(k,k')$}
\put(9.7,6.1){K}
\put(6.3,5.9){K}
\put(1.70,-1.57){\includegraphics[trim=0cm 0 0cm 0cm,clip=true,width=5.77in]{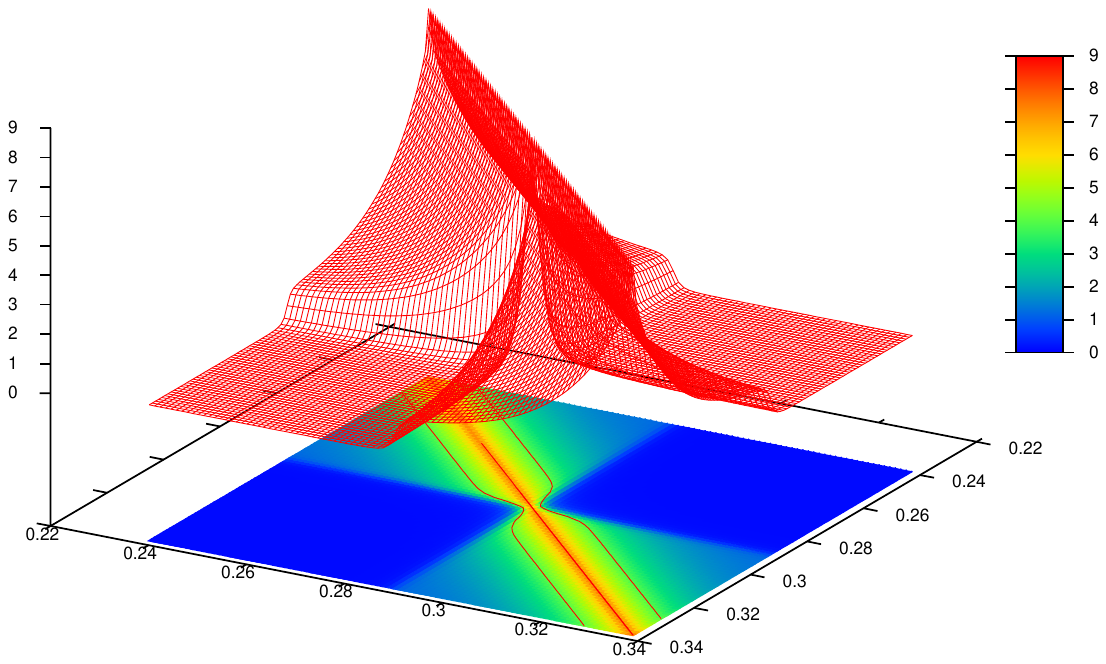}}
\put(1.70,4.02){\includegraphics[trim=0cm 0 0cm 0cm,clip=true,width=5.77in]{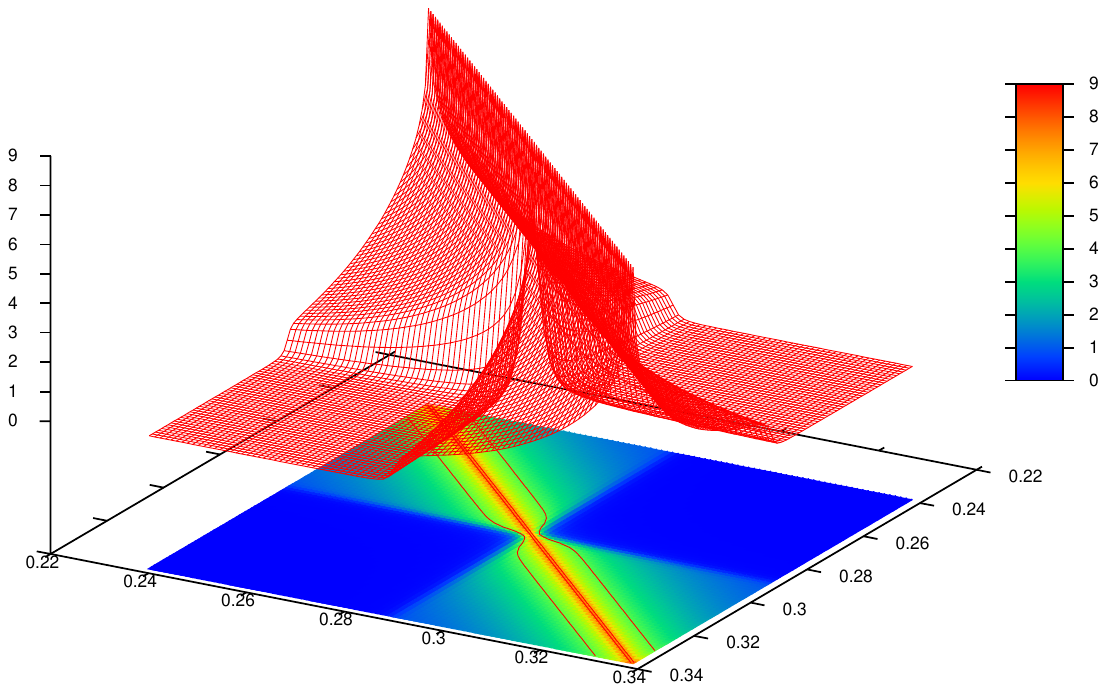}}
\put(3.4,4.2){$W^{\text{DFT}}(k,k')$}
\put(10.8,4.6){meV}
\put(9.5,-0.1){\begin{turn}{30}$k$  (units of $2\pi/a$)\end{turn}}
\put(4.5,0.3){\begin{turn}{-10}$k'$  (units of $2\pi/a$)\end{turn}}
\put(9.7,0.5){K}
\put(6.3,0.3){K}
\end{picture}
\caption*{{\bf Supplementary Fig.~6} 
Interband Coulomb matrix element
in the $(k,k')$ space in the presence
of a magnetic field.
Dominant interband Coulomb matrix element 
in the $(k,k')$ space close to the K point in the presence
of a magnetic field, with $\varphi=7.59\cdot 10^{-3}$. 
{\bf a} Effective-mass dressed matrix element $W(k,k')$, with 
$\kappa_{\text{r}}=10$ and $A_{\text{ansatz}}=50/(\pi\gamma)$.  
The isolines of the two-dimensional
contour map point to the heights of 4 and 8 meV, respectively.
{\bf b} Modulus of DFT screened matrix element $W^{\text{DFT}}(k,k')$
obtained within the random phase approximation.
Here $N=900$ and K = $0.289(2\pi)/a$.
}
\end{figure}

{\bf Effect of the magnetic field.}
The magnetic field along the tube axis adds an Aharonov-Bohm phase
to the transverse momentum, $k_{\perp}$. This breaks the chiral symmetry
${\cal{C}}$ of single-particle states,
alters the form factors of Supplementary Eq.~\eqref{eq:formfactor} 
(see Ando\cite{Ando1997}), 
and lifts the selection rule on $k$.
This is apparent from the smearing of the maps of Supplementary Fig.~6
close to the frontiers of the quadrants, $k,k'=$ K,
wheres at the same locations in Supplementary Fig.~4 (no field) 
the plots exhibit
sharp discontinuities. The cuts of Supplementary Fig.~6
along the line $k'=k_0$,
as shown in Supplementary Figs.~7a
and b for $k_0=0.289 (2\pi)/a$ and $0.28 (2\pi)/a$,
respectively, confirm the good agreement between
$W(k,k_0)$ and $W^{\text{DFT}}(k,k_0)$.

\begin{figure}
\setlength{\unitlength}{1 cm}
\begin{picture}(14.5,6.0)
\put(0.2,0.0){\includegraphics[trim=0cm 0 0cm 0cm,clip=true,width=3.1in]{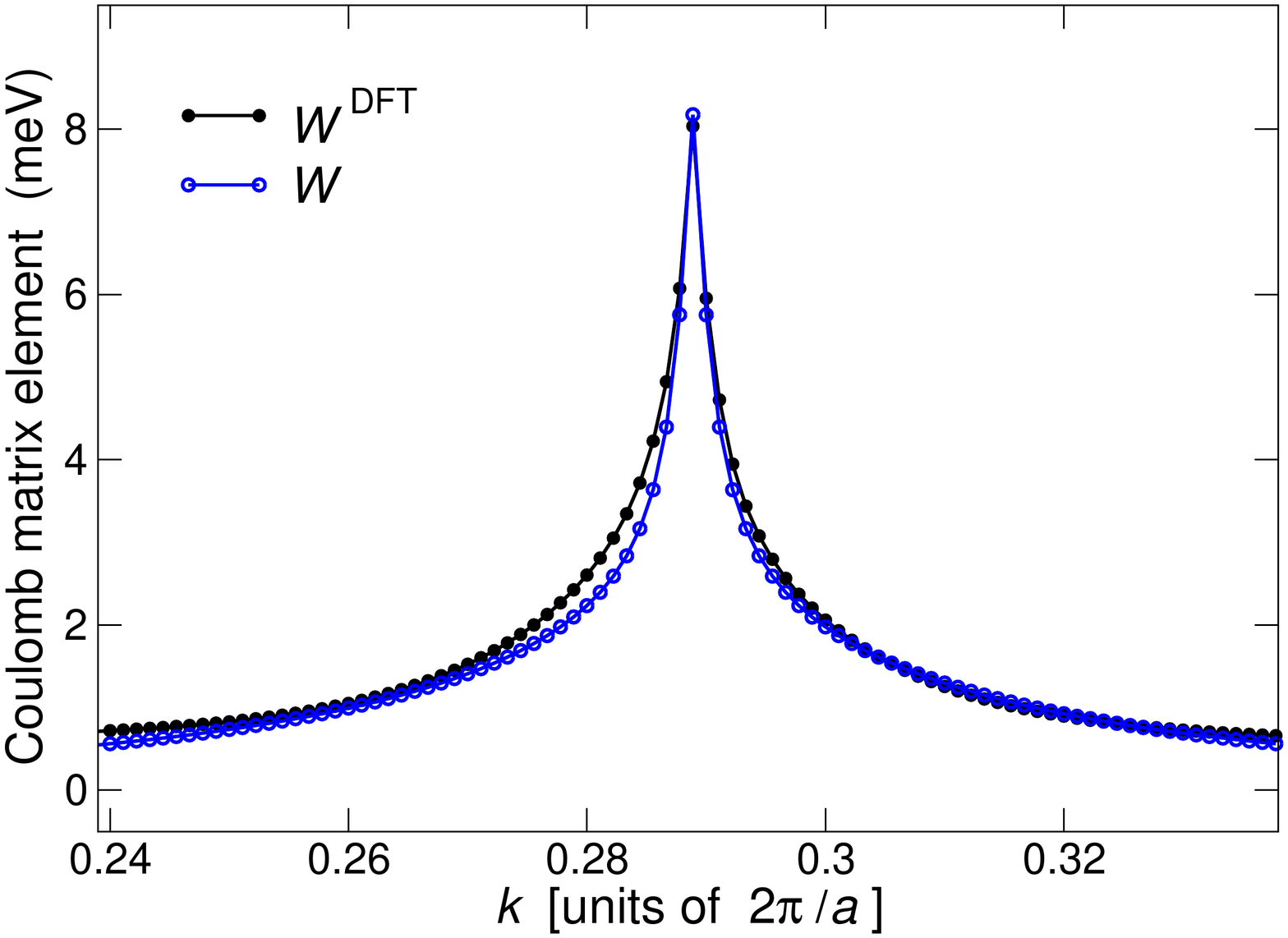}}
\put(8.2,0.0){\includegraphics[trim=0cm 0 0cm 0cm,clip=true,width=3.1in]{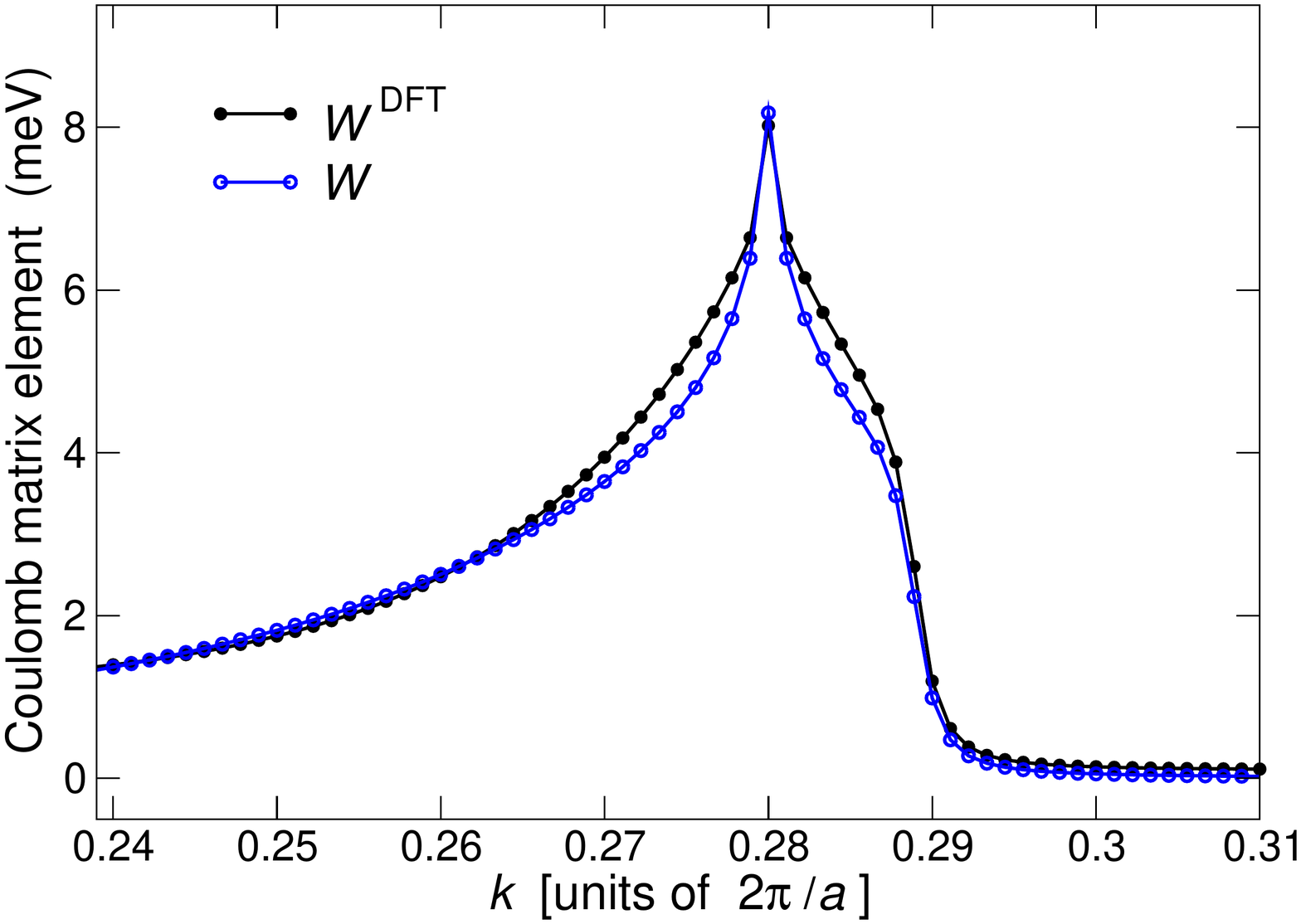}}
\put(6.2,1.4){\bf a}
\put(14.2,1.4){\bf b}
\end{picture}
\caption*{{\bf Supplementary Fig.~7} 
Coulomb matrix element vs $k$ in the presence 
of the magnetic field.
Dominant interband Coulomb matrix elements  
$W(k,k_0)$ (empty circles) and
$W^{\text{DFT}}(k,k_0)$ (filled circles) vs $k$ with fixed $k_0$ and
$\varphi=7.59\cdot 10^{-3}$. {\bf a}
$k_0=0.289(2\pi)/a$. {\bf b} $k_0=0.28(2\pi)/a$.
Lines are guides to the eye. $N=900$.
}
\end{figure}

\section*{Supplementary Note 3}

\section*{Effective mass: Bethe-Salpeter equation}\label{s:Bethe-Salpeter}

In this Note we detail the calculation of  
low-lying excitons of armchair carbon nanotubes, $\left|\text{u}\right>$,
within the effective mass theory.
The analysis of the first-principles exciton wave function 
for the (3,3) tube  
shows that the lowest conduction and highest valence
bands contribute more than 99.98\% to the spectral weight of excitons.
Therefore, according to conventional taxonomy, these excitons 
are of the $M_{00}$ type. 
Within the effective-mass approximation,
$\left|\text{u}\right>$ is written as 
\begin{equation}
\left|\text{u}\right>=\sum_{\sigma\sigma'\tau k}
\psi_{\tau}(k)\,\chi_{\sigma\sigma'} \,
\hat{c}^{\tau +}_{k,\sigma}\,
\hat{v}^{\tau }_{k,\sigma'}\! \left| 0 \right>,
\label{eq:exciton}
\end{equation}
where $\left| 0 \right>$ is the noninteracting ground state
with all valence states filled and conduction states empty,
and the operator $\hat{c}^{\tau +}_{k,\sigma}$ 
($\hat{v}^{\tau +}_{k,\sigma}$)
creates an electron in
the conduction (valence) band labeled by 
wave vector $k$, spin $\sigma$,
valley $\tau$. 
The exciton $\left|\text{u}\right>$ is a coherent
superposition of electron-hole pairs having zero center-of-mass momentum 
and amplitude $\psi_{\tau}(k)$. The latter may be regarded as
the exciton wave function in $k$ space.
The $2\times 2$ spin matrix $\chi_{\sigma\sigma'}$
is the identity for singlet excitons, $\bm{\chi}=\bm{1}_s$,
whereas for triplet excitons $\bm{\chi}=\bm{\sigma}_s\mathbf{\cdot n}$,
where $\mathbf{n}$ is the arbitrary direction
of the spin polarization ($\left|\mathbf{n}\right|=1$)
and $\bm{\sigma}_s$ is a vector made of the three
Pauli matrices. Throughout this work 
we ignore the small Zeeman term coupling the magnetic
field with electron spin, hence triplet excitons exhibit three-fold 
degeneracy. Here we use the same notation, 
$\left|\text{u}\right>$, for both
singlet and triplet excitons, as its meaning is clear from the context. 

The Bethe-Salpeter equation  
for the triplet exciton is
\begin{eqnarray}
&& E_{\text{eh}}(\tau, k)\,
\psi_{\tau}(k) 
- \frac{1}{A}\sum_{q} 
\tilde{W}^{\tau}\!( k+q,k )
\;\psi_{\tau}(k+q) \nonumber\\
&& \qquad 
- \quad \frac{1}{A}\sum_{\tau'\neq \tau}\sum_q 
\tilde{W}^{\tau\tau'}\!( k+q,k )
\;\psi_{\tau'}(k+q) 
= \varepsilon_{\text{u}}\, \psi_{\tau}(k). 
\label{eq:BSE_generic}
\end{eqnarray}
The diagonal term $E_{\text{eh}}(\tau, k)$ is the energy cost to 
create a free electron-hole pair $(\tau,c,k)(\tau,v,k)$,
\begin{equation}
E_{\text{eh}}(\tau, k) = 2\gamma
\sqrt{k_{\perp}^2 + k^2   }
+ 
\Sigma^{\tau}(k),
\end{equation}
including the sum of self-energy corrections
to electron and hole energies, $\Sigma^{\tau}(k)$, which may be
evaluated e.g.~within the $GW$ approximation.
This self-energy, which describes the dressing of electrons 
by means of the interaction with the other
electrons present in the tube, is responsible for the small 
asymmetry of the Dirac cone close to K, as shown by the
$GW$ dispersion of Supplementary Fig.~2a.
Since this asymmetry appears already at the DFT level of theory 
and is similar to the one predicted for the Dirac cones
of graphene\cite{Trevisanutto2008}, it necessarily
originates from mean-field
electron-electron interaction and it does not depend on $R$.
We take into account the effect of $\Sigma^{\tau}(k)$ onto
$E_{\text{eh}}(\tau, k)$ by explicitly considering 
different velocities (slopes of 
the linear dispersions) for respectively left- and right-moving fermions,
according to:
\begin{eqnarray}
E_{\text{eh}}(\text{K}, k) & = & 
2\gamma\left[1 + \alpha_{\text{sl}}\,\text{sign}(k)\right]
\sqrt{k_{\perp}^2 + k^2   },\nonumber\\
E_{\text{eh}}(\text{K}', k) & = & 
2\gamma\left[1 - \alpha_{\text{sl}}\, \text{sign}(k)\right]
\sqrt{k_{\perp}^2 + k^2   }.
\label{eq:asymmetry}
\end{eqnarray}
We infer the actual values of $\gamma$ and slope mismatch parameter 
$\alpha_{\text{sl}}$ from the linear fit 
to the first-principles $GW$ dispersion 
(in Supplementary Fig.~2b the solid lines 
are the fits and the dots the $GW$ data), which provides
$\gamma=5.449$ eV$\cdot${\AA} and $\alpha_{\text{sl}}=0.05929$.

\begin{figure}
\setlength{\unitlength}{1 cm}
\begin{picture}(14.5,5.8)
\put(2.2,-0.97){\includegraphics[trim=0cm 0 0cm 0cm,clip=true,width=4.2in]{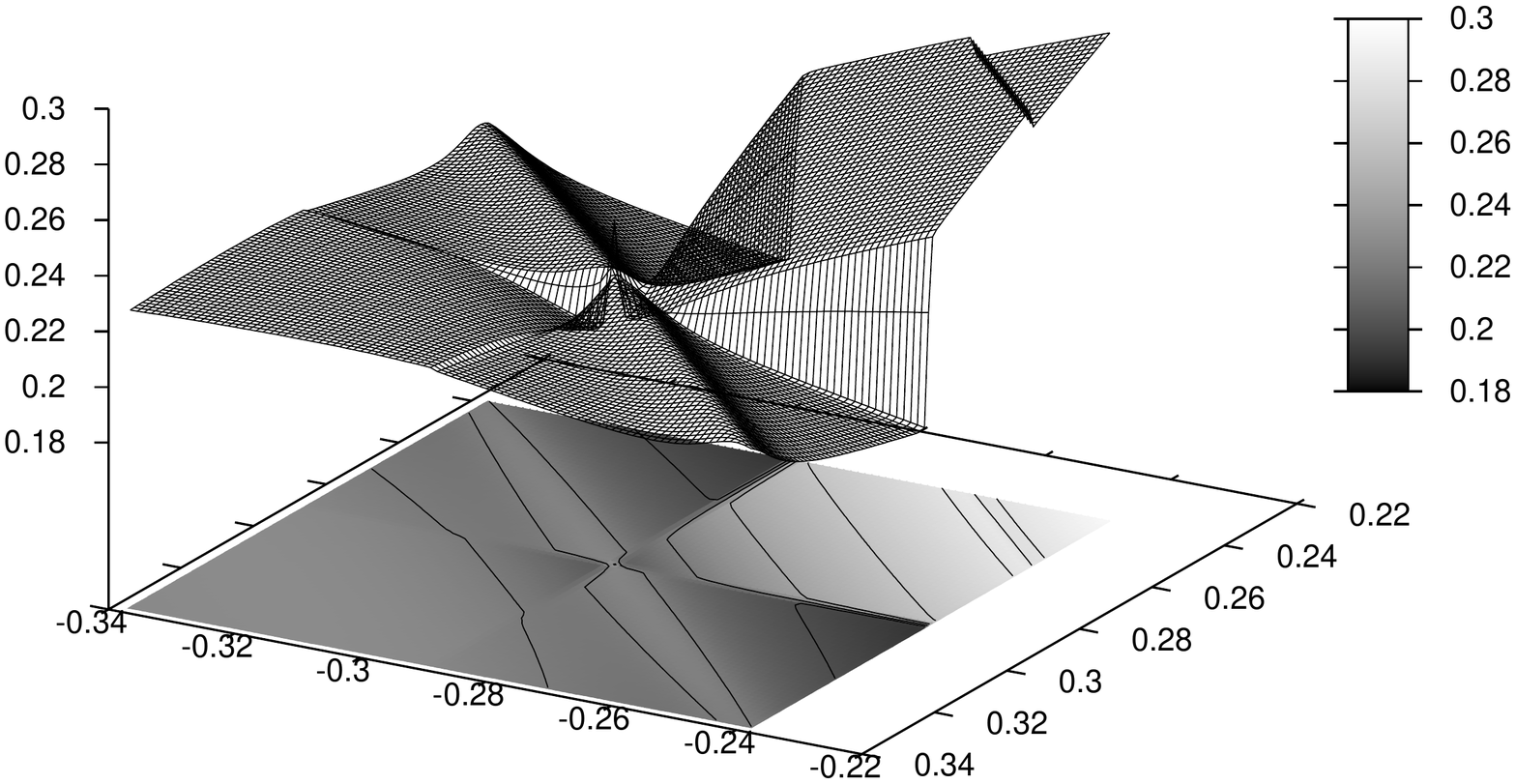}}
\put(3.4,4.2){$W^{\text{KK}'}(k,k')$}
\put(10.8,4.6){meV}
\put(9.5,-0.1){\begin{turn}{30}$k$  (units of $2\pi/a$)\end{turn}}
\put(4.0,0.1){\begin{turn}{-10}$k'$  (units of $2\pi/a$)\end{turn}}
\put(9.7,0.5){K}
\put(5.6,0.2){K$'$}
\end{picture}
\caption*{{\bf Supplementary Fig.~8} 
DFT intervalley interband Coulomb matrix element 
$W^{\text{KK}'}\!( k,k' )$
in $(k,k')$ space. 
Here $\varphi=1.52\cdot 10^{-3}$, $N=900$, $A=aN$,
and K = $=-\text{K}'= 0.289(2\pi)/a$.
}
\end{figure}

The second and third terms on the left hand side of Supplementary
Eq.~\eqref{eq:BSE_generic} involve interband Coulomb matrix elements.
The intravalley term $\tilde{W}^{\tau}$ is the dressed long-ranged 
interaction discussed in the previous Note.
The intervalley term $\tilde{W}^{\text{KK}'}$
makes electron-hole
pairs to hop between valleys.
As illustrated by the DFT map of 
$W^{\text{KK}'}\!( k,k' )= \tilde{W}^{\text{KK}'}/A$ in
$(k,k')$ space (Supplementary Fig.~8), 
this term, almost constant in reciprocal space,
is at least one order of magnitude 
smaller than $W$, as seen by
comparing the small range 0.18--0.3 meV of the energy axis of 
Supplementary Fig.~8
with the range 0--9 meV of Supplementary Figs.~4
and 6. Therefore, $W^{\text{KK}'}$
may be regarded as
a weak contact interaction that couples the valleys,
consistently
with the model by Ando\cite{Ando2006,Rontani2014},
\begin{equation}
\tilde{W}^{\text{KK}'}\!( k,k' ) = \frac{\Omega_0w_2}{4\pi R},
\end{equation}
where $\Omega_0 = (\sqrt{3}/2)a^2$ is the area of graphene unit cell
and $w_2>0$ is the characteristic energy associated with
short-range Coulomb interaction.
We reasonably reproduce first-principles results taking 
$w_2=2.6$ eV---this would be a plane located at 0.24 meV in 
Supplementary Fig.~8. 
This estimate is not far from Ando's prediction 
$w_2 = 4$ eV.
Note that the previous theory proposed by one of 
us\cite{Rontani2014} relies
on the scenario $W^{\text{KK}'} > W$, which is 
ruled out by the present study. 

The Bethe-Salpeter equation for the singlet exciton is obtained
from Supplementary Eq.~\eqref{eq:BSE_generic} by simply
adding to the kernel the bare exchange term 
\begin{equation}
+ \quad \frac{\Omega_0w_1}{2\pi R A}\sum_{\tau'}\sum_q \psi_{\tau'}(k+q),
\label{eq:J}
\end{equation}
where $w_1>0$ is a characteristic exchange energy\cite{Ando2006,Rontani2014}.
From first-principles results we estimate 
$w_1=4.33$ eV, whose magnitude is again comparable to that predicted
by Ando\cite{Ando2006}. Supplementary Eq.~(\ref{eq:BSE_generic}), 
with or without
 the exchange term, is solved numerically 
by means of standard linear algebra routines.

{\bf Minimal Bethe-Salpeter equation.}
The minimal Bethe-Salpeter equation illustrated in the main text includes
only one valley (with $\alpha_{\text{sl}} = 0$) and long-range
Coulomb interaction.
Within the effective-mass approximation, the Dirac cone indefinitely extends
in momentum space, hence one has to introduce a cutoff onto
allowed momenta, $\left|k\right| \le k_{\text{c}}$.
Supplementary Fig.~9a shows the convergence of the lowest-exciton
energy, $\varepsilon_{\text{u}}$, as a function of $k_{\text{c}}$. 
Reassuringly, $\varepsilon_{\text{u}}$
smoothly converges well within the range in which $GW$ bands are linear.
This is especially true for the screened interaction $W$ (black circles),
whereas the convergence is slower for the unscreened interaction $V$
(red circles), as it is obvious since $W(q)$ dies faster with increasing $q$.
This behavior implies that the energy scale associated with 
the exciton is intrinsic to the tube and unrelated to the cutoff, 
as we further discuss below. 

In the reported calculations we took $k_{\text{c}} = 0.05 (2\pi)/a$
as a good compromise between accuracy and computational burden
(we expect that the maximum absolute error on $\varepsilon_{\text{u}}$ 
is less than 0.1 meV). 
This corresponds to an energy cutoff of 1.4 eV for
e-h pair excitations. Whereas for these calculations, as well as for the
data of Supplementary Fig.~9a, the mesh $\Delta k$ in momentum space is 
fixed [$\Delta k = 1.43 \cdot 10^{-5}(2\pi)/a$],
Supplementary Fig.~9b shows 
the convergence of $\varepsilon_{\text{u}}$ as a 
function of the mesh, $\Delta k$.
Interestingly, $\varepsilon_{\text{u}}$ smoothly decreases with $\Delta k$
only for a very fine mesh, whereas for larger values of $\Delta k$ the
energy exhibits a non-monotonic behaviour. This is a consequence
of the logarithmic spike of the Colulomb potential 
at vanishing momentum, which requires a very fine mesh to be dealt with
accurately.

\begin{figure}
\setlength{\unitlength}{1 cm}
\begin{picture}(14.5,6.0)
\put(3.2,0.0){\includegraphics[trim=0cm 0 0cm 0cm,clip=true,width=3.6in]{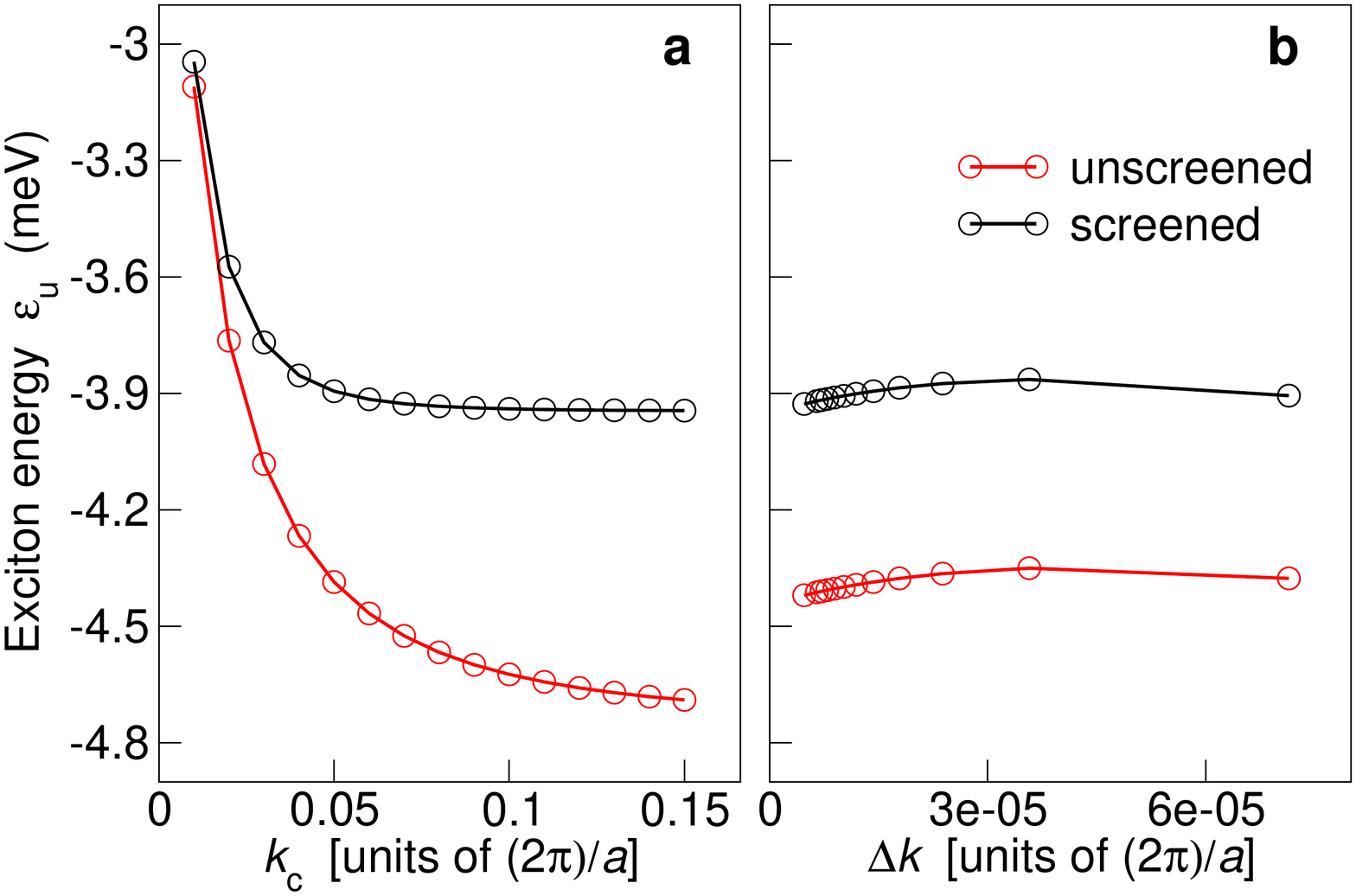}}
\end{picture}
\caption*{{\bf Supplementary Fig.~9} 
Convergence of exciton energy within a single valley in the
effective-mass approximation.
{\bf a} Excitation energy of the lowest exciton, $\varepsilon_{\text{u}}$,
vs cutoff in momentum space, $k_{\text{c}}$.
The black (red) curve is the energy obtained using the screened (unscreened)
long-range interaction, $W$ ($V$), in the Bethe-Salpeter equation for 
the triplet exciton.
Here $\Delta k = 1.43 \cdot 10^{-5}(2\pi)/a$ and
$\varphi=1.52\cdot 10^{-5}$.
{\bf b} Excitation energy of the lowest exciton,
$\varepsilon_{\text{u}}$, vs mesh in momentum space, $\Delta k$.
Here $k_{\text{c}} = 0.05 (2\pi)/a$ and
$\varphi=1.52\cdot 10^{-5}$.
}
\end{figure}

We refine the minimal effective-mass Bethe-Salpeter
equation by including:
(i) The short-range part of interaction, which
couples the two valleys as well as lifts the degeneracy of spin singlet
and triplet excitons.
(ii) The tiny difference between the e-h pair
excitation energies of left and right movers.
This eventually leads to a quantitative agreement with exciton energies and
wave functions obtained from first principles,
as shown by Fig.~3b, c  and Fig.~6a, b of main text.

{\bf Scaling properties of the Bethe-Salpeter equation.}
If a well-defined (i.e., bound and normalizable) 
solution of the Bethe-Salpeter
equation \eqref{eq:BSE_generic} exists, then it must own a characteristic 
length and energy scale---respectively the exciton Bohr
radius and binding energy\cite{Gronqvist2012}.
To check this, we introduce the scaling length $\ell$
to define the following dimensionless quantities: the wave vector 
$\kappa = k \ell$, the 
energy ${\cal{E}}_{\text{u}}=\varepsilon_{\text{u}} \ell / \gamma$,
and the exciton wave function
$\xi_{\tau}(\kappa)=\psi_{\tau}(k)/\sqrt{\ell}$.
We also define the dimensionless intravalley interaction as
$\Omega^{\tau}\!(kR,k'R)=(\kappa_{\text{r}}/e^2)\tilde{W}^{\tau}\!(k,k')$, 
to highlight that the wave vector $k$ appearing as an argument of
the interaction is always multiplied by $R$.
This is important for the exciton scaling behaviour.

Neglecting the small corrections to the exciton binding energy
due to intervalley scattering ($w_2=0$) and
cone asymmetry ($\alpha_{\text{sl}}=0$),
the dimensionless 
Bethe-Salpeter equation for armchair tubes in the absence of a magnetic flux
becomes
\begin{equation}
2\left|\kappa\right|\xi_{\tau}(\kappa)\,-\,
\frac{\alpha_{\text{graph}}}{2\pi}
\int d\kappa'\;\Omega^{\tau}[\kappa'(R/\ell),\kappa(R/\ell)]\;
\xi_{\tau}(\kappa') 
 =   {\cal{E}}_{\text{u}}\; \xi_{\tau}(\kappa),  
\label{eq:BSE_dimensionless}
\end{equation}
where $\alpha_{\text{graph}} = e^2/(\kappa_{\text{r}}\gamma)$ is graphene
fine-structure constant, 
the scaled exciton wave function must satisfy the scale invariant
normalization requirement, 
$\sum_{\tau} \int d\kappa \left|\xi_{\tau}(\kappa) \right|^2 = 1$,
and the dielectric function entering $\Omega$ takes the
dimensionless form
\begin{equation}
\varepsilon(\kappa) = 1 + 
\frac{2 A_{\text{ansatz}}}{\pi}\alpha_{\text{graph}}\kappa^2
(R/\ell)^2I_0\!\left(\left|\kappa\right|R/\ell\right)\,
K_0\!\left(\left|\kappa\right|\ell/R\right).
\label{eq:RPA_dimensionless}
\end{equation}

The only scaling length $\ell$ leaving 
Supplementary Eqs.~\eqref{eq:BSE_dimensionless}
and \eqref{eq:RPA_dimensionless}
invariant is the tube radius, $R$, wich fixes the binding energy unit,
$\gamma/R$. Supplementary Eq.~\eqref{eq:BSE_dimensionless}
shows that $\alpha_{\text{graph}}$ is the single parameter
combination affecting the scale invariant solution, whereas solutions
for different radius $R$ are related via scaling,
\begin{equation}
\varepsilon_{\text{u}}=\frac{E_0}{R},
\end{equation}
with $E_0$ being calculated once for all for the (3,3) tube radius,
$R=2$ \AA. The same conclusion holds for finite cone 
asymmetry $\alpha_{\text{sl}}$ and dimensionless magnetic flux $\varphi$.
Note that, for a fixed value of $\varphi$,
the possible values of the magnetic field $B$ scale like $1/R^2$.

The above demonstration relies on the assumption that
the parameters $\kappa_{\text{r}}$ and $A_{\text{ansatz}}$, which 
control the screening behavior of the carbon nanotube, do not depend
significantly on $R$. 
On the other hand, one might expect to recover, for large $R$,
the screening properties of graphene. This in turn would imply
that $\kappa_{\text{r}}$ would tend to smaller values and hence 
$\varepsilon_{\text{u}}$ would decay slower than $1/R$.
The first-principles investigation of this issue
is left to future work.

\section*{Supplementary Note 4}

\section*{Self-consistent mean-field theory of the excitonic insulator}

The ground-state wave function of the excitonic insulator, 
$\left|\Psi_{\text{EI}}\right>$, exhibits a BCS-like form,
\begin{equation}
\left|\Psi_{\text{EI}}\right>=\prod_{\sigma\sigma'\tau k}\left[u_{\tau k}
+ \chi_{\sigma\sigma'} v_{\tau k} e^{i \eta} \,
\hat{c}^{\tau +}_{k,\sigma}
\hat{v}^{\tau }_{k,\sigma'} \right] \left| 0 \right>,
\end{equation}
where $\eta$ is the arbitrary phase of the condensate, the e-h pairs
$\hat{c}^{\tau +}_{k,\sigma}
\hat{v}^{\tau }_{k,\sigma'}\left| 0 \right>$ 
replace the Cooper
pairs (e.g.~$\hat{c}^{\text{K} +}_{k,\sigma}
\hat{c}^{\text{K}' +}_{-k,-\sigma}\left| 0 \right>$), and the
$2\times 2$ matrix $\chi_{\sigma\sigma'}$
discriminates between
singlet and triplet spin symmetries. 
The positive variational quantities $u_{\tau k}$
and $v_{\tau k}$ are the population amplitudes
of valence and conduction levels, respectively,
which are determined at once with the excitonic 
order parameter, $\Delta(\tau k)$.
Explicitly, one has
\begin{eqnarray}
u_{\tau k}^2  & = &
\frac{1}{2}\left(1 + \frac{E_{\text{eh}}(\tau,k)/2}
{\left[ E^2_{\text{eh}}(\tau,k)/4+ 
\left|\Delta(\tau k)\right|^2\right]^{1/2}}\right),
\nonumber \\
v_{\tau k}^2  & = &  1 - u_{\tau k}^2,
\label{eq:uv_def}
\end{eqnarray}
plus the self-consistent equation for $\Delta$ 
[equivalent to Eq.~(2) of main text],
\begin{equation}
\left|\Delta(\tau k)\right|
 =  \frac{1}{A}\sum_{\tau' k'} 
\tilde{W}^{\tau\tau'}\!(k,k')\frac{
\left|\Delta(\tau' k')\right|}
{2\left[ E^2_{\text{eh}}(\tau',k')/4+ 
\left|\Delta(\tau' k')\right|^2\right]^{1/2}} .
\label{eq:gap_equation}
\end{equation}
The symbol $\tilde{W}^{\tau\tau'}\!(k,k')$
in Supplementary Eq.~(\ref{eq:gap_equation}) 
is a shorthand for both intra and intervalley Coulomb interaction 
matrix elements. 
For the spin-triplet EI [$\chi_{\sigma\sigma'} =
(\bm{\sigma}_s\mathbf{\cdot n})_{\sigma\sigma'}$],
which is the absolute ground state, one has, for $\tau=\tau'$, 
the long-range intravalley term,   
$\tilde{W}^{\tau\tau}\!(k,k')=\tilde{W}^{\tau}\!(k,k')$, 
and, for $\tau \neq \tau'$, the 
short-range intervalley term, 
$\tilde{W}^{\tau\tau'}\!(k,k')$.  
For the spin singlet 
($\chi_{\sigma\sigma'}=
\delta_{\sigma\sigma'}$), 
the unscreened direct term must be
subtracted from the dressed interaction, 
$\tilde{W}^{\tau\tau'}\!(k,k') \rightarrow 
\tilde{W}^{\tau\tau'}\!(k,k') - \Omega_0w_1/2\pi R$. 
Supplementary Eq.~(\ref{eq:gap_equation}) allows for a scaling
analysis similar to that for the exciton binding energy.

If interaction matrix elements $\tilde{W}$ were constant, 
then Supplementary Eq.~(\ref{eq:gap_equation}) 
would turn into the familiar 
gap equation of BCS theory, 
with $\Delta$ constant as well. Since the long-range
part of interaction is singular, 
the dependence of $\Delta(\tau k)$ on $\tau$ and $k$ cannot be neglected
and hence the solution is not obvious.
It is convenient to rewrite Supplementary 
Eq.~(\ref{eq:gap_equation}) as a pseudo
Bethe-Salpeter equation,
\begin{equation}
2\left[ E^2_{\text{eh}}(\tau,k)/4+ 
\left|\Delta(\tau k)\right|^2\right]^{1/2}\varphi(\tau k)
 -  \frac{1}{A}\sum_{\tau' k'} 
\tilde{W}^{\tau\tau'}\!(k,k')\; \varphi(\tau' k') = 0,
\label{eq:pseudo_Bethe}
\end{equation}
with the pseudo exciton wave function defined as
\begin{equation}
\varphi(\tau k) = \frac{
\left|\Delta(\tau k)\right|}
{2\left[ E^2_{\text{eh}}(\tau,k)/4+ 
\left|\Delta(\tau k)\right|^2\right]^{1/2}}.
\label{eq:phik}
\end{equation}
This shows that, at the onset of the EI phase, when $\Delta(\tau k)$ 
is infinitesimal---at the critical magnetic field---the 
exciton wave function for 
$\varepsilon_{\text{u}}=0$ is the same as $\varphi$ apart from a constant,
$\varphi(\tau k)\sim  \psi_{\tau}(k)$. This observation
suggests to use $\psi_{\tau}(k)$ at all values of the field as
a good ansatz to start the self-consistent cycle of 
Supplementary Eq.~(\ref{eq:pseudo_Bethe}),
which is numerically implemented as a matrix product having the form
${\bm{\varphi}}_{\text{new}} = {\bm{W}}\cdot {\bm{\varphi}}_{\text{old}}$.
Taking at the first iteration $\left|\Delta_{\text{old}}(\tau k)\right|
= 2^{-1}\left[E_{\text{eh}}(\tau,k)-\varepsilon_{\text{u}}\right]
\left|\psi_{\tau}(k)/\psi_{\tau}(0)\right|$ and building
$\varphi_{\text{old}}(\tau k)$ according to 
Supplementary Eq.~(\ref{eq:phik}),
we obtain numerical convergence within a few cycles, 
${\bm{\varphi}}_{\text{new}} = {\bm{\varphi}}_{\text{old}}$,
with the number of iterations increasing with decreasing $\Delta$. 
At finite temperatures, the self-consistent equation for 
$\Delta$ takes the form
\begin{eqnarray}
\left|\Delta(\tau k)\right|
& = & \frac{1}{A}\sum_{\tau' k'} 
\tilde{W}^{\tau\tau'}\!(k,k')\frac{
\left|\Delta(\tau' k')\right|}
{2\left[ E^2_{\text{eh}}(\tau',k')/4+ 
\left|\Delta(\tau' k')\right|^2\right]^{1/2}} \nonumber \\
& \times & \tanh{\left\{\frac{1}{2k_{\text{B}}T}
\left[ E^2_{\text{eh}}(\tau',k')/4+ 
\left|\Delta(\tau' k')\right|^2\right]^{1/2}\right\}} ,
\label{eq:gap_equation_T}
\end{eqnarray}
where $T$ is the temperature and $k_{\text{B}}$ is Boltzmann constant.

The quasiparticles of the EI are 
the free electrons and holes. 
For example, in the simplest case of the spin-singlet EI 
($\chi_{\sigma\sigma'} = \delta_{\sigma\sigma'}$), 
the electron quasiparticle wave function $\left|\Psi_{\text{EI}}^{\tau 
k\uparrow}\right>$ 
differs from the ground state
$\left|\Psi_{\text{EI}}
\right>$ as the conduction electron state
labeled by $(\tau, k,\uparrow)$ is occupied with
probability one as well as
the corresponding valence state:
\begin{equation}
\left|\Psi_{\text{EI}}^{\tau k\uparrow}\right> =
\hat{c}^{\tau +}_{k,\uparrow}
\left[u_{\tau k}
+ v_{\tau k} e^{i\eta} \, 
\hat{c}^{\tau +}_{k,\downarrow}
\hat{v}^{\tau }_{k,\downarrow} \right]
{\prod_{\sigma\tau' k'}}^{'}\left[u_{\tau' k'}
+ v_{\tau' k'} e^{i\eta} \,
\hat{c}^{\tau' +}_{k',\sigma}
\hat{v}^{\tau' }_{k',\sigma} \right] \left| 0 \right>.
\end{equation}
Here the symbol $\prod^{'}$ means that the dummy indices $\tau' k'$
take all values but $\tau k$.
The quasiparticle energy dispersion is 
\begin{equation}
E(\tau k)  =    
\sqrt{ E_{\text{eh}}^2(\tau,k) /4 
+ \left| \Delta(\tau k) \right|^2    ,              } 
\label{eq:Eqp}
\end{equation}
with the reference chemical potential being zero, as
for the noninteracting undoped ground state.
$E(\tau k)$ is 
increased quadratically by the amount $\left|\Delta(\tau k)\right|$
with respect to the noninteracting energy,
$\varepsilon(\tau k)=E_{\text{eh}}(\tau,k)/2$.
This extra energy cost is a collective effect
reminescent of the exciton binding energy, since now 
the exciton condensate must be ionized to unbind one e-h pair and hence have
a free electron and hole. 

\section*{Supplementary Note 5}

\section*{Inversion symmetry breaking in the excitonic insulator phase}

Carbon nanotubes inherit from graphene fundamental
symmetries such as time reversal and spatial inversion.
Time reversal $\hat{\mathbb{T}}$ swaps K and K$'$ valleys
whereas the inversion $\hat{\mathbb{I}}$ is a $\pi$ rotation 
around an axis perpendicular to the tube surface and 
located in the origin of one of the frames shown in 
Supplementary Fig.~1. 
This swaps the valleys as well as the A and B sublattices.
Whereas the noninteracting ground state $\left|0\right>$ is invariant under
both inversion and time reversal,
$\hat{\mathbb{T}}\left|0\right>=\left|0\right>$ and
$\hat{\mathbb{I}}\left|0\right>=\left|0\right>$,
the EI ground state breaks the inversion symmetry\cite{Portengen1996b}.
Here we consider a spin-singlet exciton condensate ($\chi_{\sigma\sigma'}
=\delta_{\sigma\sigma'}$) with 
$\hat{\mathbb{T}} \; \left|\Psi_{\text{EI}}\right>=
\left|\Psi_{\text{EI}}\right>$, hence the excitonic order parameter 
is real, $\eta=0,\pi$ (otherwise the EI ground state would
exhibit transverse orbital currents).

To see that the inversion symmetry of the EI ground state
is broken we use the following transformations 
(whose details are given in Supplementary Note 7):
\begin{eqnarray}
\hat{\mathbb{I}} \; \hat{v}^{\tau }_{k,\sigma} & = &
-i \, \text{sign}(k)
 \,\hat{v}^{-\tau }_{-k,\sigma},\nonumber \\
\hat{\mathbb{I}} \; \hat{c}^{\tau }_{k,\sigma} & = &
i\, \text{sign}(k)
 \,\hat{c}^{-\tau }_{-k,\sigma},
\end{eqnarray}
where the shorthand $-\tau$ labels the valley other than $\tau$.
The transformed ground state is
\begin{equation}
\hat{\mathbb{I}} \; \left|\Psi_{\text{EI}}\right> =
\prod_{\sigma\tau k}\left[u_{\tau k}
- v_{\tau k} e^{i \eta}  \,
\hat{c}^{\tau +}_{k,\sigma}
\hat{v}^{\tau }_{k,\sigma} \right] \left| 0 \right>,
\end{equation}
where we have used the fact that
$u_{\tau k}=u_{\tau k}^*=u_{-\tau -k}$ and
$v_{\tau k}=v_{\tau k}^*=v_{-\tau -k}$, as a consequence of
time reversal symmetry.
The original and transformed ground states are orthogonal
in the thermodynamic limit,
\begin{equation}
\left< \Psi_{\text{EI}}\right | \hat{\mathbb{I}} \left|\Psi_{\text{EI}}\right>
= 2\prod_{\tau k}\left(u_{\tau k}^2 - v^2_{\tau k}\right)
\rightarrow 0,
\end{equation}
since $u^2-v^2<1$.
On the contrary,
$\left<0\right | \hat{\mathbb{I}} \left|0\right>=1$.
Therefore, the symmetry of the EI ground state
is lower than that of the noninteracting ground state so the EI phase has
broken inversion symmetry, i.e., charge is displaced from A to B sublattice 
or vice versa.

\section*{Supplementary Note 6}

\section*{Charge displacement between A and B sublattices}

In this section we compute the 
charge displacement between A and B carbon sublattices
in the EI ground state.
To this aim we must average over the ground state
the space-resolved charge density
\begin{equation}
\varrho(\mathbf{r})=e \sum_i \delta(\mathbf{r}-\mathbf{r}_i),
\end{equation}
where the sum runs over all electrons in the Dirac valleys.
The explicit form of the charge density, in second quantization, is
\begin{eqnarray}
\hat{\varrho}(\mathbf{r}) 
& = &e \sum_{\tau k\tau' k' \sigma}\Big[
\varphi^*_{c\tau k}(\mathbf{r})\varphi_{c\tau' k'}(\mathbf{r})
\hat{c}^{\tau +}_{k,\sigma}\hat{c}^{\tau'}_{k',\sigma}
+ \varphi^*_{v\tau k}(\mathbf{r})\varphi_{v\tau' k'}(\mathbf{r})
\hat{v}^{\tau +}_{k,\sigma}\hat{v}^{\tau'}_{k',\sigma} \nonumber \\
& + & \varphi^*_{c\tau k}(\mathbf{r})\varphi_{v\tau' k'}(\mathbf{r})
\hat{c}^{\tau +}_{k,\sigma}\hat{v}^{\tau'}_{k',\sigma}
+ \varphi^*_{v\tau k}(\mathbf{r})\varphi_{c\tau' k'}(\mathbf{r})
\hat{v}^{\tau +}_{k,\sigma}\hat{c}^{\tau'}_{k',\sigma}
\Big].
\label{eq:rho_long}
\end{eqnarray}
We recall that the states of conduction ($\alpha=c$) and 
valence ($\alpha=v$) bands appearing in Supplementary
Eq.~(\ref{eq:rho_long}),
$\varphi_{\alpha\tau k}(\mathbf{r})$, are products of the
envelope functions $F$ times the Bloch states $\psi_{\tau}$
at Brillouin zone corners
$\tau = $ K, K$'$,
\begin{equation}
\varphi_{\alpha\tau k}(\mathbf{r}) = F^{\tau A}_{\alpha k}(\mathbf{r})\,
\psi_{\tau A}(\mathbf{r}) +
F^{\tau B}_{\alpha k}(\mathbf{r})\,
\psi_{\tau B}(\mathbf{r}),
\end{equation}
where $\psi_{\tau A}(\mathbf{r})$ [$\psi_{\tau B}(\mathbf{r})$]
is the component  
on the A (B) sublattice. Neglecting products
of functions localized on different
sublattices, like $\psi^*_{\tau A} \psi_{\tau B}$, 
as well as products of operators non diagonal in $\tau$ and $k$ indices,
which are immaterial when averaging over the ground state, one obtains:
\begin{eqnarray}
\hat{\varrho}(\mathbf{r}) 
& = &
\frac{e}{2AL}
\sum_{\tau}
\left[ \left|\psi_{\tau A}(\mathbf{r})\right|^2
+ \left|\psi_{\tau B}(\mathbf{r})\right|^2 \right] 
\sum_{k \sigma}
\left( \hat{v}^{\tau +}_{k,\sigma}
\hat{v}^{\tau}_{k,\sigma}  +
\hat{c}^{\tau +}_{k,\sigma}
\hat{c}^{\tau}_{k,\sigma} \right) \nonumber \\
& + &
\frac{e}{2AL}
\sum_{\tau}
\left[ \left|\psi_{\tau A}(\mathbf{r})\right|^2
- \left|\psi_{\tau B}(\mathbf{r})\right|^2 \right] 
\sum_{k \sigma}
\left( \hat{c}^{\tau +}_{k,\sigma}
\hat{v}^{\tau}_{k,\sigma}  +
\hat{v}^{\tau +}_{k,\sigma}
\hat{c}^{\tau}_{k,\sigma} \right) .
\label{eq:rho_explicit}
\end{eqnarray}

The first and second line on the right hand side of 
Supplementary Eq.~(\ref{eq:rho_explicit}) are 
respectively the intra and interband contribution to the charge density.
Only the intraband contribution survives when averaging
$\hat{\varrho}$ over $\left|0\right>$, providing the noninteracting
system with the uniform
background charge density $\varrho_0 (\mathbf{r}) $,
\begin{equation}
{\varrho}_0(\mathbf{r}) 
 =
\left<0\right|\hat{\varrho}(\mathbf{r}) \left|0\right>
=
\frac{e}{aL}
\sum_{\tau}
\left[ \left|\psi_{\tau A}(\mathbf{r})\right|^2
+ \left|\psi_{\tau B}(\mathbf{r})\right|^2 \right] ,
\end{equation}
with $\sum_k 1=A/a$. Since
$ \left|\psi_{\text{K} A}(\mathbf{r})\right|
= \left|\psi_{\text{K}' A}(\mathbf{r})\right| =
\left|\psi_{A}(\mathbf{r})\right| $, and similarly for B, this
expression may be further simplified as
\begin{equation}
{\varrho}_0(\mathbf{r})  =
\frac{2e}{aL}
\left[ \left|\psi_{A}(\mathbf{r})\right|^2
+ \left|\psi_{B}(\mathbf{r})\right|^2 \right] .
\end{equation}
It is clear from this equation that ${\varrho}_0$ is 
obtained by localizing the two $\pi$-band electrons 
uniformly on each sublattice site.
When averaging $\hat{\varrho}$ over $\left|\Psi_{\text{EI}}\right>$,
the charge density $\varrho (\mathbf{r}) $ 
exhibits an additional interband contribution,
\begin{eqnarray}
\varrho (\mathbf{r}) & = & \left< \Psi_{\text{EI}}\right|
\hat{\varrho}(\mathbf{r}) 
\left|\Psi_{\text{EI}}\right> = {\varrho}_0(\mathbf{r})
+ 
\frac{2e\cos{\eta}}{AL}
\left[ \left|\psi_{A}(\mathbf{r})\right|^2
- \left|\psi_{B}(\mathbf{r})\right|^2 \right] 
\sum_{\tau k} u_{\tau k}v_{\tau k},
\end{eqnarray} 
which is proportional to $\sum_{\tau k} u_{\tau k}v_{\tau k}$ 
and hence related to the EI order parameter.
This term, whose origin is similar to that of the transition density 
shown in Fig.~3d of main text, as it takes into account the 
polarization charge fluctuation between $\left|0\right>$ 
and a state with one or more
e-h pairs excited, is driven from the long-range excitonic correlations. 
Importantly, the charge displacement is uniform 
among all sites of a given sublattice and changes sign with sublattice,
the sign depending on the phase of the exciton condensate, $\eta$.
The charge displacement per electron, $\Delta e/e$, on---say---each A
site is
\begin{equation}
\frac{\Delta e}{e} = \frac{a \cos \eta }{A} \sum_{\tau k} u_{\tau k}v_{\tau k},
\end{equation}
which is the same as Eq.~(3) of main text.
In order to evaluate numerically $\Delta e/e$, 
for the sake of simplicity we neglect the exchange terms 
splitting the triplet and singlet order parameters
(i.e., we assume $w_1=0$). 
The quantum Monte Carlo
order parameter $\varrho_{\text{AB}}$ 
defined in the main text is, in absolute value, 
twice $\left|\Delta e/e\right|$ 
as there are two relevant electrons per site.

\section*{Supplementary Note 7}

\section*{Reference frame and symmetry operations}\label{s:frame}

The reference frame of the armchair carbon nanotube 
shown in Supplementary Fig.~1a 
is obtained by rigidly translating the frame used by   
Ando in a series of papers\cite{Ando1997,Ando2005,Ando2006},
recalled in Supplementary Fig.~1b. 
In Ando's frame the origin is placed on an atom of the B sublattice
and the $y$ axis is parallel to the tube axis, after a rotation
by the chiral angle $\alpha$ with respect to the $y^{\prime}$ axis of
graphene. 
On the basis of primitive
translation vectors of graphene $\mathbf{a}$ and $\mathbf{b}$
displayed in Supplementary Fig.~1b,
the chiral vector takes the form
$\mathbf{L}=-m\mathbf{a} -(n+m)\mathbf{b}$ when expressed in  
terms of the conventional\cite{Dresselhaus1998} 
chiral indices $(n,m)$.
For an equivalent choice of $\mathbf{L}$, one has $\alpha=\pi/6$
for $(n,n)$ armchair tubes and $\alpha=0$ for $(n,0)$
zigzag tubes. 

In the frame of Supplementary Fig.~1a 
used throughout this Supplementary Information, 
the vectors locating the sites of A and B sublattices
are respectively
\begin{equation}
\mathbf{R}^A_{n_a,n_b} = \mathbf{R}_0^A + n_a \mathbf{a} 
+ n_b \mathbf{b}
\end{equation}
and 
\begin{equation}
\mathbf{R}^B_{n_a,n_b} = \mathbf{R}_0^B + n_a \mathbf{a} + n_b \mathbf{b},
\end{equation}
where $(n_a,n_b)$ is a couple of integers and 
$\mathbf{R}_0^A$ ($\mathbf{R}_0^B$) is the basis vector pointing to 
the origin of the A (B) sublattice.
Besides, one has $\mathbf{a}\equiv a(\sqrt{3}/2,-1/2)$,
$\mathbf{b}\equiv a(0,1)$, $\mathbf{R}_0^A \equiv a(1/\sqrt{3},1/2)$,
$\mathbf{R}_0^B \equiv a(1/(2\sqrt{3}),0)$, where $a=2.46$ {\AA} is the 
lattice constant of graphene.
Among the equivalent
corners of graphene first Brillouin zone,
we have chosen as Dirac points
$\text{\bf{K}}\equiv \frac{2\pi}{a} (1/\sqrt{3}, 1/3)$
and $\text{\bf{K}}' = - \text{\bf{K}}$.
The corresponding Bloch states are:
\begin{eqnarray}
\psi_{\text{K}A}(\mathbf{r}) & = & \frac{1}{\sqrt{N}}\sum_{n_a,n_b}
e^{i\text{\bf{K}}\cdot \mathbf{R}^A_{n_a,n_b}}\,\phi_{\pi}
(\mathbf{r} - \mathbf{R}^A_{n_a,n_b}), \nonumber\\
\psi_{\text{K}B}(\mathbf{r}) & = & - e^{i\pi/6}\,
\omega \frac{1}{\sqrt{N}}\sum_{n_a,n_b}
e^{i\text{\bf{K}}\cdot \mathbf{R}^B_{n_a,n_b}}\,\phi_{\pi}
(\mathbf{r} - \mathbf{R}^B_{n_a,n_b}), \nonumber\\
\psi_{\text{K}'A}(\mathbf{r}) & = & 
e^{i\pi/6}\, \omega^{-1}
\frac{1}{\sqrt{N}}\sum_{n_a,n_b}
e^{i\text{\bf{K}}'\cdot \mathbf{R}^A_{n_a,n_b}}\,\phi_{\pi}
(\mathbf{r} - \mathbf{R}^A_{n_a,n_b}), \nonumber\\
\psi_{\text{K}'B}(\mathbf{r}) & = &  \omega\frac{1}{\sqrt{N}}\sum_{n_a,n_b}
e^{i\text{\bf{K}}'\cdot \mathbf{R}^B_{n_a,n_b}}\,\phi_{\pi}
(\mathbf{r} - \mathbf{R}^B_{n_a,n_b}),
\label{eq:Bloch_armchair}
\end{eqnarray}
where $N$ is the number of sublattice sites,
$\phi_{\pi}(\mathbf{r})$ is the $2p_z$ carbon
orbital perpendicular to the graphene plane, normalized as in 
Secchi \& Rontani\cite{Secchi2010},
and $\omega=\exp{(i2\pi/3)}$.

The relative phase between the two sublattice
components of Bloch states within each valley, shown in 
Supplementary Eq.~(\ref{eq:Bloch_armchair}), is determined
by symmetry considerations\cite{Slonczewski1958}.
Specifically, the sublattice pseudospinor 
transforms as a valley-specific 
irreducible representation of the symmetry point group
of the triangle, $C_{3v}$:
\begin{equation}
\bm{F}_{\text{K} \alpha k} \sim 
{  x - iy \choose  x + iy  } ,\qquad
\bm{F}_{\text{K}' \alpha k} \sim 
{  x + iy \choose  -x + iy  } .
\label{eq:transf} 
\end{equation}

In addition, the relative phase between Bloch states of
different valleys is fixed by exploiting the additional $C_2$
symmetry. The latter consists of a rotation of a $\pi$ angle
around the axis perpendicular 
to the graphene
plane and intercepting the frame origin.
This rotation, which in the $xy$ space is equivalent to the 
inversion $\hat{\mathbb{I}}$, swaps K and 
K$'$ valleys as well as A
and B sublattices. With the choice of phases explicited in  
Supplementary Eq.~(\ref{eq:Bloch_armchair})
the inversion operator $\hat{\mathbb{I}}$ takes the 
form 
\begin{equation}
\hat{\mathbb{I}} = - {\bm{\sigma}}_y \otimes {\bm{\tau}}_y\, \hat{R},
\label{eq:inversion}
\end{equation}
where $\hat{R}$ is the inversion operator in the $xy$ space.
In contrast, the time-reversal operator $\hat{\mathbb{T}}$ 
swaps valleys but not sublattices,
\begin{equation}
\hat{\mathbb{T}} = {\bm{\sigma}}_z \otimes {\bm{\tau}}_x\, \hat{K},
\label{eq:time_reversal}
\end{equation}
where $\hat{K}$ is the complex-conjugation operator.
The orthogonal time-reversal of 
Supplementary Eq.~(\ref{eq:time_reversal}) should
not be confused with the symplectic transformation\cite{Suzuura2002},
which does not exchange valleys. 

The magnetic field along the tube
axis breaks both $\hat{\mathbb{I}}$ and $\hat{\mathbb{T}}$ symmetries.
However, the reflection symmetry $y\rightarrow -y$ 
along the tube axis still swaps the valleys (but
not sublattices), as it may be easily seen from a judicious choice
of K and K$'$ Dirac points. This protects the degeneracy of states
belonging to different valleys in the presence of a magnetic field.

\section*{Supplementary Discussion}

\section*{Effects of Dirac cone asymmetry and magnetic field
on the exciton wave function}

The origin of the 
asymmetry of the exciton wave function in $k$ space, illustrated 
by Fig.~3b of main text, may be understood within the effective mass 
model applied to a single Dirac valley---say K.
In the presence of a
vanishing gap, electrons 
(and excitons) acquire a chiral
quantum number, $\cal{C}$, which was defined above.
With reference to the noninteracting ground state, $\left|0\right>$,
the e-h pairs 
$\hat{c}^{K \dagger}_{k,\sigma}\hat{v}^K_{k,\sigma'}\left|0\right>$
have chiral quantum number $\Delta {\cal{C}}=+2$ for positive $k$ 
and $\Delta {\cal{C}}=-2$ for negative $k$. Since long-range 
Coulomb interaction conserves chirality, 
we expect the wave function of a chiral exciton to live only 
on one semi-axis in $k$ space,
either $\psi_K(k)=0$ for $k<0$ and $\Delta {\cal{C}}=+2$, or
$\psi_K(k)=0$ for $k>0$ and $\Delta {\cal{C}}=-2$.

Supplementary Fig.~10 plots $\psi_K(k)$ by comparing 
the case of a perfectly symmetric Dirac cone (panel a, 
$\alpha_{\text{sl}}=0$) with the case of a distorted cone, mimicking
the first-principles $GW$ band dispersion (panel b, 
$\alpha_{\text{sl}}=0.05929$). This analysis is of course possible
only within the effective mass model, as no free parameter
such as $\alpha_{\text{sl}}$ may be changed in the first-principles
calculation.
In the symmetric case (Supplementary Fig.~10a)
$\psi_K(k)$ is even in $k$ since nothing prevents 
the numerical diagonalization routine from mixing the two degenerate
amplitude distributions with $\Delta {\cal{C}}=\pm 2$. Hovever, as
the Dirac cone symmetry under axis inversion, $k\rightarrow -k $, 
is lifted by energetically favoring e-h pairs with 
$\Delta {\cal{C}}=-2$ (Supplementary Fig.~10b),
the wave function weight collapses on the negative side of the axis.
Therefore, the asymmetry of the exciton wave function is explained
by the combined effects of chiral symmetry and cone distortion.

\begin{figure}
\setlength{\unitlength}{1 cm}
\begin{picture}(14.5,8.5)
\put(3.6,0.0){\includegraphics[trim=0cm 0 0cm 0cm,clip=true,width=3.2in]{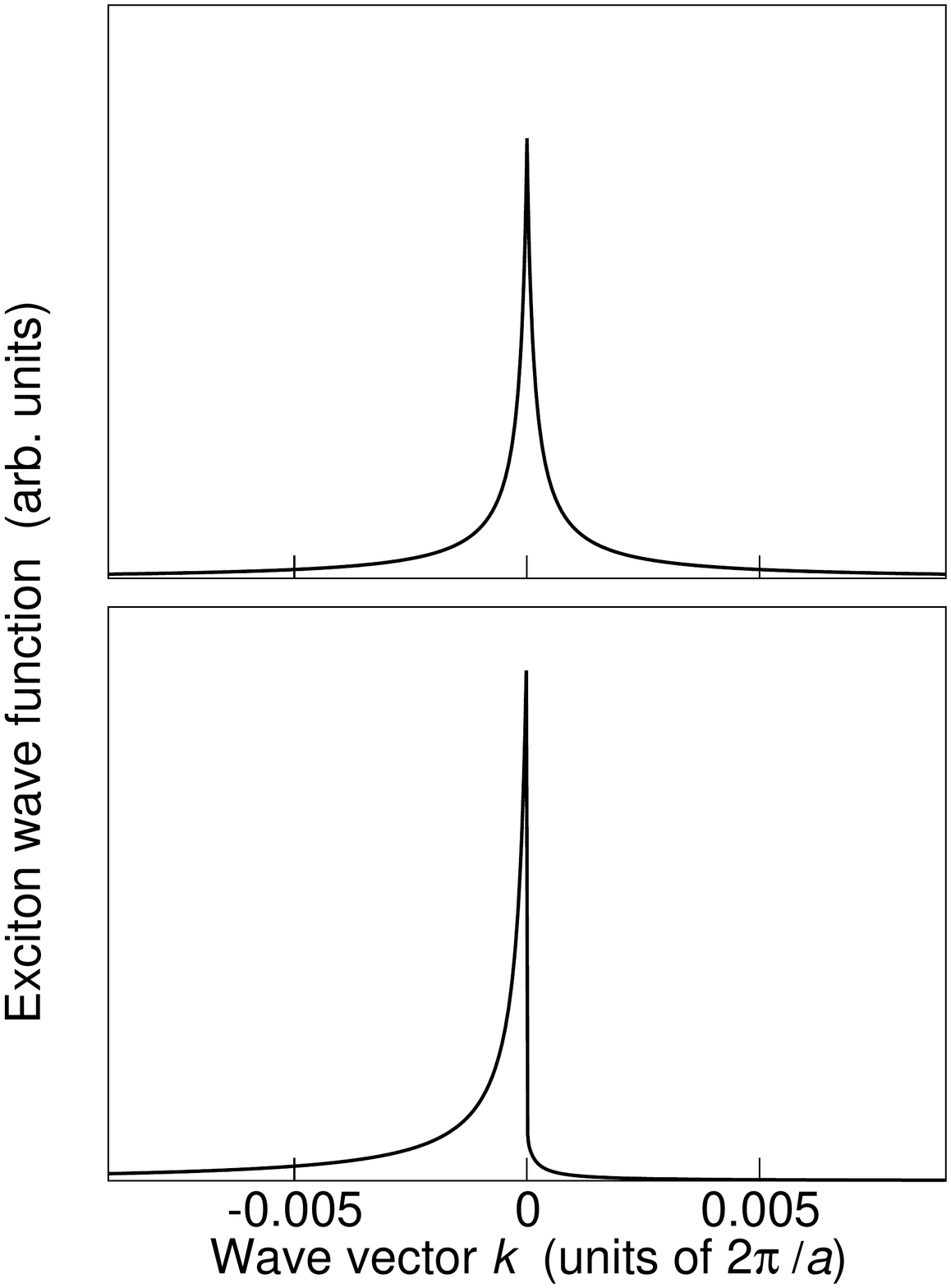}}
\put(8.8,8.0){$\alpha_{\text{sl}}=0$}
\put(8.8,3.6){$\alpha_{\text{sl}}=0.059$}
\put(5.3,9.5){\bf a}
\put(5.3,5.0){\bf b}
\end{picture}
\caption*{{\bf Supplementary Fig.~10} 
Effect of the asymmetry of the Dirac cone 
on the exciton wave function
within the effective mass approximation.
Wave function of the lowest-energy
exciton within a
single valley, $\psi_{\text{K}}(k)$, vs wave vector, $k$. 
{\bf a} The slope asymmetry
parameter has a vanishing value, $\alpha_{\text{sl}}=0$, hence the 
Dirac cone is symmetric under axis inversion, $k\rightarrow -k$.
{\bf b} $\alpha_{\text{sl}}=0.05929$.
Here $w_2=w_1=0$ and $\varphi=1.52\cdot 10^{-5}$.
}
\end{figure}

\begin{figure}
\setlength{\unitlength}{1 cm}
\begin{picture}(14.5,9.0)
\put(3.0,0.0){\includegraphics[trim=0cm 0cm 0cm 7cm,clip=true,width=3.9in]{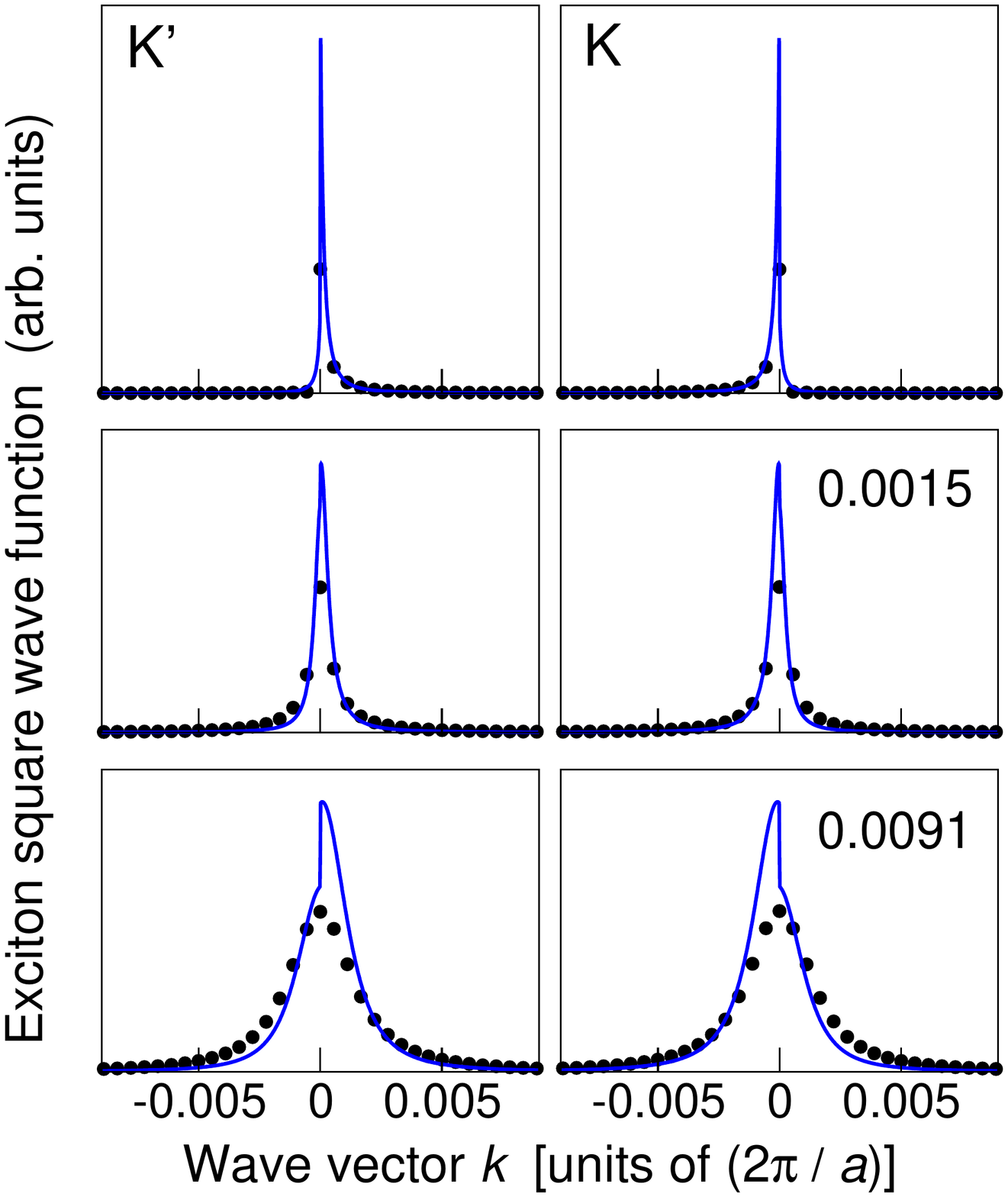}}
\put(10.56,10.1){$\phi/\phi_0=0$}
\put(4.96,7.7){\bf a}
\put(4.96,4.85){\bf c}
\put(4.96,2.0){\bf e}
\put(8.83,7.7){\bf b}
\put(8.83,4.85){\bf d}
\put(8.83,2.0){\bf f}
\end{picture}
\caption*{{\bf Supplementary Fig.~11} 
Effect of the magnetic field on the exciton wave function
in $k$ space.
Square modulus of the wave function of the lowest triplet exciton,
$\left|\psi_{\tau}(k)\right|^2$,
vs wave vector, $k$, for increasing values of the magnetic flux,
$\phi/\phi_0$. {\bf a, b} $\phi/\phi_0=$ 0. 
{\bf c, d} $\phi/\phi_0=$ 0.0015. 
{\bf e, f} $\phi/\phi_0=$ 0.0091.
Panels a, c, e (b, d, f) refer to valley K$'$ (K).
The first-principles data for the (3,3) tube (black dots) are compared
with the effective-mass predictions (blue curves).
As the field increases
the weight distribution becomes broader and more
symmetrical around the Dirac point.
}
\end{figure}

As the chiral symmetry is destroyed by piercing the tube with a magnetic flux,
the exciton wave function $\psi$ becomes symmetrically distibuted
around the origin of $k$ axis. This is shown in Supplementary Fig.~11
from both first-principles (dots) and effective-mass (solid curves)
calculations of $\left|\psi_{\tau}(k)\right|^2$
for increasing values of the dimensionless magnetic flux $\phi/\phi_0$
(for panels a,b $\phi/\phi_0=$ 0, for c, d $\phi/\phi_0=$ 0.0015, 
for e, f $\phi/\phi_0=$ 0.0091).
As the gap increases with the field, the excitons becomes massive and
more similar to the conventional Wannier excitons reported
in the literature\cite{Ando1997,Maultzsch2005,Spataru2004}: 
the weight distribution in $k$ space
is broader and its peak more rounded, with a Gaussian-like
shape identical in both valleys (respectively valley K$'$ in panels a, c, e
and valley K in panels b, d, f).
The agreement between first-principles (dots) and 
effective-mass (solid lines)
predictions is very good, further validating the model.     
However, at high field (panels e and f, $\phi/\phi_0=0.0091$),
the effective-mass curve becomes discontinuous at the Dirac cone
whereas the first-principles curve is smooth.
This is an artefact of the effective-mass model as the high-field
functional form of the distorted Dirac cone shown
by Supplementary Eq.~(\ref{eq:asymmetry})
exhibits a step at $k=0$ that increases with $k_{\perp}$.
This crude modelization may be cured rather simply: however,
its drawbacks do not affect the results presented in this paper in any 
significant way.

\section*{The EI mean-field wave function as specialization of 
the QMC variational wave function}

The QMC variational wave function, 
$\left|\Psi_{\text{QMC}}\right>$, is the 
zero-gap state, $\left|0\right>$, multiplied by the Jastrow factor,
$J=J_1J_2$,
which accounts for one- and two-body correlations
encoding the variational degrees of freedom.
In this section we 
focus on a relevant specialization of the pair Jastrow factor,
$J_2=\prod_{i<j} \exp{\!\left[u(\mathbf{r}_i,\mathbf{r}_j)\right]}$, 
showing that a proper choice of
the two-body term $u(\mathbf{r},\mathbf{r}')$
allows to 
recover the mean-field EI wave function
to first order in $u$,
i.e., 
$\left|\Psi_{\text{QMC}}\right>$  takes the form
\begin{equation}
\left|\Psi_{\text{EI}}\right>=\prod_{k}(u_{k}
+ v_{k} \,
\hat{c}^{\dagger}_{k}
\hat{v}_{k} ) \left| 0 \right>.
\end{equation}
Note that the first-order restriction is consistent with the 
range of validity of EI mean-field theory\cite{Kohn1967b}. 
Throughout this section we take $J_1=1$ and suppress
spin and valley indices, as they may be included
straightforwardly in the derivation,
as well as we assume positive order parameter 
for the sake of clarity ($\eta=0$). 

To first order in the two-body factor $u$,
the QMC wave function is
\begin{equation}
\Psi_{\text{QMC}}(\mathbf{r}_1,\mathbf{r}_2,\ldots,\mathbf{r}_{N_{\text{e}}})
= \left[ 1 + 
\sum_{i<j} u(\mathbf{r}_i,\mathbf{r}_j)\right] 
\Phi_0(\mathbf{r}_1,\mathbf{r}_2,\ldots,\mathbf{r}_{N_{\text{e}}}),
\label{eq:VQMCpert}
\end{equation}
where $N_{\text{e}}$ is the number of electrons. 
The Slater determinant
$\Phi_0$ in real space is obtained 
by projecting $\left|0\right>$ onto
\begin{equation}
\hat{\psi}^{\dagger}\!(\mathbf{r}_1)\hat{\psi}^{\dagger}\!(\mathbf{r}_2)
\ldots \hat{\psi}^{\dagger}\!(\mathbf{r}_{N_{\text{e}}})\left|\text{vac}\right>,
\end{equation}
where $\left|\text{vac}\right>$ is the vacuum with no electrons present.
The Fermi field annihilation
operator $\hat{\psi}$ is spanned by the basis of conduction and
valence band operators,
\begin{equation}
\hat{\psi}(\mathbf{r}) = \hat{\psi}_c(\mathbf{r})+\hat{\psi}_v(\mathbf{r}),
\end{equation}
with
\begin{equation}
\hat{\psi}_c(\mathbf{r})=\sum_k\varphi_{ck}(\mathbf{r})\,\hat{c}_k
\label{eq:basis_c}
\end{equation}
and
\begin{equation}
\hat{\psi}_v(\mathbf{r})=\sum_k\varphi_{vk}(\mathbf{r})\,\hat{v}_k,
\label{eq:basis_v}
\end{equation}
where the explicit effective-mass form of Bloch states $\varphi_{ck}$
and $\varphi_{vk}$ was given in Supplementary Note 1.

Similarly, we work out the form of $\Psi_{\text{EI}}$ in real
space, 
\begin{eqnarray}
&& \Psi_{\text{EI}}(\mathbf{r}_1,\mathbf{r}_2,\ldots,\mathbf{r}_{N_{\text{e}}})
=\left<\text{vac}\right|\hat{\psi}(\mathbf{r}_{N_{\text{e}}})
\hat{\psi}(\mathbf{r}_{N_{\text{e}} - 1})\ldots \hat{\psi}(\mathbf{r}_1) \nonumber\\
&&\times\quad \prod_{k} u_{k}\left(1 + g_k\, \hat{c}^{\dagger}_{k}
\hat{v}_{k} \right)\hat{v}^{\dagger}_{k_1}\hat{v}^{\dagger}_{k_2} 
\ldots \hat{v}^{\dagger}_{k_{N_{\text{e}}}} \left| \text{vac}\right>,
\end{eqnarray}
where the valence band states $k_1$, $k_2$, $\ldots$, $k_{N_{\text{e}}}$, 
are filled up to the Dirac point in $\left|0\right>$ and 
we defined $g_k = v_k / u_k$.
To first order in $g_k$, $\Psi_{\text{EI}}$ reads
\begin{eqnarray}
&& \Psi_{\text{EI}}(\mathbf{r}_1,\mathbf{r}_2,\ldots,\mathbf{r}_{N_{\text{e}}})
= B\, \Phi_0(\mathbf{r}_1,\mathbf{r}_2,\ldots,\mathbf{r}_{N_{\text{e}}}) \nonumber \\
&& + \quad B \sum_k g_k \left<\text{vac}\right|\hat{\psi}(\mathbf{r}_{N_{\text{e}}})
\hat{\psi}(\mathbf{r}_{N_{\text{e}} - 1})\ldots \hat{\psi}(\mathbf{r}_1)\;
\hat{c}^{\dagger}_{k}
\hat{v}_{k}\; \hat{v}^{\dagger}_{k_1}\hat{v}^{\dagger}_{k_2} 
\ldots \hat{v}^{\dagger}_{k_{N_{\text{e}}}} \left| \text{vac}\right>,
\end{eqnarray}
where $B=\prod_{k} u_{k}$ is a constant. 
After expanding the field operators $\hat{\psi}$
in the second row onto the basis of $\hat{v}$ and $\hat{c}$
[cf.~(\ref{eq:basis_v}) and (\ref{eq:basis_c})], 
we observe that the only non-vanishing contributions
consist in products of $N_{\text{e}}-1$ operators
$\hat{v}$ times a single operator $\hat{c}_k$. Since 
$\hat{c}_k$ occurs $N_{\text{e}}$ times in the $\hat{\psi}(\mathbf{r}_i)$'s, with
$i=1,\ldots,N_{\text{e}}$, we may write
\begin{eqnarray}
&& \Psi_{\text{EI}}(\mathbf{r}_1,\mathbf{r}_2,\ldots,\mathbf{r}_{N_{\text{e}}})
= B\, \Phi_0(\mathbf{r}_1,\mathbf{r}_2,\ldots,\mathbf{r}_{N_{\text{e}}})  
\quad + \quad B \sum_k g_k \sum_{i=1}^{N_{\text{e}}} 
\sum_{k_1'}\cdots
\sum_{k_{i-1}'}\sum_{k_{i+1}'}\cdots \sum_{k_{N_{\text{e}}}'}
\nonumber \\
&& \times \quad 
\varphi_{ck}(\mathbf{r}_i) \; \varphi_{vk_1'}(\mathbf{r}_1)\ldots
\varphi_{vk_{i-1}'}(\mathbf{r}_{i-1}) \varphi_{vk_{i+1}'}(\mathbf{r}_{i+1})
\ldots \varphi_{vk_{N_{\text{e}}}'}(\mathbf{r}_{N_{\text{e}}})
\nonumber \\
&&\quad \times\quad \left<\text{vac}\right|\hat{v}_{k_{N_{\text{e}}}'}\ldots
\hat{v}_{k_{i+1}'}\hat{c}_k\hat{v}_{k_{i-1}'}  \ldots \hat{v}_{k_{1}'}\;
\hat{c}^{\dagger}_{k}
\hat{v}_{k}\; \hat{v}^{\dagger}_{k_1}\hat{v}^{\dagger}_{k_2} 
\ldots \hat{v}^{\dagger}_{k_{N_{\text{e}}}} \left| \text{vac}\right>.
\label{eq:monstre}
\end{eqnarray}

To make progress, we consider 
the generic operator identity
\begin{equation}
\hat{\psi}(\mathbf{r}) \hat{\psi}^{\dagger}(\mathbf{r}')
+ \hat{\psi}^{\dagger}(\mathbf{r}') \hat{\psi}(\mathbf{r}) = 
\delta{(\mathbf{r}-\mathbf{r}')}.
\end{equation}
Since electrons are mainly localized at honeycomb lattice sites $\mathbf{R}$
and there is---on the average---one electron per site ($N_{\text{e}}=2N$),
this identity may approximately be expressed as 
\begin{equation}
\hat{\psi}_v(\mathbf{r}_i) \hat{\psi}_v^{\dagger}(\mathbf{r}_j)
+ \hat{\psi}_v^{\dagger}(\mathbf{r}_j) \hat{\psi}_v(\mathbf{r}_i) \approx 
\frac{\delta_{\mathbf{r}_i,\mathbf{r}_j}}{aLN},
\end{equation}
which provides a useful representation of the identity operator $\hat{I}$
for any position of the $i$th electron:  
\begin{equation}
aLN \sum_{j=1}^{N_{\text{e}}}\left[ \hat{\psi}_v(\mathbf{r}_i) 
\hat{\psi}_v^{\dagger}(\mathbf{r}_j)
+ \hat{\psi}_v^{\dagger}(\mathbf{r}_j) \hat{\psi}_v(\mathbf{r}_i)\right] \approx
\hat{I}.
\end{equation}
Furthermore, in the spectral representation of $\hat{I}$
we single out the contribution of momentum $k$,
\begin{equation}
\hat{I}  \approx  aLN \sum_{j=1}^{N_{\text{e}}} \left[ \varphi_{vk}(\mathbf{r}_i)
\varphi_{vk}^*(\mathbf{r}_j)\left( \hat{v}_k \hat{v}^{\dagger}_k
+ \hat{v}^{\dagger}_k \hat{v}_k \right) 
 + \sum_{k'\neq k} 
\varphi_{vk'}(\mathbf{r}_i)
\varphi_{vk'}^*(\mathbf{r}_j) \left( \hat{v}_{k'} \hat{v}^{\dagger}_{k'}
+ \hat{v}^{\dagger}_{k'} \hat{v}_{k'} \right) \right],
\label{eq:spectral}
\end{equation}
which we plug into Supplementary Eq.~(\ref{eq:monstre}).
Note that,
unless $\mathbf{r}_i=\mathbf{r}_j$, the contribution originating
from the second addendum between square
brackets of Supplementary Eq.~(\ref{eq:spectral}) is
much smaller than the one linked to the first addendum because
terms that are summed over $k'$ 
cancel out as they have random phases,
being proportional to $\exp{[ik'(y_i - y_j)]}$. 
The outcome is 
\begin{eqnarray}
&& \Psi_{\text{EI}}(\mathbf{r}_1,\mathbf{r}_2,\ldots,\mathbf{r}_{N_{\text{e}}})
\quad = \quad B\, \Phi_0(\mathbf{r}_1,\mathbf{r}_2,\ldots,\mathbf{r}_{N_{\text{e}}})  
\quad + \quad B \sum_k g_k \sum_{i,j=1}^{N_{\text{e}}} 
\sum_{k_1'}\cdots
\sum_{k_{i-1}'}\sum_{k_{i+1}'}\cdots \sum_{k_{N_{\text{e}}}'}
\nonumber \\
&& \times \quad aLN
\varphi_{ck}(\mathbf{r}_i) \varphi_{vk}^*(\mathbf{r}_j) 
\; \varphi_{vk_1'}(\mathbf{r}_1)\ldots
\varphi_{vk_{i-1}'}(\mathbf{r}_{i-1}) 
\varphi_{vk}(\mathbf{r}_i)
\varphi_{vk_{i+1}'}(\mathbf{r}_{i+1})
\ldots \varphi_{vk_{N_{\text{e}}}'}(\mathbf{r}_{N_{\text{e}}})
\nonumber \\
&&\quad \times\quad \left<\text{vac}\right|\hat{v}_{k_{N_{\text{e}}}'}\ldots
\hat{v}_{k_{i+1}'}\;
\hat{v}_k \hat{v}^{\dagger}_k
\hat{c}_k\;
\hat{v}_{k_{i-1}'}  \ldots \hat{v}_{k_{1}'}\;
\hat{c}^{\dagger}_{k}
\hat{v}_{k}\; \hat{v}^{\dagger}_{k_1}\hat{v}^{\dagger}_{k_2} 
\ldots \hat{v}^{\dagger}_{k_{N_{\text{e}}}} \left| \text{vac}\right>\nonumber \\
&& \quad \quad + \quad (\text{contact term}),
\label{eq:monstre_bis}
\end{eqnarray}
where the last contact term is negligible unless two electrons touch.
Importantly, the e-h pair wave function 
\begin{equation}
\varphi_{ck}(\mathbf{r}) \varphi_{vk}^*(\mathbf{r}') 
= \chi^{cv}_k( \mathbf{r} - \mathbf{r}') 
\end{equation}
occurring in the second row of (\ref{eq:monstre_bis})
depends on
$\mathbf{r} - \mathbf{r}'$  
only, which allows to decouple the sums over
$\mathbf{r}_i$ and 
$\mathbf{r}_i - \mathbf{r}_j$, respectively.  
Then Supplementary Eq.~(\ref{eq:monstre_bis})
may be rearranged as
\begin{eqnarray}
&& \Psi_{\text{EI}}(\mathbf{r}_1,\mathbf{r}_2,\ldots,\mathbf{r}_{N_{\text{e}}})
\quad = \quad B\, \Phi_0(\mathbf{r}_1,\mathbf{r}_2,\ldots,\mathbf{r}_{N_{\text{e}}})  
\quad + \quad B \sum_{\ell=1}^{N_{\text{e}}}
aLN \sum_k g_k \;   \chi^{cv}_k(\mathbf{r}_{\ell})
\nonumber \\
&& \times \quad 
\sum_{k_1'}\cdots
\sum_{k_{i-1}'}\sum_{i=1}^{N_{\text{e}}}\sum_{k_{i+1}'}\cdots 
\sum_{k_{N_{\text{e}}}'}
\; \varphi_{vk_1'}(\mathbf{r}_1)\ldots
\varphi_{vk_{i-1}'}(\mathbf{r}_{i-1}) 
\varphi_{vk}(\mathbf{r}_i)
\varphi_{vk_{i+1}'}(\mathbf{r}_{i+1})
\ldots \varphi_{vk_{N_{\text{e}}}'}(\mathbf{r}_{N_{\text{e}}})
\nonumber \\
&&\quad \times\quad \left<\text{vac}\right|\hat{v}_{k_{N_{\text{e}}}'}\ldots
\hat{v}_{k_{i+1}'}\;
\hat{v}_k 
\hat{v}_{k_{i-1}'}  \ldots \hat{v}_{k_{1}'}\;
\hat{v}^{\dagger}_{k_1}\hat{v}^{\dagger}_{k_2} 
\ldots \hat{v}^{\dagger}_{k_{N_{\text{e}}}} \left| \text{vac}\right>
 \quad + \quad (\text{contact term}),
\label{eq:monstre_ter}
\end{eqnarray}
where, among all addenda of the mixed sum over momenta $k'$ and index
$i$, the only non-vanishing contributions are those
permutating the annihilation operators applied to $\left|0\right>$
that belong to the set
$\{\hat{v}_{k_1}, \hat{v}_{k_2}, \ldots, \hat{v}_{k_{N_{\text{e}}}}\}$.

The final result is
\begin{equation}
\Psi_{\text{EI}}(\mathbf{r}_1,\mathbf{r}_2,\ldots,\mathbf{r}_{N_{\text{e}}})
 =  B\left[ 1 +  \sum_{\ell= 1}^{N_{\text{e}}} 
\Phi_{\text{exc}}(\mathbf{r}_{\ell})
\right] \Phi_0(\mathbf{r}_1,\mathbf{r}_2,\ldots,\mathbf{r}_{N_{\text{e}}})
+ (\text{contact term}),
\label{eq:final}
\end{equation}
with the exciton wave function $\Phi_{\text{exc}}$ being defined as
\begin{equation}
\Phi_{\text{exc}}(\mathbf{r}) = aLN \sum_{\text{filled }k}
g_k\; \chi^{vc}_k(\mathbf{r}),
\end{equation}
where the sum over $k$ is limited to those levels that are filled in 
$\left|0\right>$
and $\mathbf{r}$ is the electron-hole distance.
Supplementary Eq.~(\ref{eq:final}) is a non trivial result, as it shows 
that the EI wave function in real space is the product of the Slater
determinant $\Phi_0$---a conventional fermionic state---times 
the sum over $\ell$ of bosonic wave functions 
$\Phi_{\text{exc}}(\mathbf{r}_{\ell})$---the exciton wave function 
integrated over the whole range of possible e-h distances.
The significance of $\Phi_{\text{exc}}$ relies on its
Fourier transform in reciprocal space,
$g_k$, which is the ratio of those variational factors
that solve the gap equation, $v_k$ and $u_k$. The 
gap equation may be regarded as the many-exciton
counterpart of the Bethe-Salpeter equation.

Supplementary Eq.~(\ref{eq:final}) should be compared with
Supplementary Eq.~(\ref{eq:VQMCpert}):
When no pairs of electrons are in contact, QMC and mean-field EI 
wave functions coincide apart from a normalization factor,
provided that $u(\mathbf{r},\mathbf{r}')=
2\Phi_{\text{exc}}(\mathbf{r}-\mathbf{r}')/N$.
When two electrons touch, say $\mathbf{r}_i=\mathbf{r}_j$, a discrepancy
arises, which is expected since
the QMC wave function enforces the cusp condition whereas 
the mean-field ansatz does not.

\section*{Detection of Peierls charge density wave 
through the order parameter $\varrho_{\text{Transl}}$}

The QMC analysis of main text introduces
the order parameter
$\varrho_{\text{Transl}}$ 
as a measure of the 
charge displacement between adjacent unitary cells along the tube axis.
If the ground state is a charge density wave (CDW) with 
period $2a$ (the characteristic wave vector is $q=\pi/a$), 
then the quantum average of $\varrho_{\text{Transl}}$ extrapolated to the
thermodynamic limit is finite.  
In this section we discuss whether the order parameter
$\varrho_{\text{Transl}}$ 
may also detect a Peierls CDW with nesting vector $q=2k_{\text{F}}$, 
the Fermi wave vector being located at Dirac point K.

\begin{figure}
\setlength{\unitlength}{1 cm}
\begin{picture}(14.5,8.0)
\put(0.0,-0.5){\includegraphics[trim=0cm 0cm 0cm 3cm,clip=true,width=2.7in]{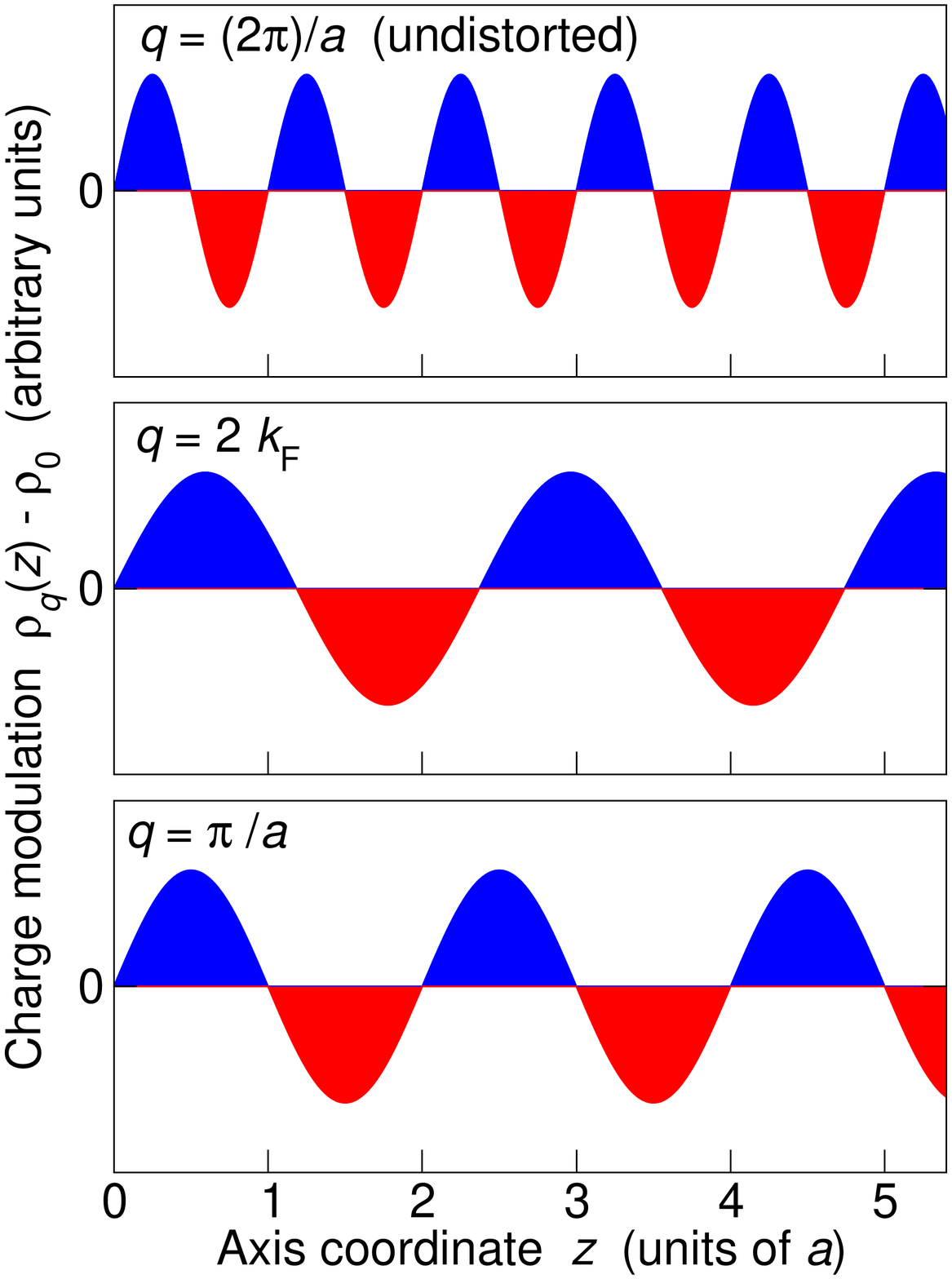}}
\put(7.1,1.0){\includegraphics[trim=0cm 0cm 0cm 0cm,clip=true,width=3.6in]{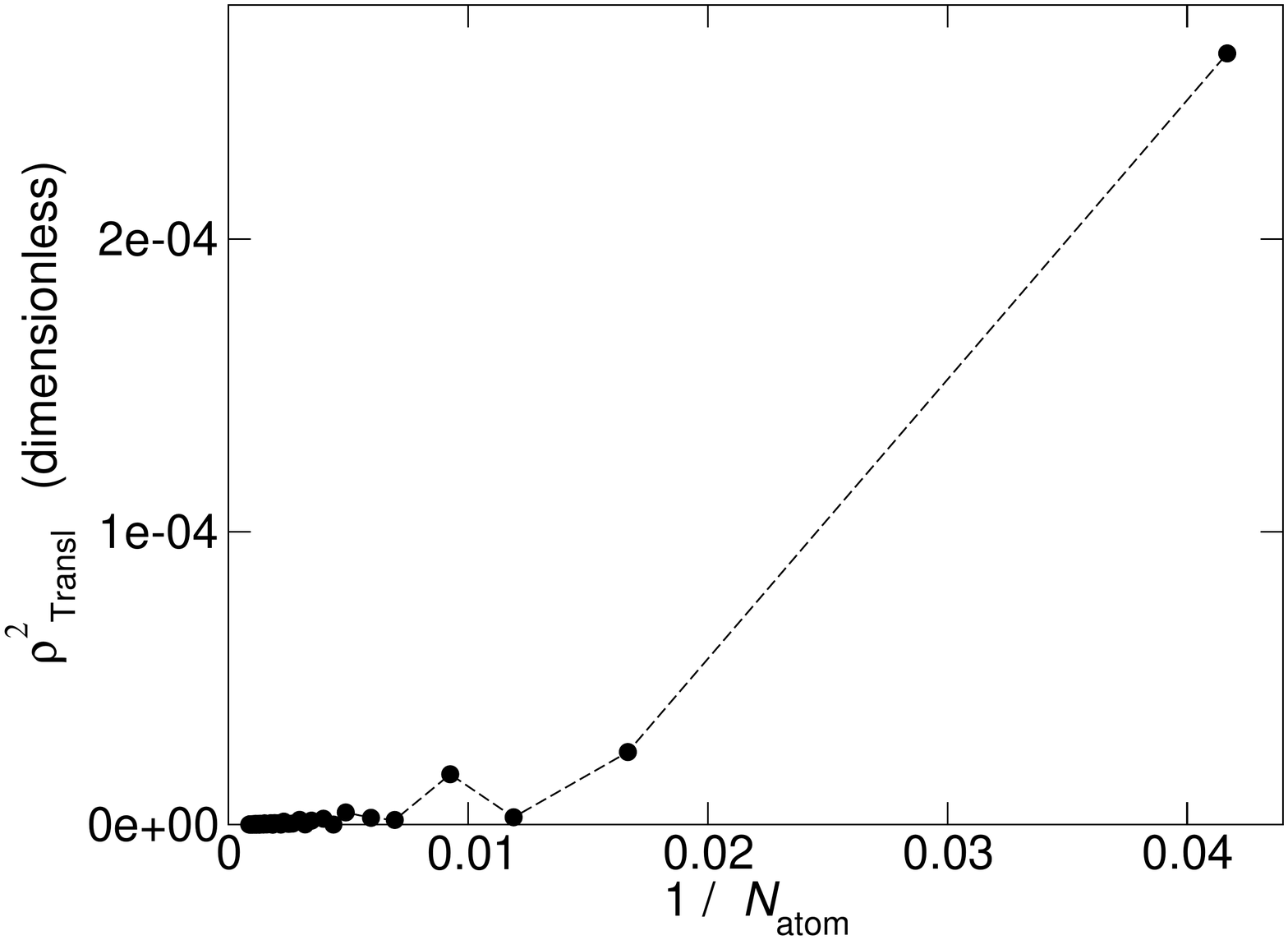}}
\put(6.0,5.9){\bf a}
\put(6.0,3.4){\bf b}
\put(6.0,0.9){\bf c}
\put(14.6,2.2){\bf d}
\end{picture}
\caption*{{\bf Supplementary Fig.~12} 
Model charge density wave ground state. {\bf a-c}
Model charge density along the axis, $n_q(z) - n_0$,
vs axial coordinate, $z$.  
The density is reference from its average value,
$n_0$, and the blue (red) colour stands for positive (negative)
charge deviation. The wave vector $q$ identifies the period
of the charge density wave as $(2\pi)/\left|q\right|$:
undistorted structure, $q=(2\pi)/a$ (panel a); 
charge density wave \`a la Peierls, $q=2k_{\text{F}}$, 
with $k_{\text{F}}\equiv$ K (panel b); dimerized charge density wave,
$q=\pi/a$ (panel c). {\bf d} Square 
order parameter, 
$\left(\varrho^{\text{model}}_{\text{Transl}}\right)^2$,
evaluated over the 
Peierls charge density wave model ground state,
vs inverse number of atoms in the supercell, $1/N_{\text{atom}}$.
Dashed lines are guides to the eye.
}
\end{figure}

A first issue is the commensurability of the QMC supercell
with respect to the period of Peierls CDW.
According to DFT calculation $k_{\text{F}} = 0.289(2\pi)/a$, hence  
$q=2k_{\text{F}}=0.422 (2\pi)/a$ (folded back to first Brillouin zone) 
and the period is 2.37 $a$ 
(Supplementary Fig.~12b). 
This implies that the size of the
commensurate supercell exceeds our computational capability.
On the other hand, the size of a smaller supercell may
approximately match a multiple of the Peierls CDW period.
This is the case e.g.~of a supercell made of seven units, 
whose length compares with three times the period, $7.11\, a$.

The key issue is the finite-size scaling of $\varrho_{\text{Transl}}$
averaged over the Peierls CDW ground state.
To gain a better understanding, we introduce a simple model for a generic 
CDW. The charge density profile, $n_q(z)$,
is a sinusoidal modulation of wave vector $q$ along the axis $z$,
\begin{equation}
n_q(z) = n_{\text{mod}}\sin{qz} + n_0,
\label{eq:CDW_model}
\end{equation}
where  
$n_{\text{mod}}$ 
is the modulation amplitude, 
$n_0$ is the homogeneous background, and we ignore
the relaxation of the ground state occurring in a finite-size supercell. 
The order parameter $\varrho^{\text{model}}_{\text{Transl}}$  
that fits to the model (\ref{eq:CDW_model}) is
\begin{equation}
\varrho^{\text{model}}_{\text{Transl}} = \frac{1}{N_{\text{cell}}}
\sum_{\ell=1}^{N_{\text{cell}}}
(-1)^{\ell-1} \int_{a(\ell-1)}^{a\ell}\!\!\!dz \,
\left[ n_q(z) - n_0\right], 
\label{eq:rho_Transl}
\end{equation}
where 
$N_{\text{cell}}$ 
is a number of unitary cells such that
$N_{\text{cell}}a$ is approximately commensurate
with the CDW period, $2\pi/q$. 
If $N_{\text{cell}}$ is even, then, except for a prefactor,
$\varrho^{\text{model}}_{\text{Transl}}$
is equivalent to $\varrho_{\text{Transl}}$  
as defined in the main text.
 
The extrapolated value of
$\varrho^{\text{model}}_{\text{Transl}}$
in the thermodynamic limit, $N_{\text{cell}}\rightarrow \infty$,
is trivial in two cases.
For the undistorted structure,
$\varrho_{\text{Transl}}=0$ as the integral of 
$n_q-n_0$ over the unitary cell vanishes. 
This is illustrated in  
Supplementary Fig.~12a, where the blue (red) colour stands for positive
(negative) charge deviation, $n_q(z)-n_0$.
Second, for
the dimerized CDW of period $2a$, which is discussed in the
main text, 
any cell with $N_{\text{cell}}$ 
even is commensurate 
and hence $\varrho^{\text{model}}_{\text{Transl}} =
2an_{\text{mod}}/\pi$,
the integral of 
$n_q - n_0$ over the unitary cell exhibiting alternate sign 
between adjacent cells
(Supplementary Fig.~12c).
In the following, in order to compare with the VQMC
extrapolated order parameter $\varrho_{\text{AB}}$
discussed in the main text,
we take
$n_{\text{mod}}a/\pi$ = $\varrho_{\text{AB}}/2$ = 0.00825.

We now focus on the Peierls case of nesting vector $q=2k_{\text{F}}$
(Supplementary Fig.~12b).
We assume that
$N_{\text{cell}}$ takes only those integer values
closest to $(2.37) m$, with $m=1,2,\ldots$, which ensures
that supercell and CDW periods are
approximately commensurate.
As illustrated by Supplementary Fig.~12d,
$\left(\varrho^{\text{model}}_{\text{Transl}} 
\right)^2$ 
exhibits a complex, non-monotonic dependence on the
inverse number of atoms  
before vanishing as $1/N_{\text{atom}}\rightarrow 0$
[here $N_{\text{atom}}=12 N_{\text{cell}}$
as the (3,3) nanotube has twelve atoms per cell].
This trend should be compared with the perfectly linear 
vanishing behavior exhibited by $\varrho^2_{\text{Transl}}$ 
in Fig.~4a of main text.
We infer that, if the Peierls CDW 
were the actual ground state,
than $\varrho^2_{\text{Transl}}$ evaluated through QMC
would show some deviation from linearity, which is not observed.
In conclusion, we rule out the Peierls CDW ground state.

\newpage

\begin{table}[h]
\begin{center}
\begin{tabular}{ c c c c }
\hline
 C atom label  & $x$ ({\AA}) & $y$ ({\AA}) & $z$ ({\AA}) \\
\hline
1  &  {\tt 2.101836417}   & {\tt 0.002803388} & {\tt -1.230783688}  \\
2 &  {\tt 1.607960625}    & {\tt 1.353495674} & {\tt -1.230783688}   \\
3 &  {\tt 1.048473311}    & {\tt 1.821401270} & {\tt 0.000000000} \\
4 &  {\tt -0.368220898}   & {\tt 2.068523780} & {\tt 0.000000000} \\
5 &  {\tt -1.053234936}   & {\tt 1.818369640} & {\tt -1.230783688} \\
6 &  {\tt -1.976339544}   & {\tt 0.715641791} & {\tt -1.230783688} \\
7 &  {\tt -2.102053467}   & {\tt -0.002752850} & {\tt 0.000000000} \\
8 &  {\tt -1.607747246}   & {\tt -1.353292444} & {\tt 0.000000000} \\
9 &  {\tt -1.048327785}   & {\tt -1.821179199} & {\tt -1.230783688} \\
10 & {\tt 0.368284950}   & {\tt  -2.068831149} & {\tt -1.230783688} \\
11 & {\tt 1.053260216}   & {\tt  -1.818547251} & {\tt  0.000000000} \\
12 & {\tt 1.976108358}   & {\tt  -0.715632650} & {\tt  0.000000000} \\
\hline
\end{tabular}
\end{center}
\caption*{{\bf Supplementary Table 1} 
Equilibrium coordinates
of the twelve atoms making the unitary cell of the (3,3) carbon
nanotube, after structural DFT optimization.
The cell size along the tube axis, parallel to $z$, is 2.461566 {\AA}.
}
\end{table}


\newpage

\section*{References}





\begin{addendum}
 \item This work was supported in part by European Union H2020-EINFRA-2015-1
  programme under grant agreement No.~676598 project
 `MaX--Materials Design at the Exascale'.
 S.S.~acknowledges computational resouces provided through
 the HPCI System Research Project No.~hp160126 on the K computer
 at RIKEN Advanced Institute for Computational Science.
 D.V., E.M.~\&~M.R. acknowledge PRACE for awarding them access to
 resource Marconi based in Italy at CINECA (Grant No.~Pra14\_3622).
 \item[Contributions] M.R. and E.M. initiated this project,
D.V., E.M., and M.R. designed a comprehensive strategy
to tackle the instability problem by means of different methods,
D.V. developed the many-body perturbation theory
calculations and analysis, 
D.V. and D.S. optimized the Yambo code for the calculation
in the presence of the magnetic field, S.S. and M.B. developed 
the quantum Monte Carlo calculations and analysis, 
M.R. developed the effective-mass 
theory and wrote the paper, all authors contributed 
to the analysis of data and critically discussed the paper.
 \item[Competing Interests] The authors declare that they have no
 competing financial interests.
 \item[Correspondence] Correspondence and requests for materials
 should be addressed to M.R.~(email: massimo.rontani@nano.cnr.it).
\end{addendum}


\end{document}